\documentclass[useAMS,usenatbib,usegraphicx,onecolumn]{mn2e}

\usepackage{amssymb,amsfonts,amsmath,amstext,amsgen,amsopn,amsxtra,indentfirst,lscape,times,rotating}

\title[Atmospheric circulation of tidally-locked exoplanets]{Atmospheric circulation of tidally-locked exoplanets: \\ a suite of benchmark tests for dynamical solvers}
\author[Heng, Menou \& Phillipps]{Kevin Heng$^{1,2}$\thanks{E-mail: kheng@phys.ethz.ch (KH)}, 
Kristen Menou$^{3,4}$\thanks{E-mail: kristen@astro.columbia.edu (KM)} 
and Peter J. Phillipps$^{5}$\thanks{E-mail: Peter.Phillipps@noaa.gov (PJP)}\\
$^{1}$Zwicky Fellow, ETH Z\"{u}rich, Institute for Astronomy, Wolfgang-Pauli-Strasse 27, CH-8093, Z\"{u}rich, Switzerland\\
$^{2}$Frank \& Peggy Taplin Member, Institute for Advanced Study, School of Natural Sciences, Einstein Drive, Princeton, NJ 08540, U.S.A.\\
$^{3}$Department of Astronomy, Columbia University, 550 West 120th Street, New York, NY 10027, U.S.A.\\
$^{4}$Perimeter Institute for Theoretical Physics, 31 Caroline Street, North Waterloo, Ontario, N2L 2Y5, Canada\\
$^{5}$Geophysical Fluid Dynamics Laboratory, 201 Forrestal Road, Princeton, NJ 08540, U.S.A.}

\begin{document}

\date{Submitted 2010 October 6.  Re-submitted 2010 December 24.  Accepted 2011 January 10.}

\pagerange{\pageref{firstpage}--\pageref{lastpage}} \pubyear{2010}

\maketitle

\label{firstpage}

\begin{abstract}
The rapid pace of extrasolar planet discovery and characterization
is legitimizing the study of their atmospheres via three-dimensional
numerical simulations.  The complexity of atmospheric modelling and
its inherent non-linearity, together with the limited amount of data
available, motivate model intercomparisons and benchmark tests.  In
the geophysical community, the Held-Suarez test is a standard
benchmark for comparing dynamical core simulations of the Earth's
atmosphere with different solvers, based on statistically-averaged
flow quantities.  In the present study, we perform analogues of the
Held-Suarez test for tidally-locked exoplanets with the GFDL-Princeton
{\it Flexible Modeling System} \texttt{(FMS)} by subjecting both the
spectral and finite difference dynamical cores to a suite of tests,
including the standard benchmark for Earth, a hypothetical
tidally-locked Earth, a ``shallow'' hot Jupiter model and a ``deep''
model of HD 209458b.  We find qualitative and quantitative agreement
between the solvers for the Earth, tidally-locked Earth and shallow hot
Jupiter benchmarks, but the agreement is less than satisfactory for
the deep model of HD 209458b.  Further investigation reveals that
closer agreement may be attained by arbitrarily adjusting the values
of the horizontal dissipation parameters in the two solvers, but it
remains the case that the \emph{magnitude} of the horizontal
dissipation is not easily specified from first principles.
Irrespective of radiative transfer or chemical composition
considerations, our study points to limitations in our ability to
accurately model hot Jupiter atmospheres with meteorological solvers
at the level of ten percent for the temperature field and several tens
of percent for the velocity field.  Direct wind measurements should
thus be particularly constraining for the models. Our suite of
benchmark tests also provides a reference point for researchers
wishing to adapt their codes to study the atmospheric circulation
regimes of tidally-locked Earths/Neptunes/Jupiters.

\end{abstract}


\section{Introduction}

The nascent field of extrasolar planets is rapidly expanding, as
evidenced by the flood of discoveries made in the past decade alone
\citep[e.g.,][]{us07,sd10}.  The abundance of exoplanetary data has
legitimized several new fields of inquiry.  Among these is the
theoretical study of exoplanetary atmospheres via detailed numerical
simulations (for reviews see \citealt{showman08,showman10}), which
describe the atmospheric dynamics, its radiative transfer, as well as ---
in principle --- the chemistry and the cloud physics
\citep{sg02,cho03,cho08,cs05,cs06,ll08,mr09,showman09,burrows10,rm10,tc10,dd10}.
The complexity of atmospheric modelling and its inherent non-linearity
motivate clean comparisons between studies that either utilize
different methods of solution or even implement the same methods
differently \citep{held05}.

In the geophysical fluid dynamics community, it has long been
suggested by \cite{hs94} that a useful comparison is between
``dynamical cores'', which are codes that deal with the essential
dynamics of an atmosphere and omit details such as radiative transfer.
In a pair of benchmark calculations, \cite{hs94} demonstrated that
Earth-like simulations using two different methods --- spectral and
finite difference solvers --- produced quantitatively similar profiles
of the temporally-averaged, zonal-mean temperature and zonal wind, as
functions of vertical height (or pressure).  Attempts have been made
(e.g., \citealt{rm10} versus \citealt{cs06}) to compare hot Jupiter
models, but these were performed via different simulation platforms.
In this study, we generalize Held-Suarez-type benchmarks to include
tidally-locked exoplanets, using a single simulation platform --- namely
the dynamical cores of the {\it Flexible Modeling System}
\texttt{(FMS)}.

Our main finding is that while we can achieve qualitative and
quantitative agreement for the Earth (and shallow hot Jupiter) tests, noticeable differences
appear when we simulate the deep atmospheric circulation of the hot
Jupiter HD 209458b.  Closer agreement may be attained by specifying
arbitrary values for the horizontal dissipation parameters --- by
trial and error --- but it remains the case that the \emph{magnitude}
of the horizontal dissipation cannot be rigorously specified.
Dynamical uncertainties at the level of $\gtrsim 10\%$ therefore exist both between simulations utilizing different methods of solutions and also within the same method of solution,
which may ultimately have implications for studies attempting to match
observed versus simulated atmospheres of extrasolar planets.

Operationally, we implement both the spectral and finite difference
cores of the \texttt{FMS} and subject them to a battery of tests,
including the Held-Suarez benchmark for Earth (\S\ref{subsect:hs94}),
a hypothetical tidally-locked Earth (\S\ref{subsect:tidalearth};
\citealt{ms10}), a ``shallow'' hot Jupiter model
(\S\ref{subsect:mr09}; \citealt{mr09}) and a ``deep'' model for HD
209458b (\S\ref{subsect:hd209458b}; \citealt{cs05,cs06,rm10}).  In
\S\ref{sect:equations}, we discuss the governing equations handled by
meteorological solvers such as the \texttt{FMS}.  In \S\ref{sect:fms},
we briefly describe the \texttt{FMS}.  Our results are collectively
stated in \S\ref{sect:tests} and we discuss their implications in
\S\ref{sect:discussion}.  Table \ref{tab:params} lists the parameters
and commonly used symbols in our study, while Table
\ref{tab:resolution} describes the resolutions of the simulations.
Appendices \ref{append:levels}, \ref{append:fits} and \ref{append:efold} contain technical
details and useful fitting functions relevant to simulating the
atmospheric circulation on the hot Jupiter HD 209458b.

\begin{table*}
\centering
\scriptsize
\caption{Table of Parameters and Commonly Used Symbols}
\label{tab:params}
\begin{tabular}{lccccc}
\hline\hline
\multicolumn{1}{c}{Quantity (Units)} & \multicolumn{1}{c}{Description} & \multicolumn{1}{c}{Earth (Held-Suarez)} & \multicolumn{1}{c}{Earth (Menou-Rauscher)} & \multicolumn{1}{c}{Shallow Hot Jupiter (Menou-Rauscher)} & HD 209458b \\
\hline
\vspace{2pt}
Resolution & conventional shorthand & T63/G72 & T63/G72$^\spadesuit$ & T63/G72$^\spadesuit$ & T63/G72$^\spadesuit$ \\
$N_v$ & vertical resolution & 20 & 20$^\spadesuit$ & 20$^\spadesuit$ & 33$^\diamondsuit$ \\
$\Delta t$ (s) & computational time step & 1200 & 1200$^\clubsuit$ & 120$^\clubsuit$ & 120$^\clubsuit$ \\
$t^{-1}_\nu$ (s$^{-1}$) & hyperviscous dissipation rate$^\dagger$ & $1.15741 \times 10^{-4}$ & $1.15741 \times 10^{-4}$ & $0.33422537$ & $0.32785918$ \\
$t_\nu$ & hyperviscous dissipation time$^\dagger$ & 0.1 day & 0.1 day & $10^{-5}$ hot Jupiter day & $10^{-5}$ HD 209458b day \\
${\cal K}$ & horizontal mixing coefficient$^\ast$ & 0.35 & 0.35 & 0.35 & 0.1--1 \\
$\Theta$ & longitude & 0--360$^\circ$ & 0--360$^\circ$ & 0--360$^\circ$ & 0--360$^\circ$ \\ 
$\Phi$ & latitude & -90$^\circ$--90$^\circ$ & -90$^\circ$--90$^\circ$ & -90$^\circ$--90$^\circ$ & -90$^\circ$--90$^\circ$ \\
$P_s$ (bar) & mean surface pressure & 1 & 1 & 1 & 220 \\
$\tau_{\rm fric}$ (day) & Rayleigh friction time & 1 &  1 & $\infty$ & $\infty$ \\
$\sigma_b$ & planetary boundary layer & 0.7 & 0.7$^\clubsuit$ & --- & --- \\
$\tau_{\rm rad}$ (day) & Newtonian relaxation time & 4--40 & 15 & $\pi/\Omega_p \approx 1.731$ & equation (\ref{eq:trad_hd209458b}) \\
$T_{\rm init}$ (K) & initial temperature & 264$^\clubsuit$ & 264$^\clubsuit$ & 1800$^\clubsuit$ & 1759 \\
\hline
$z_{\rm stra}$ (m) & height of tropopause & --- & $1.2 \times 10^4$ & $2 \times 10^6$ & $\ddagger$ \\
$\sigma_{\rm stra}$ & location of tropopause & --- & $\approx 0.22$ & $\approx 0.12$ & $\ddagger$ \\
$T_{\rm surf}$ (K) & surface temperature at equator & 315 & 288 &  1600 & $\ddagger$ \\
$T_{\rm stra}$ (K) & stratospheric temperature & 200 & 212 & 1210 & $\ddagger$ \\
$\Delta T_{\rm EP}$ (K) & equator-to-pole temperature difference & 60 & 60 & 300 & $\ddagger$ \\
$\Delta T_{\rm stra}$ (K) & tropopause temperature increment & --- & 2 & 10 & $\ddagger$ \\
$\Delta T_z$ (K) & stability parameter & 10 & --- & --- & $\ddagger$ \\
\hline
$c_p$ (J kg$^{-1}$ K$^{-1}$) & specific heat capacity at constant pressure & 1004.64  & 1004.64 & 13226.5 & 14308.4 \\
${\cal R}$ (J kg$^{-1}$ K$^{-1}$) & ideal gas constant & 287.04 & 287.04 &  3779 & 4593 \\
$\kappa \equiv {\cal R}/c_p$ & --- & 2/7 & 2/7 &  2/7 & 0.321 \\
$\Omega_p$ (s$^{-1}$) & planetary rotation rate & $7.292 \times 10^{-5}$ & $7.292 \times 10^{-5}$ &  $2.1 \times 10^{-5}$ & $2.06 \times 10^{-5}$ \\
$g_p$ (m s$^{-2}$) & planetary surface gravity & 9.80 & 9.80 & 8 & 9.42 \\
$R_p$ (m) & planetary radius & $6.371 \times 10^6$ & $6.371 \times 10^6$ & $10^8$ & $9.44 \times 10^7$ \\
\hline
\hline
\end{tabular}\\
Note: unless otherwise stated, ``day'' refers to an Earth day (86400 seconds). \\
$\dagger$: Spectral models only.  $\ast$: Finite difference models only.\\
$\spadesuit$: Value(s) used is different from in original publication.\\
$\clubsuit$: Value not explicitly specified in original publication.\\
$\diamondsuit$: Vertical levels are logarithmically spaced.\\
$\ddagger$: Thermal forcing of HD 209458b is given by equation (\ref{eq:forcing_hd209458b}).
\end{table*}

\section{The Primitive Equations of Meteorology}
\label{sect:equations}

The study of (terrestrial) meteorology involves solving the Navier-Stokes and thermodynamic equations on a rotating sphere (e.g., Chapter 14 of \citealt{kundu04}).  Such an endeavour is usually inefficient or even intractable without invoking some simplifications, which results in a set of equations known as the ``primitive\footnote{From a historical viewpoint, the term ``primitive" is a misnomer, since it means ``full" rather than ``simple" (see Chapter 3.2 of \citealt{wp05}).} equations of meteorology'' (e.g., \citealt{sma63,sma64}; Chapter 3 of \citealt{wp05}; Chapter 2 of \citealt{vallis06}).  The first simplification involves the assumption of vertical hydrostatic equilibrium,
\begin{equation}
\frac{\partial P}{\partial z} = - \rho g ~\Longleftrightarrow ~\frac{\partial \phi}{\partial \ln P} = - {\cal R} T,
\label{eq:hydrostatic}
\end{equation}
where $P$ denotes the pressure, $z$ is the vertical/radial coordinate, $\rho$ is the mass density of the fluid, $g$ is the acceleration due to the gravity of the planet, $\phi \equiv gz$ is the geopotential, ${\cal R}$ is the ideal gas constant and $T$ is the temperature.  The hydrostatic approximation filters out vertically propagating sound waves, but allows for vertically propagating gravity and Rossby waves as well as horizontally propagating waves in general.  On large scales, hydrostatic equilibrium is a good approximation because the vertical pressure scale height $H$ is much less than the planetary radius $R_p$,
\begin{equation}
\frac{H}{R_p} = \frac{k_{\rm B} T}{\bar{m} g R_p} \approx 6 \times 10^{-3} \left(\frac{T}{1000 \mbox{ K}}\right) \left(\frac{\bar{m}}{2m_{\rm H}} \frac{g}{10 \mbox{ m s}^{-2}} \frac{R_p}{R_{\rm J}} \right)^{-1},\\
\end{equation}
where $k_{\rm B}$ is the Boltzmann constant, $\bar{m}$ is the mean molecular mass, $m_{\rm H}$ is the mass of a hydrogen atom and $R_{\rm J} \approx 71492$ km is the (mean) radius of Jupiter.  Such an assumption precludes the explicit treatment of small-scale, three-dimensional turbulence, which may be a non-negligible source of dissipation \citep{goodman09,lg10}.

Consider the quantity $r = R_p + z$.  The second approximation then replaces $r$ with $R_p$ in the equations of motion except where the former is used as the differentiating argument.  The third approximation neglects the Coriolis terms in the horizontal momentum equation involving the vertical velocity.  These approximations are collectively made such that angular momentum and energy conservation are ensured \citep{vallis06}.

Let $v_\Theta$ and $v_\Phi$ denote the zonal (east-west) and meridional (north-south) components of the flow, respectively.  The equations of momentum and mass conservation are
\begin{equation}
\begin{split}
&\frac{D v_\Theta}{Dt} = 2\Omega v_\Phi \sin\Phi + \frac{v_\Theta v_\Phi \tan\Phi}{R_p} - \frac{1}{\rho R_p \cos\Phi} \frac{\partial P}{\partial \Theta},\\
&\frac{D v_\Phi}{Dt} = -2\Omega v_\Theta \sin\Phi - \frac{v^2_\Theta \tan\Phi}{R_p} - \frac{1}{\rho R_p} \frac{\partial P}{\partial \Phi},\\
&\frac{\partial}{\partial P} \left( \frac{D P}{Dt} \right) + \nabla .\vec{v} = 0,\\
\end{split}
\label{eq:primitive}
\end{equation}
where $\vec{v}$ denotes the velocity vector.  In a departure from traditional notation, we denote the latitude and longitude by $\Phi$ and $\Theta$, respectively.  Equations (\ref{eq:hydrostatic}) and (\ref{eq:primitive}) are augmented by the first law of thermodynamics,
\begin{equation}
\frac{D T}{Dt} = \frac{\kappa T}{P} \frac{D P}{Dt} + \frac{Q}{c_p},
\label{eq:thermodynamics}
\end{equation}
where $\kappa \equiv {\cal R}/c_p$ and $c_p$ denotes the specific heat capacity at constant pressure.  The diabatic heating is denoted by $Q$.  For an ideal gas, $c_p = c_v + {\cal R}$, where $c_v$ is the specific heat capacity at constant volume.  \cite{goodman09} has remarked that equations (\ref{eq:hydrostatic}), (\ref{eq:primitive}) and (\ref{eq:thermodynamics}) collectively describe a frictionless heat engine, where no viscous terms exist to convert mechanical energy back into heat.

Solving the equations explicitly with $z$ is computationally awkward, especially when dealing with non-uniform topography.  Instead, $P$ is used in place of $z$ such that the temporal derivative following the flow is
\begin{equation}
\frac{D}{Dt} = \frac{\partial}{\partial t} + \frac{v_\Theta}{R_p \cos\Phi} \frac{\partial}{\partial \Theta} + \frac{v_\Phi}{R_p} \frac{\partial}{\partial \Phi} + \frac{D P}{Dt} \frac{\partial}{\partial P}.
\end{equation}
In addition, the pressure is normalized by the instantaneous surface pressure $P_s$, such that 
\begin{equation}
\sigma \equiv \frac{P}{P_s}.
\end{equation}
This is also known as Phillips' $\sigma$-coordinate and was designed to deal with mountainous terrain in geophysical calculations \citep{p57}.  By definition, the $\sigma=1$ level tracks the (exo)planet's orography (if any).

\section{The GFDL-Princeton {\it Flexible Modeling System}}
\label{sect:fms}

The {\it Flexible Modeling System} \texttt{(FMS)} is an open source, parallel simulation platform developed at the Geophysical Fluid Dynamics Laboratory (GFDL) of Princeton University.  The \texttt{FMS} has three core options: finite difference, spectral and finite volume.  In this study, we implement the \texttt{Memphis} release of \texttt{FMS} and utilize both the spectral and finite difference (``B-grid'')\footnote{We note that the B-grid finite difference scheme is an old one and is known to be less accurate than the more commonly-used ``C-grid" scheme, e.g., as employed by \cite{hs94}, \cite{cs05,cs06} and \cite{showman09}.} dynamical cores.  As the \texttt{FMS} utilizes MKS units, some of the discussion in the paper will follow suit.  In this section, we describe some salient features of the \texttt{FMS}.  Readers interested in more technical details may consult \texttt{http://www.gfdl.noaa.gov/fms}.  For an overview of the various simulation platforms which are publicly available, please refer to Chapter 5 of \cite{wp05}.

\subsection{Spectral Core}
\label{subsect:spectralcore}

In the spectral\footnote{Strictly speaking, the code is pseudo-spectral because only the linear terms in the governing equations are transformed to the spectral domain, while the non-linear terms are computed on a finite difference grid.  This statement is independent of the method of solution for the vertical coordinate.} dynamical core of the \texttt{FMS}, the hydrodynamic variables are described as a sum of spherical harmonics truncated at $N_{\rm h}$ terms \citep{gs82}.  Triangular truncation is used in our implementation of the \texttt{FMS}, such that the truncation is rotationally symmetric --- a function and its rotated counterpart are both expressible within this truncation (see \S13.6.2 of \citealt{holton}).  On a sphere, the number of zonal and meridional waves retained are $N_{\rm h}$ and $N_{\rm h} + 1$, respectively, to prevent aliasing.  The corresponding number of longitudinal grid points ($N_{\rm lon}$) is always twice that of the latitudinal grid points ($N_{\rm lat}$).  Domain decomposition is 1D in the spectral core: the number of processors allocatable to computing a given model is $N_{\rm lat}/2$.

A key aspect of any spectral model is spectral blocking, which is the
accumulation of numerical noise --- specifically, enstrophy --- at the
smallest grid scales, since spectral codes are intrinsically
non-dissipative \citep{s94}.  Numerical ``hyperviscosity'' is needed
to mimick enstrophy dissipation at the smallest length scales,
analogous to a two-dimensional turbulent cascade \citep{shapiro71}.  The hyperviscous
term takes the form,
\begin{equation}
{\cal D}_{\rm hyper} = -\nu \left(-1\right)^{n_\nu} \nabla^{2n_\nu} \left( \nabla \times \vec{v} \right)_z,
\end{equation}
where $n_\nu$ is the hyperviscosity damping order and $( \nabla \times \vec{v} )_z$ is the relative vorticity.  Following \cite{hs94}, \cite{mr09} and \cite{rm10}, we adopt $n_\nu=4$.  Within the \texttt{FMS}, one may either specify the hyperviscosity coefficient ($\nu$) or the dissipation rate ($\sim \nu \nabla^{2n_\nu}$); we will discuss this issue further in \S\ref{subsect:damping}.  The spectral core has an optional switch to ensure global energy conservation, which we enforce for all of our simulations.

The conventional shorthand notation used to describe the resolution of the spectral models is T$N_{\rm h}$L$N_{\rm v}$, where $N_{\rm h}$ is the horizontal resolution while $N_{\rm v}$ is the number of vertical levels.  The fiducial resolution we will adopt for our spectral simulations is T63, which corresponds to $N_{\rm lon} = 192$ and $N_{\rm lat} = 96$.  Finite differencing is used for the vertical grid by employing the Simmons-Burridge scheme \citep{sb81}.  For example, the lowest layer modelled within such a scheme has $\sigma=0.95$--1; the boundaries between the layer are called the ``half levels" (i.e., $\sigma=0.95$ and 1).  A noteworthy feature of the Simmons-Burridge scheme is that the simulation output is not exactly at the midpoint between the half levels (i.e., not at $\sigma=0.975$ in this example).  Therefore, it should be noted that when we present our results, we usually quote the pressure level $P$ as the larger of the pair of half level values (e.g., for $P=0.95$--1 bar layer, we label it ``$P=1$ bar").

A key advantage of the spectral method described here is that it does not require special (damping) treatment at the poles.  This is not the case for the finite difference core.

\subsection{Finite Difference Core}

The finite difference dynamical core of the \texttt{FMS} uses a ``Arakawa B-grid'' (see Chapter 4.2 of \citealt{wp05}) for the horizontal coordinates, which belongs to a family of finite difference grids where the temperature and velocity are solved at staggered points \citep{wyman96,anderson04}.  The vertical grid uses a hybrid $\sigma$-$P$ coordinate system; the labelling of the different model layers again follows the larger of the pair of half level values (see \S\ref{subsect:spectralcore}).  Finite differencing is used for both the horizontal and vertical grids.  Analogous to the case of the spectral core, small-scale noise accumulates in the B-grid \citep{shapiro70} and has to be damped via a ``horizontal mixing''\footnote{Also termed ``horizontal diffusion".} algorithm \citep{rvs80}.  In the finite difference core, the second-order operator for horizontal mixing is defined as
\begin{equation}
\hat{{\cal H}} \left( {\cal F} \right) = \frac{1}{A_{\cal F} ~\Delta P_j} \sum_{i} \hat{{\cal I}}_i \left[ {\cal K}_i ~\hat{{\cal I}}_i \left({\cal F}\right) \right],
\label{eq:hmix}
\end{equation}
where ${\cal F}$ denotes the temperature or zonal/meridional velocity components (i.e., $T$ or $\vec{v}$), $A_{\cal F}$ is the area of each grid box (for either the temperature or velocity fields),
\begin{equation}
{\cal K}_i \equiv {\cal K} ~\Delta_i ~\hat{{\cal J}}_i \left(A_{\cal F}\right) ~\hat{{\cal J}}_i \left(\Delta P_j\right),
\end{equation}
and the index $i=\Phi, \Theta$.  Denoting an arbitrary quantity by ${\cal Q}$, the operators $\hat{{\cal I}}_i ({\cal Q})$ and $\hat{{\cal J}}_i ({\cal Q})$ yield the difference and average between adjacent grid points along the $i$-axis, respectively.  The difference in pressure between half-levels at an index $j$ is
\begin{equation}
\Delta P_j \equiv P_{j+1/2} - P_{j-1/2}.
\end{equation}
The quantity $\Delta_i$ is a constant that describes the strength of the horizontal mixing with latitude and must satisfy the numerical stability condition: $\Delta_i {\cal K} \le 1/8$.  It is important to note that the operator defined in equation (\ref{eq:hmix}) is essentially a Laplacian and is applied \emph{twice} to the temperature and velocity fields,
\begin{equation}
\frac{\partial {\cal F}}{\partial t} \approx - \frac{1}{\Delta t} \hat{{\cal H}}\left[ \hat{{\cal H}} \left( {\cal F} \right) \right],
\end{equation}
where $\Delta t$ denotes the time step, implying that the horizontal mixing scheme is fourth order in nature.\footnote{The \texttt{Memphis} release of the \texttt{FMS} uses defaults of second and fourth order for the wind and temperature horizontal mixing schemes, respectively.  We have performed two separate suites of simulations where the wind scheme is set to second or fourth order and find little difference between the Held-Suarez statistics generated.}  The horizontal mixing coefficient has a range of values of $0 \le {\cal K} \le 1$.  Its default value within the \texttt{FMS} is ${\cal K}=0.35$, which we will adopt throughout unless otherwise stated.  Damping is increased for $\vert \Phi \vert > 80^\circ$ but is uniform with longitude.

The common problem faced by any finite difference code which solves the fluid equations on a sphere using the longitude-latitude coordinate system is that, for numerical convergence to be attained, the minimum time step needed is proportional to the zonal grid spacing $\Delta \Theta$ via the Courant-Fredricks-Levy (CFL) condition.  This implies that $\Delta t \rightarrow 0$ towards the poles.  For example, \cite{dd10} truncate their latitudinal grid at $\Phi = \pm 70^\circ$ in their simulations.  To alleviate this problem, a technique known as ``polar filtering" is applied at high latitudes to damp the shortest resolvable waves such that a non-zero time step can be taken \citep{shapiro71,ass72,tb83}.  Like in the spectral core, the finite difference core has a switch to ensure global energy conservation, which we set to ``on'' for all of our simulations.\footnote{We note that the application of horizontal mixing and polar filtering result in small violations to the conservation of mass and energy.}

The shorthand notation used for resolution is G$N_{\rm h}$L$N_{\rm v}$
where $N_{\rm h}$ now refers to \emph{half} of the number of grid
points in longitude (i.e., around a latitude circle).  Alternatively,
one can use the notation N$N^\prime_{\rm h}$L$N_{\rm v}$ where
$N^\prime_{\rm h}$ is the number of latitudinal points between the
north/south pole and equator.  Domain decomposition is 2D in the
finite difference core.  The fiducial resolution for our finite
difference simulations is G72/N45, which corresponds to $N_{\rm
  lon}=144$ and $N_{\rm lat}$=90.

Finally, we note that an alternative approach within grid-based
methods is to adopt a ``cubed-sphere'' grid, which circumvents the
problems at the poles at the price of dealing with a non-orthogonal
grid (e.g., \citealt{ad04,showman09}).

\subsection{Horizontal Dissipation}
\label{subsect:damping}

It is important to note that the horizontal dissipation schemes
described above are reasonably well-motivated but nevertheless
non-rigourous.  Hyperviscosity and horizontal mixing are numerical tools unsupported by any fundamental physical theory,
yet are routinely used by research groups studying terrestrial and
exoplanetary atmospheric circulation.  There is no rigourous way to
choose their magnitudes.  \emph{On Earth, the magnitude of horizontal
dissipation can be calibrated on the basis of the known flow, but this
is not (yet) --- and may never be --- the case for extrasolar planets.}
The use of horizontal dissipation is related to the notion that turbulent cascades of hydrodynamical quantities (energy,
enstrophy) are only partially modelled.  As such, both $\nu$ and
${\cal K}$ should be regarded as free parameters in any model of
atmospheric circulation (e.g., \citealt{s94}).

In practice, the pragmatic aim is to dissipate small-scale numerical noise within a
fraction of a planetary rotation (i.e., one day on a tidally-locked planet).
For the spectral core, the dissipation time on the scale of a
resolution element is
\begin{equation}
t_\nu \sim \frac{1}{\nu} \left( \frac{R_p}{N_{\rm lat}} \right)^{2 n_\nu},
\label{eq:tnu}
\end{equation}
where $n_\nu=4$ is usually adopted.  Spectral simulations with smaller values of $t_\nu$ are generally more dissipative.  To meaningfully compare spectral simulations with different numerical resolutions, we need to keep the hyperviscosity $\nu$ fixed by using equation (\ref{eq:tnu}) to scale the dissipation rate\footnote{From an operational standpoint, we note that when the dissipation rate is too small, the simulations will crash even when very small time steps (e.g., $\Delta t = 1$ s) are taken.  Therefore, there is a practical upper limit to the value of $t_\nu$ assumed.} assumed,
\begin{equation}
t^{-1}_\nu = \left( \frac{N_{\rm lat}}{96} \right)^8 t^{-1}_{\nu,{\rm T63}}.
\label{eq:tnu_scale}
\end{equation}
For example, the dissipation rate used in the T63L33 run, for the deep model of HD 209458b, is $t_{\nu,{\rm T63}}^{-1} \approx 0.33$ s$^{-1}$ ($\approx 10^{-5}$ HD 209458b day).  Therefore, a T31L33 run would use $t_\nu \approx 3 \times 10^{-3}$ HD 209458b day.  Within the \texttt{FMS}, specifying the dissipation rate as an input parameter is thus termed a ``resolution dependent" run.  Alternatively, specifying the hyperviscosity $\nu$ constitutes a ``resolution independent" run.  In general, we find that dealing with a dissipation rate (with units of s$^{-1}$) is somewhat more intuitive than having to vary $\nu$ (with units of m$^8$ s$^{-1}$).

In the original \cite{hs94} T63L20 ($N_{\rm lat}=96$) spectral
simulations, the dissipation rate was chosen to be $1.15741 \times
10^{-4}$ s$^{-1}$ which corresponds to a dissipation time of about 0.1
Earth day.  For their T42L15 ($N_{\rm lat}=64$) Earth-like
simulations, \cite{mr09} use $\nu = 1.18 \times 10^{37}$ m$^8$
s$^{-1}$, which corresponds to $t_\nu \sim 9 \times 10^{-3}$ Earth
day.  For their T42L15 hot Jupiter simulations, \cite{mr09} use
$\nu=6.28 \times 10^{47}$ m$^8$ s$^{-1}$, which corresponds to $t_\nu
\sim 2 \times 10^{-4}$ hot Jupiter day.  For their T31L33 ($N_{\rm
  lat}=48$) simulations of HD 209458b, \cite{rm10} use $\nu=8.54
\times 10^{47}$ m$^8$ s$^{-1}$ which is equivalent to $t_\nu \sim 9
\times 10^{-4}$ HD 209458b day.  For our Earth-like simulations, we
choose $t_\nu = 0.1$ day following \cite{hs94}.  For our hot Jupiter
simulations, we choose $t_\nu = 10^{-5}$ hot Jupiter day such that to
within a factor of a few, our chosen value for $t_\nu$ is consistent
with those used by \cite{mr09} and \cite{rm10}.  Table \ref{tab:params} lists our choices for
$t^{-1}_\nu$, which were made to match the values in the original
publications as closely as possible, while bearing in mind that all of
these choices have no strict justification beyond the requirement
that the model can be integrated without the detrimental accumulation
of small-scale noise.

For the finite difference (B-grid) core, the horizontal mixing coefficient ${\cal K}$ plays the analogous role of $t_\nu$ --- and not $\nu$ --- in the spectral core (see Appendix \ref{append:efold}).  Its default value within the \texttt{FMS} is ${\cal K}=0.35$, which we adopt unless otherwise stated.  \emph{Varying ${\cal K}$ is in effect changing the value of the analogue of $t_\nu$.}  Conversely, keeping ${\cal K}$ fixed and varying the resolution of the simulation effectively varies the value of the analogue of $\nu$.  We are unable to write down a simple analytical expression relating ${\cal K}$ and $\nu$, but note that it is possible to measure the analogue of $t_\nu$ in a finite difference simulation (see Appendix \ref{append:efold}).  While a correspondence between the dissipation parameters in the spectral and finite difference cores may exist, we do not consider it to be straightforward.  Our main intention is to demonstrate that it is possible to find equivalent pairs of values for $t_\nu$ (or $\nu$) and ${\cal K}$ by trial and error (see \S\ref{subsect:hd209458b}), which has implications for researchers wishing to adapt existing simulation platforms implementing different solution methods (and numerical dissipation schemes) to study the atmospheric circulation on exoplanets.

\subsection{Initial Conditions}

The default initialization in both the spectral and finite difference
cores uses the simplest assumption: isothermality with no wind.  Every
temperature point on the solution grid is set to $T=T_{\rm init}$
where $T_{\rm init}$ is an initial temperature and may be regarded as
a free parameter.  The tests we will describe in \S\ref{sect:tests}
use values of $T_{\rm init}$ which are tabulated in Table
\ref{tab:params}.  We will see that the active ($\tau^{-1}_{\rm rad}
\ne 0$) layers of the atmosphere, where the temperature rapidly
relaxes towards $T_{\rm eq}$, are somewhat insensitive to the choice
of $T_{\rm init}$ and produce results that are broadly consistent with
previous studies.  The simulations are started with a small initial
perturbation in the vorticity field.

\cite{tc10} have argued that initializing the simulations with non-zero winds can produce both qualitative and quantitative differences in the results, because the applied thermal forcing is projected differently onto the normal modes of the atmosphere under different initial wind conditions.  We consider this issue to be beyond the scope of the present study.

\begin{table*}
\centering
\caption{Table of Simulation Resolutions}
\label{tab:resolution}
\begin{tabular}{lccc}
\hline\hline
\multicolumn{1}{c}{Simulation} & \multicolumn{1}{c}{Spatial Resolution}  & \multicolumn{1}{c}{Angular Resolution} & \multicolumn{1}{c}{Examples (3D)}\\
\hline
\vspace{2pt}
T21 & 64$\times$32 & $5.625^\circ$ & HMP \\
T31 & 96$\times$48 & $3.75^\circ$ & \cite{burrows10,rm10}; HMP \\
T42 & 128$\times$64 & $2.8125^\circ$ & \cite{mr09,tc10}$^\dagger$ \\
T63 & 192$\times$96 & $1.875^\circ$ & \cite{hs94}; HMP \\
\hline
G24 & 48$\times$30 & $\left(7.5^\circ, 6.0^\circ \right)$ & HMP \\
G36 & 72$\times$45 & $\left(5.0^\circ, 4.0^\circ \right)$ & \cite{cs05,cs06}; HMP \\
G48 & 96$\times$60 & $\left(3.75^\circ, 3.0^\circ \right)$ & --- \\
G72 & 144$\times$90 & $\left(2.5^\circ, 2.0^\circ \right)$ & \cite{hs94,dd10}$^\ddagger$; HMP \\
\hline
\hline
\end{tabular}\\
Note: The acronym ``HMP" refers to the present study.\\
$\dagger$: \cite{mr09} and \cite{tc10} presented mainly T42 models \\ but also examined T85--T170 and T21--T85 ones, respectively, for convergence tests.\\
$\ddagger$: \cite{dd10} used a resolution similar to G72 for their simulations \\ ($160 \times 64$; $2.25^\circ \times 2.1875^\circ$) but truncate their latitudinal grid at $\Phi = \pm 70^\circ$. 
\end{table*}

\section{Atmospheric Dynamical Cores: Tests for Earth and Hot Jupiters}
\label{sect:tests}

\subsection{Held-Suarez Benchmark Test}
\label{subsect:hs94}

\begin{figure}
\begin{center}
\includegraphics[width=0.45\columnwidth]{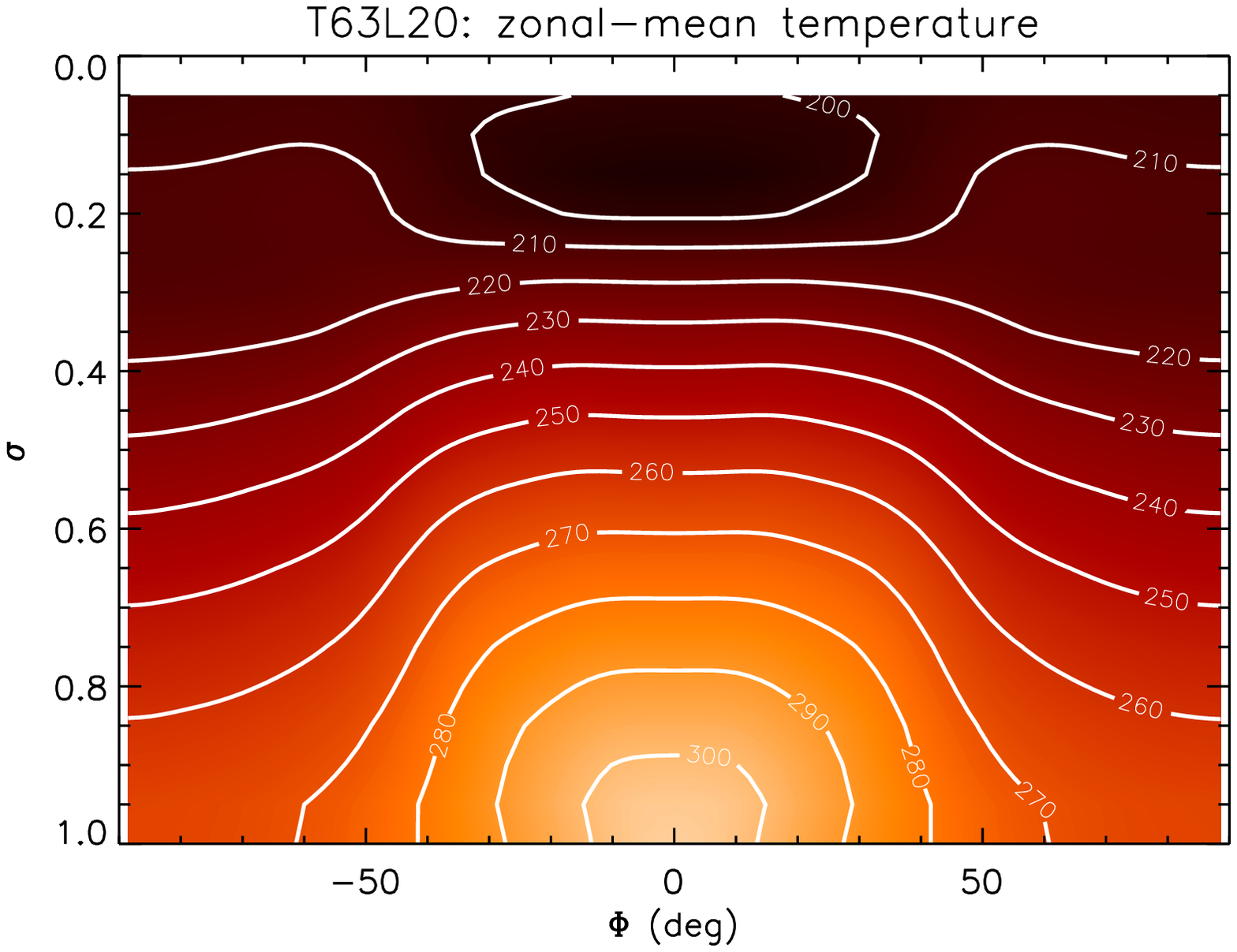}
\includegraphics[width=0.45\columnwidth]{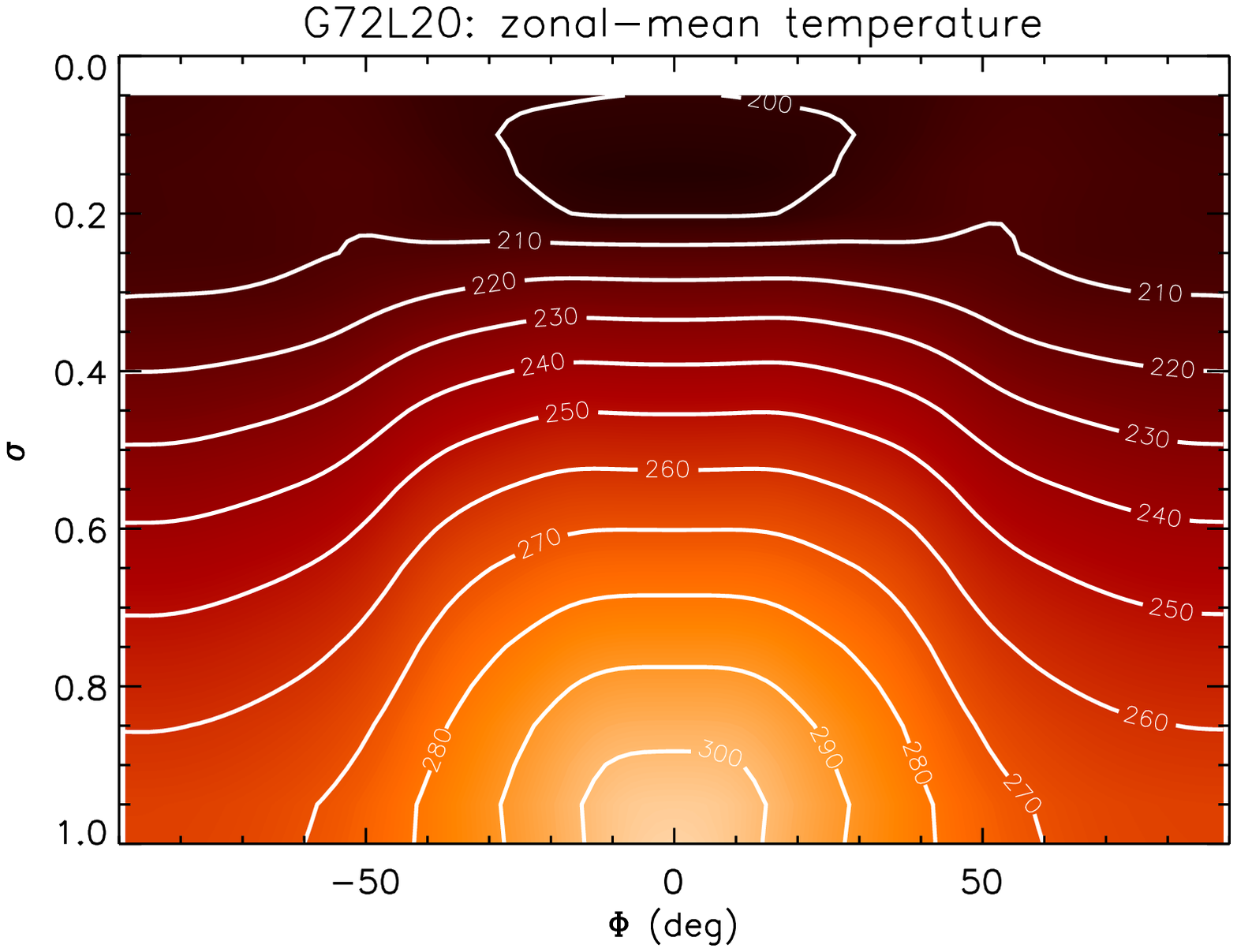}
\end{center}
\vspace{-0.2in}
\caption{Zonal-mean temperature, temporally averaged over 1000 days, for the Held-Suarez benchmark test for Earth.  Contour levels are in units of K; $P_s = 1$ bar.  Left: T63L20 spectral model.  Right: G72L20 finite difference model.}
\label{fig:temperature}
\end{figure}

\begin{figure}
\begin{center}
\includegraphics[width=0.45\columnwidth]{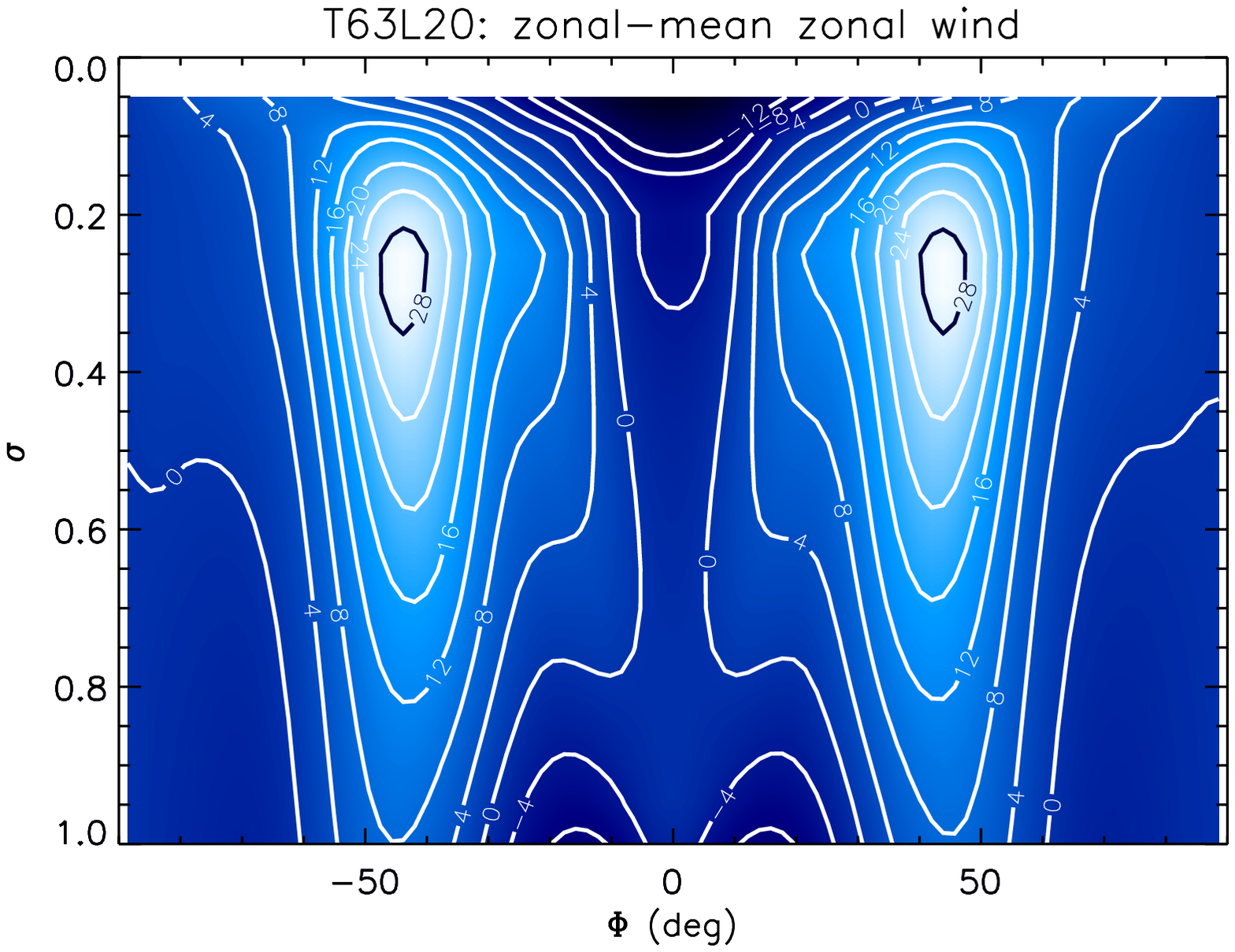}
\includegraphics[width=0.45\columnwidth]{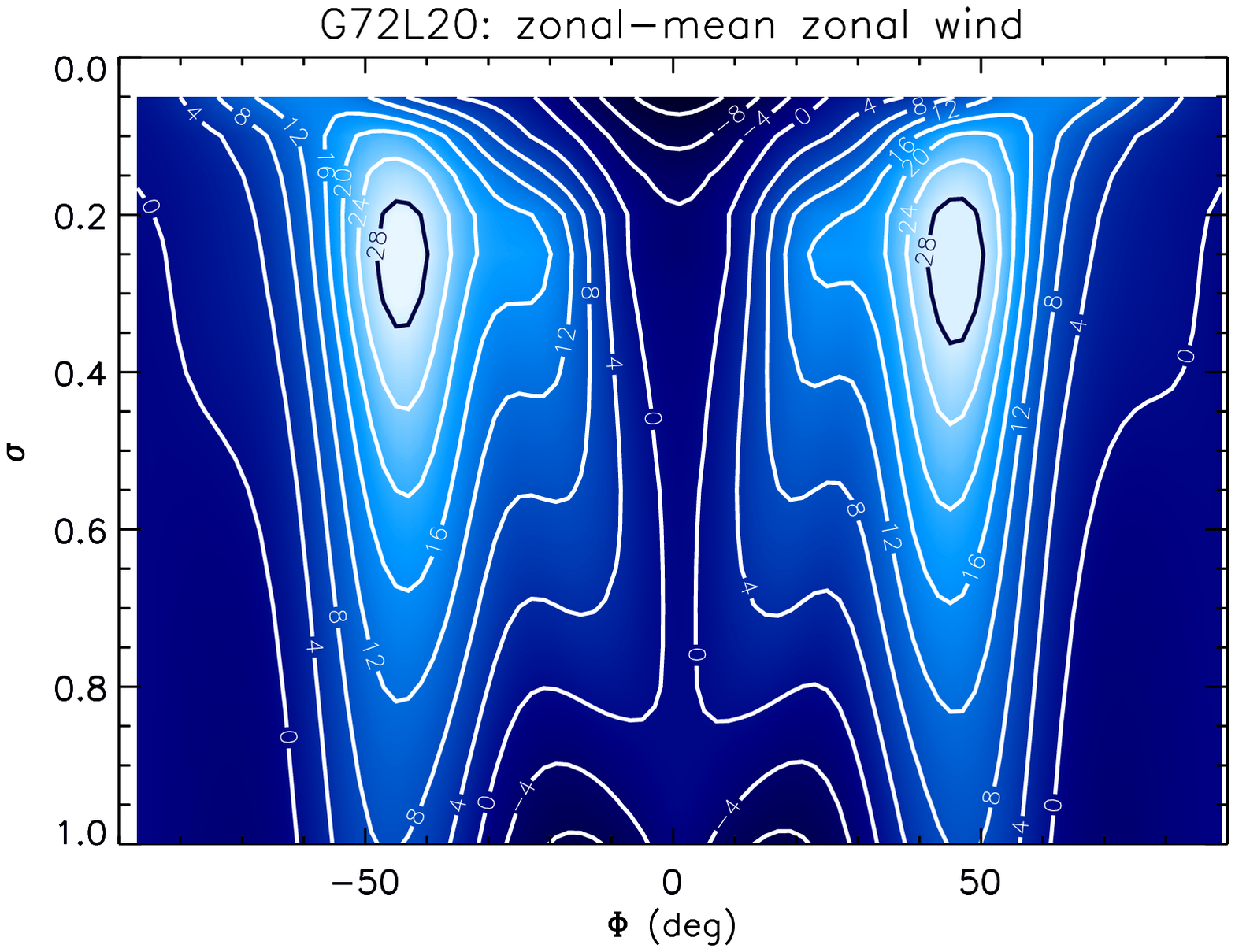}
\end{center}
\vspace{-0.2in}
\caption{Zonal-mean zonal wind, temporally averaged over 1000 days, for the Held-Suarez benchmark test for Earth.  Contour levels are in units of m s$^{-1}$; $P_s = 1$ bar.  Left: T63L20 spectral model.  Right: G72L20 finite difference model.}
\label{fig:zonal_wind}
\end{figure}

As a first check on our computational setup, we reproduced the \cite{hs94} benchmark test with both the spectral and finite difference versions of the dynamical core.  The effects of stellar irradiation, geometry, etc --- known as the ``thermal forcing" --- are encapsulated in the ``equilibrium temperature" function,
\begin{equation}
T_{\rm eq} = \mbox{max}\left\{ T_{\rm stra}, T_{\rm HS} \right\},
\label{eq:hs_forcing}
\end{equation}
where $T_{\rm stra} = 200$ K is the stratospheric temperature,
\begin{equation}
T_{\rm HS} \equiv \left[ T_{\rm surf} - \Delta T_{\rm EP} \sin^2\Phi - \Delta T_z \ln{\left(\frac{P}{P_0} \right)} \cos^2\Phi \right] \left( \frac{P}{P_0} \right)^\kappa,
\label{eq:ths}
\end{equation}
$T_{\rm surf} = 315$ K is the surface temperature at the equator and $\Delta T_{\rm EP} = 60$ K is the equator-to-pole temperature difference.  The parameters in the preceding equation specific to the Held-Suarez forcing are set to be 
\begin{equation}
\begin{split}
&\Delta T_z = 10 \mbox{ K},\\
&P_0 = 1 \mbox{ bar},\\
\end{split}
\end{equation}
while the other parameters of the test are described in Table \ref{tab:params}.  The initial temperature is not specified in \cite{hs94}, but the default value in the spectral code is $T_{\rm init}=264$ K; we will adopt this value for both the spectral and finite difference simulations.

The \texttt{FMS} implements a simple Newtonian relaxation of the temperature field, where a damping coefficient,
\begin{equation}
{\cal D}_{\rm Newton} = \frac{1}{\tau_{\rm rad,d}} + 
\begin{cases}
0, & \sigma \le \sigma_b, \\
\left(\frac{1}{\tau_{\rm rad,u}} - \frac{1}{\tau_{\rm rad,d}} \right) \left(\frac{\sigma-\sigma_b}{1-\sigma_b}\right) \cos^4\Phi, & \sigma > \sigma_b, \\
\end{cases}
\end{equation}
is applied to the temperature field relative to equilibrium $(T_{\rm eq}-T)$.  In the original implementation of Held-Suarez forcing, we have $\tau_{\rm rad,u} = 4$ days, $\tau_{\rm rad,d} = 40$ days and $\sigma_b = 0.7$ denoting the top of the planetary boundary layer in $\sigma$-coordinates.  Later, we will also implement only a single value of the Newtonian relaxation time, i.e., $\tau_{\rm rad} = \tau_{\rm rad,u} = \tau_{\rm rad,d}$, such that ``Newtonian cooling'' is represented by the term
\begin{equation}
Q_{\rm Newton} = \frac{T_{\rm eq}-T}{\tau_{\rm rad}}.
\end{equation}

Low-level winds are damped on a time scale $\tau_{\rm fric}$ using the damping coefficient,
\begin{equation}
{\cal D}_{\rm Rayleigh} =
\begin{cases}
0, & \sigma \le \sigma_b, \\
\frac{\sigma-\sigma_b}{\tau_{\rm fric} \left(1-\sigma_b\right)}, & \sigma > \sigma_b. \\
\end{cases}
\end{equation}
Such a prescription is known as ``Rayleigh friction'' (or drag) and mimicks boundary-layer friction between the atmosphere and the surface of the Earth.  In the \texttt{FMS}, Rayleigh friction is applied to the velocity field: $-{\cal D}_{\rm Rayleigh} \vec{v}$.  Note that $\sigma_b \ne \sigma_{\rm stra}$ in general, where $\sigma_{\rm stra}$ is the location of the tropopause, the transition layer between the troposphere and stratosphere.  We will later implement thermal forcings which are different from equation (\ref{eq:hs_forcing}).

For the spectral model, the resolution used for the published results
of \cite{hs94} is T63L20.  The corresponding longitude versus latitude
grid for this resolution is $192 \times 96$, which allows the model
to be computed on up to 48 processors simultaneously.  The default
setting for the \texttt{FMS} Held-Suarez benchmark uses a hyperviscous
dissipation rate of $1.15741 \times 10^{-4}$ s$^{-1}$ ($\approx 0.1$
day).  The normalized pressure ($0 < \sigma \le 1$) is equally spaced
with 20 vertical levels.

For the finite-difference model, the resolution used in the published results of \cite{hs94} is G72L20 ($144 \times 90$).  The vertical levels are treated with a hybrid $\sigma$-$P$ coordinate system: the terrain-following $\sigma$-coordinate is used near the planetary surface and transitions to the $P$-coordinate well above the surface.  The default setting for the horizontal mixing coefficient is  ${\cal K}=0.35$.

Figures \ref{fig:temperature} and \ref{fig:zonal_wind} show the zonally-averaged (or zonal-mean) temperature (in K) and zonal wind speed (in m s$^{-1}$), respectively.  Following \cite{hs94}, we ran both sets of simulations for $t_{\rm total}=1200$ days, but discarded the first $t_{\rm discard}=200$ days in order to eliminate features due to the different initialization schemes.  We see that our results are consistent with those presented in Figures 1 and 2 of \cite{hs94}; the spectral and finite difference results are also in broad quantitative agreement.  \emph{For the rest of the paper, we refer to the quantities shown in Figures \ref{fig:temperature} and \ref{fig:zonal_wind} (i.e., 1000-Earth-day averages of zonal-mean profiles) as the ``Held-Suarez statistics''.}

We conclude that our implementation of the Held-Suarez benchmark test for Earth is successful.

\subsection{Hypothetical Tidally-Locked Earth Benchmark Test}
\label{subsect:tidalearth}

\begin{figure}
\begin{center}
\includegraphics[width=0.45\columnwidth]{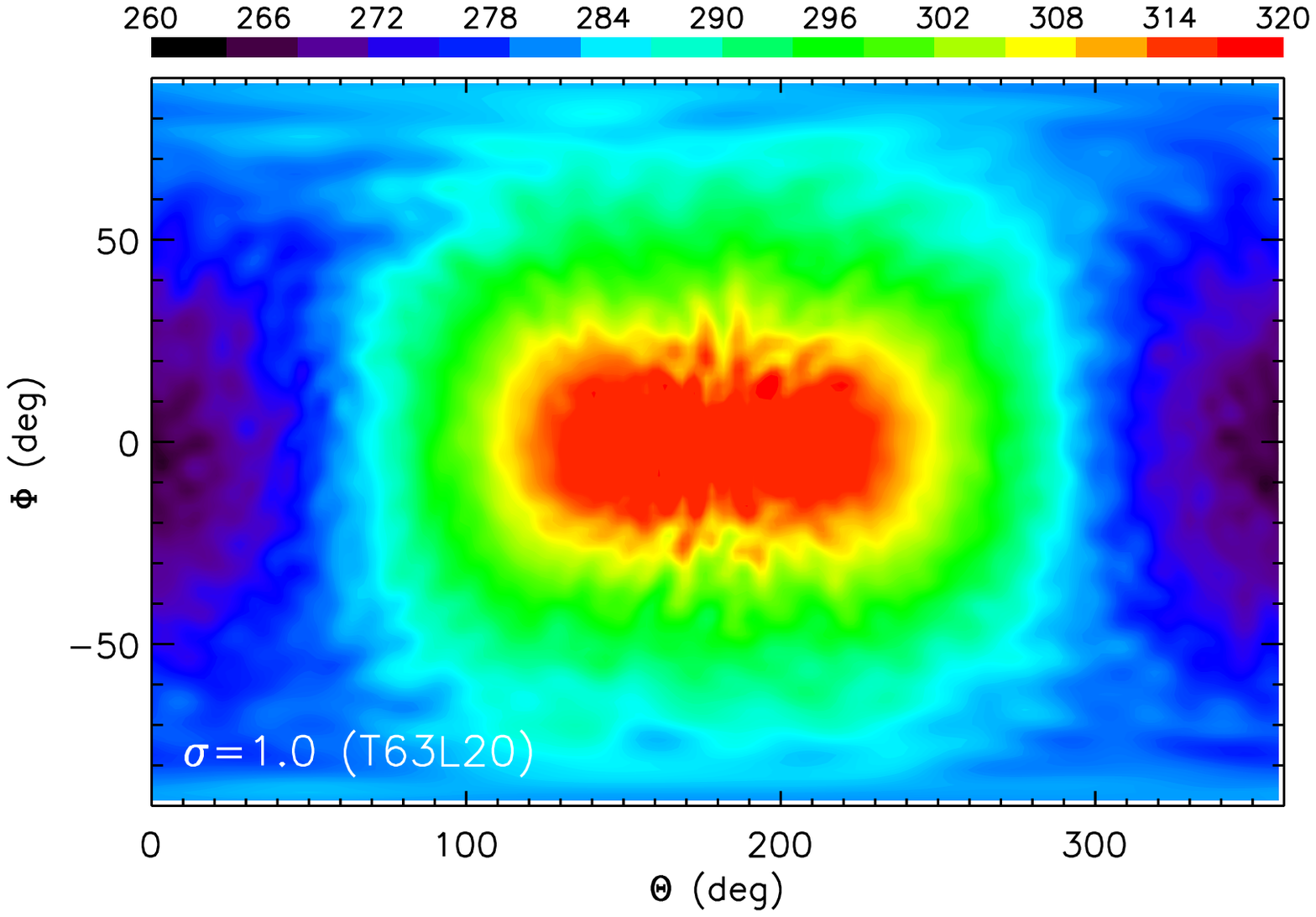}
\includegraphics[width=0.45\columnwidth]{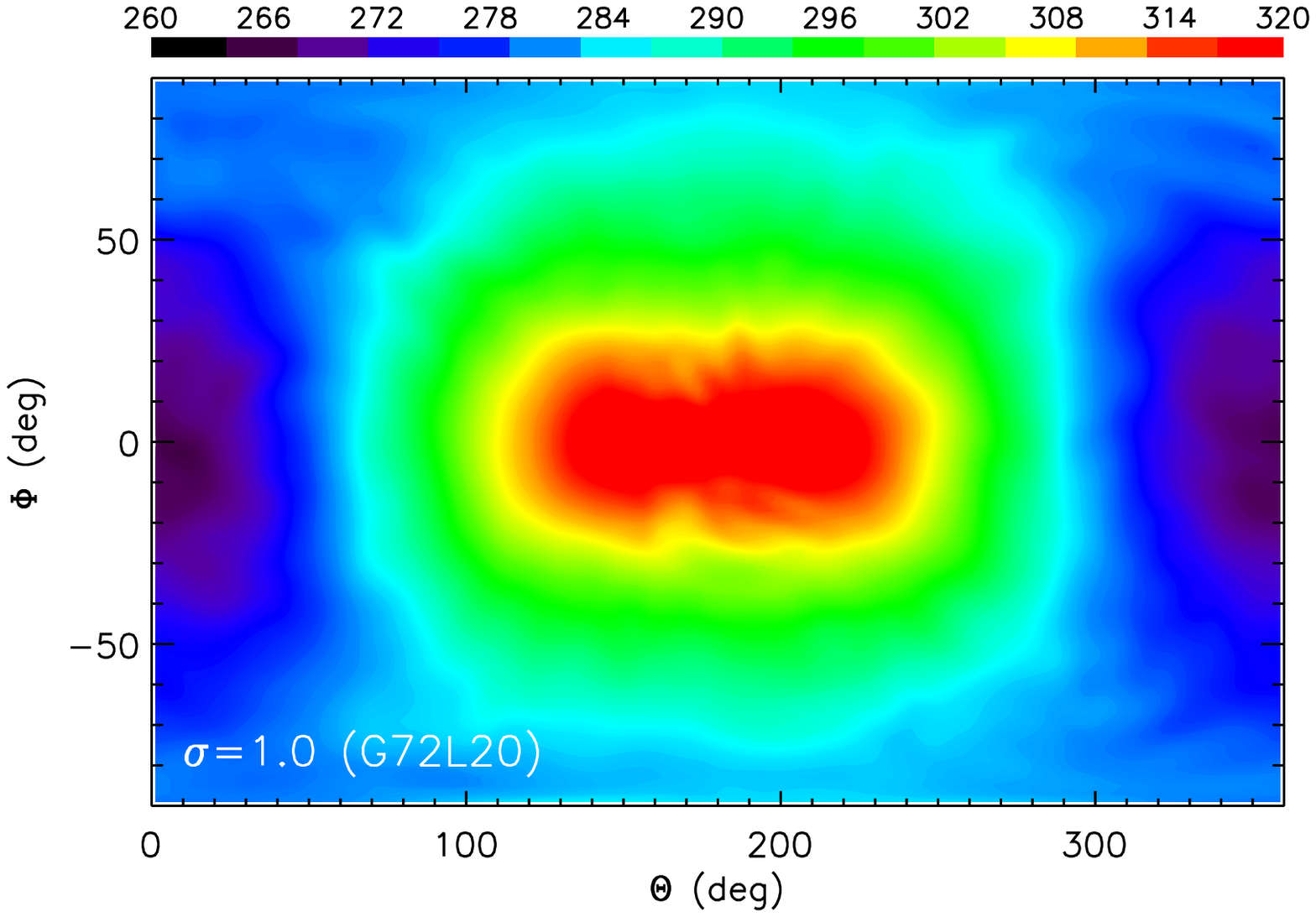}
\includegraphics[width=0.45\columnwidth]{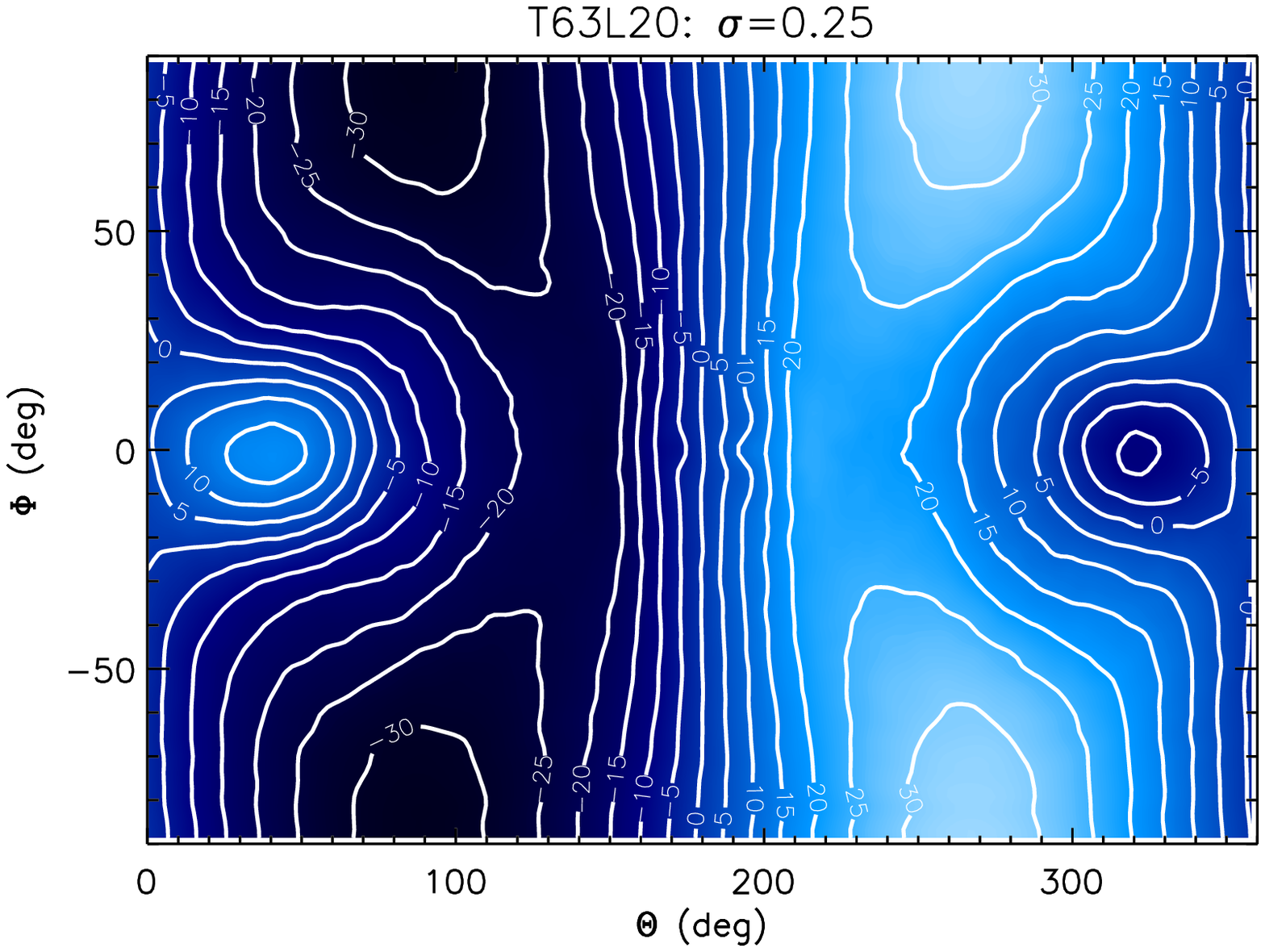}
\includegraphics[width=0.45\columnwidth]{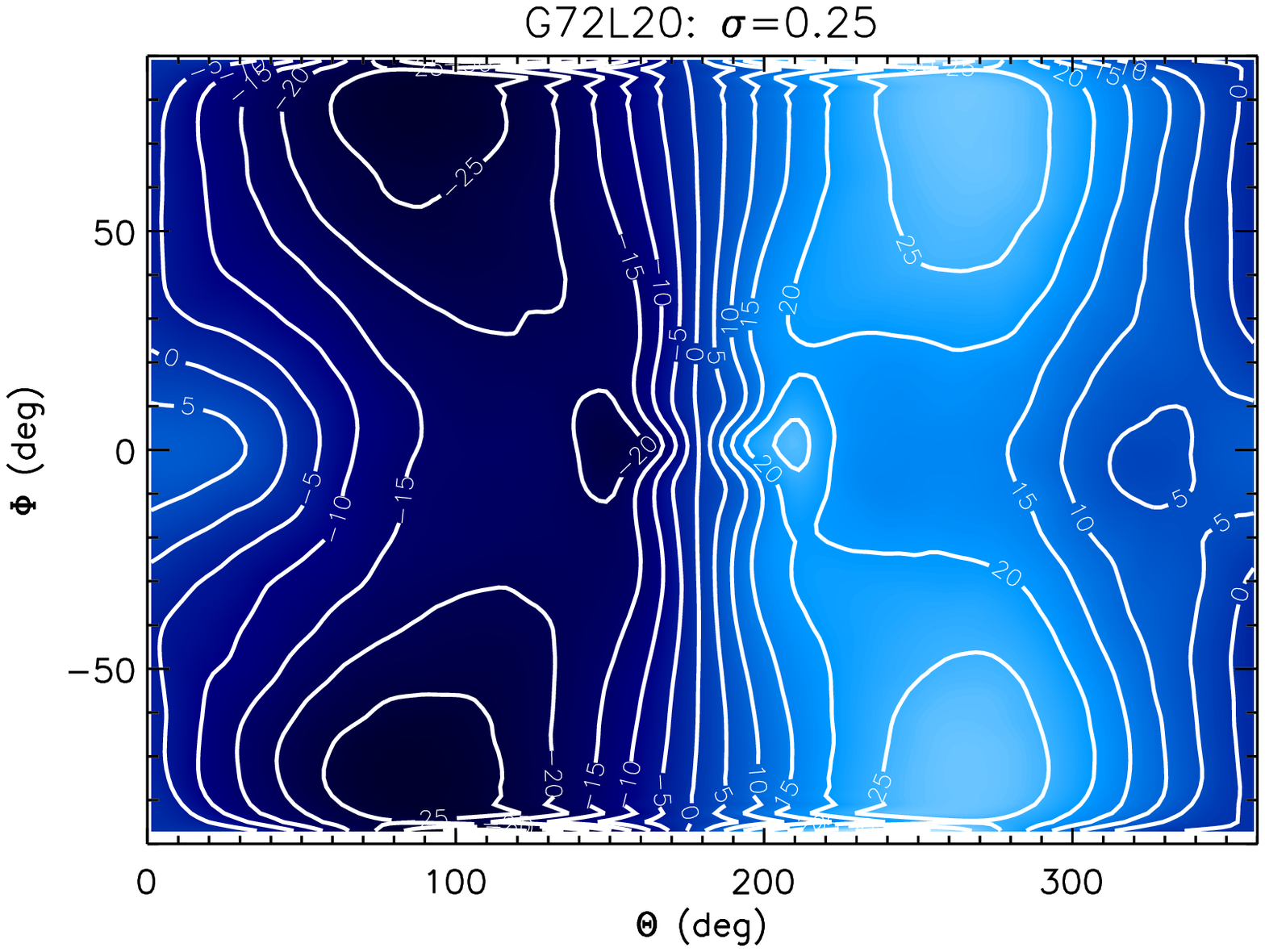}
\includegraphics[width=0.45\columnwidth]{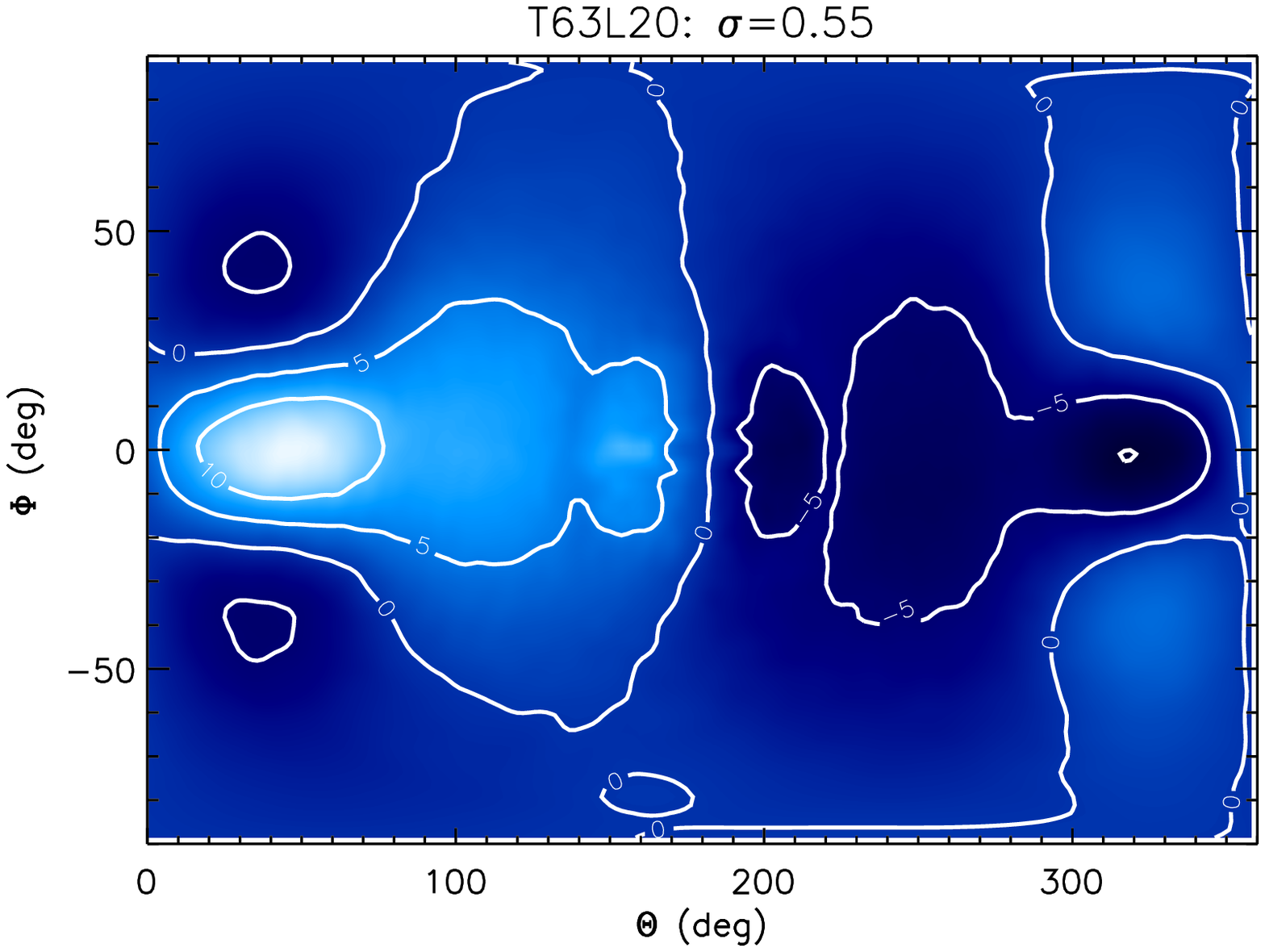}
\includegraphics[width=0.45\columnwidth]{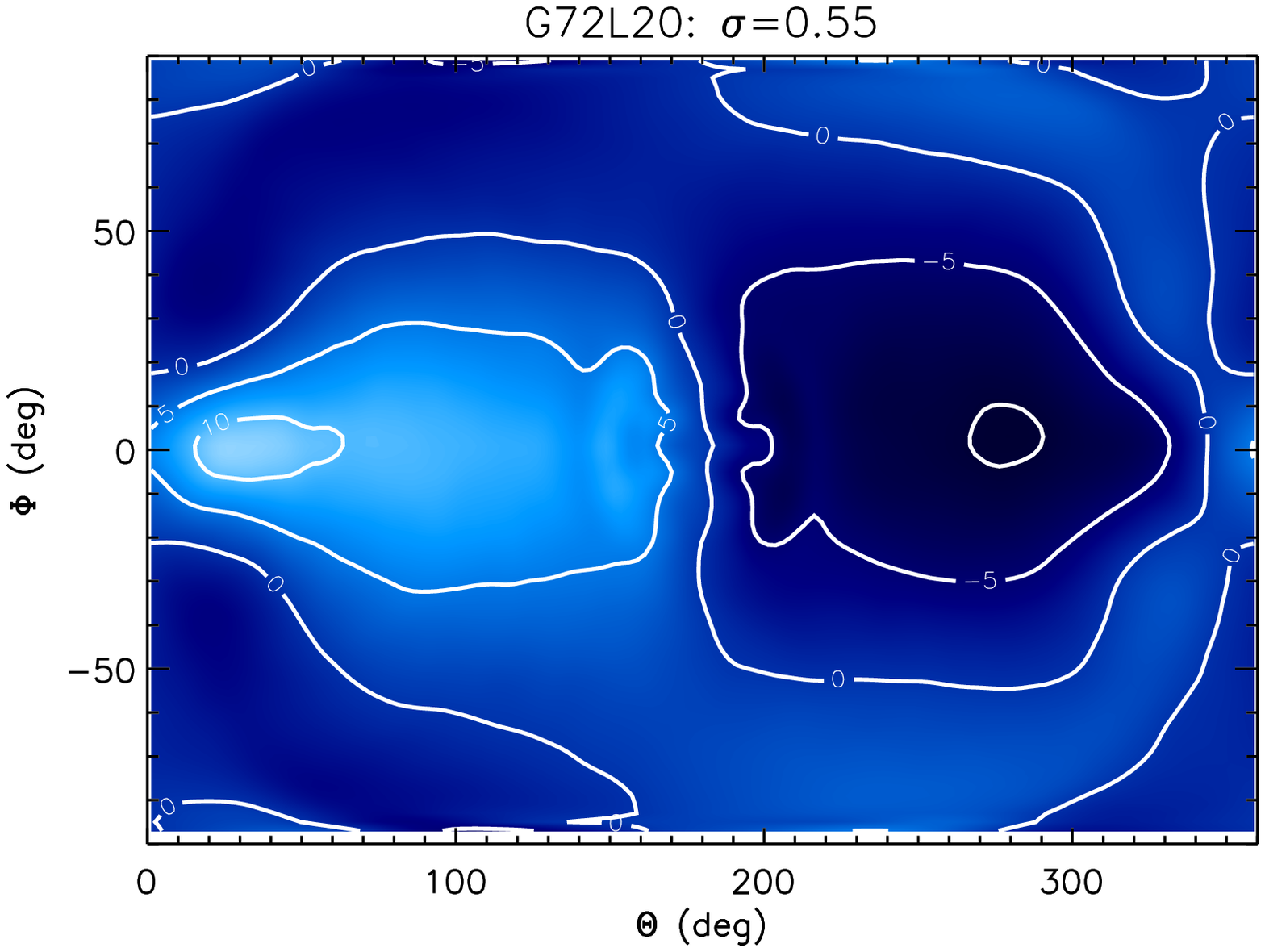}
\includegraphics[width=0.45\columnwidth]{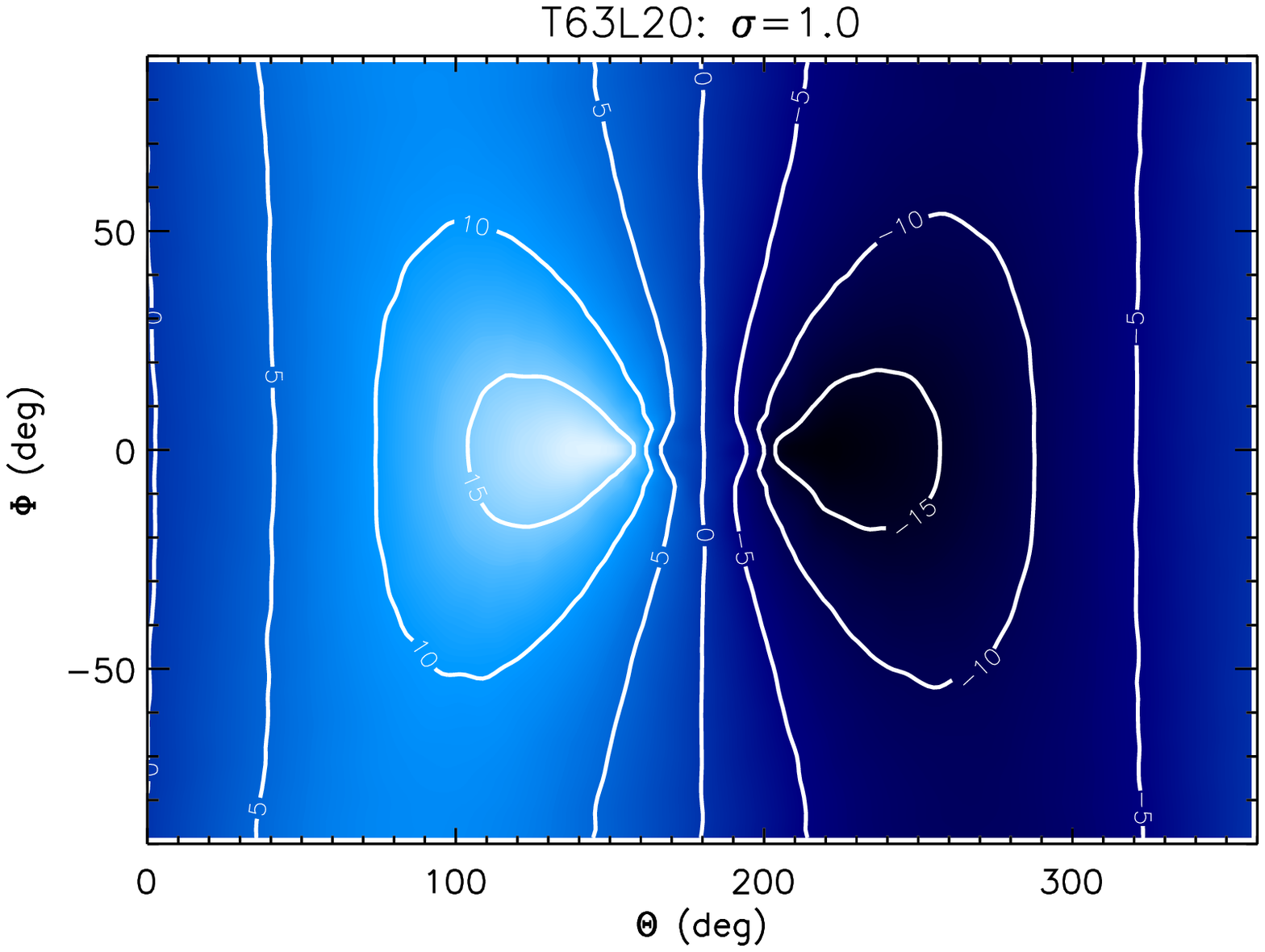}
\includegraphics[width=0.45\columnwidth]{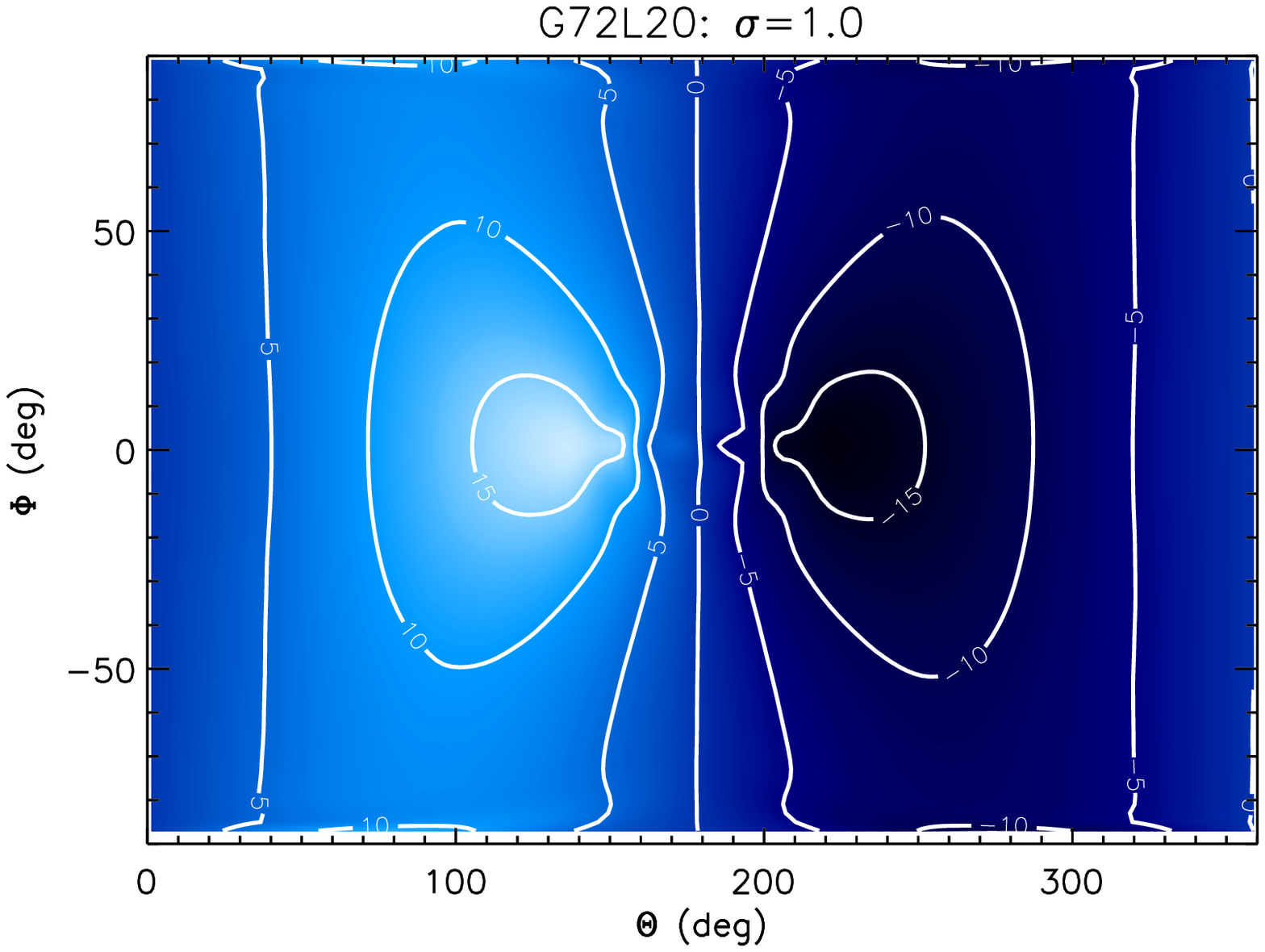}
\end{center}
\vspace{-0.2in}
\caption{Simulation of tidally-locked Earth. Top row: snapshot of the temperature (represented by colours) field at 1200 Earth days and $\sigma=1.0$.  The second, third and fourth rows are the temporally averaged zonal wind profiles at $\sigma=0.25$, 0.55 and 1.0, respectively.  The left and right columns are for the spectral (T63L20) and finite difference (G72L20) simulations, respectively.  Temperatures are in K and wind speeds are in m s$^{-1}$.}
\label{fig:tidal_zonal}
\end{figure}

\begin{figure}
\begin{center}
\includegraphics[width=0.45\columnwidth]{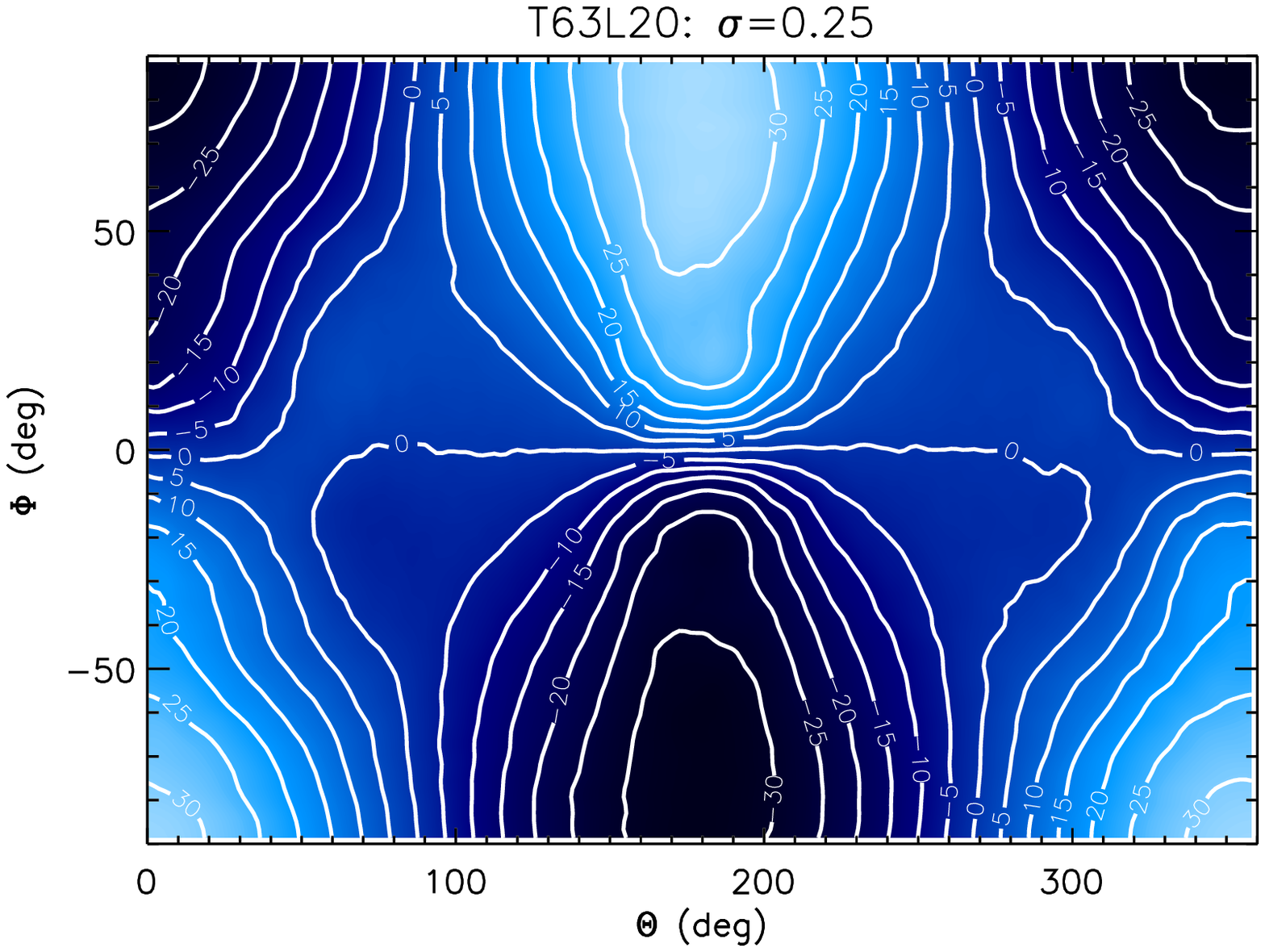}
\includegraphics[width=0.45\columnwidth]{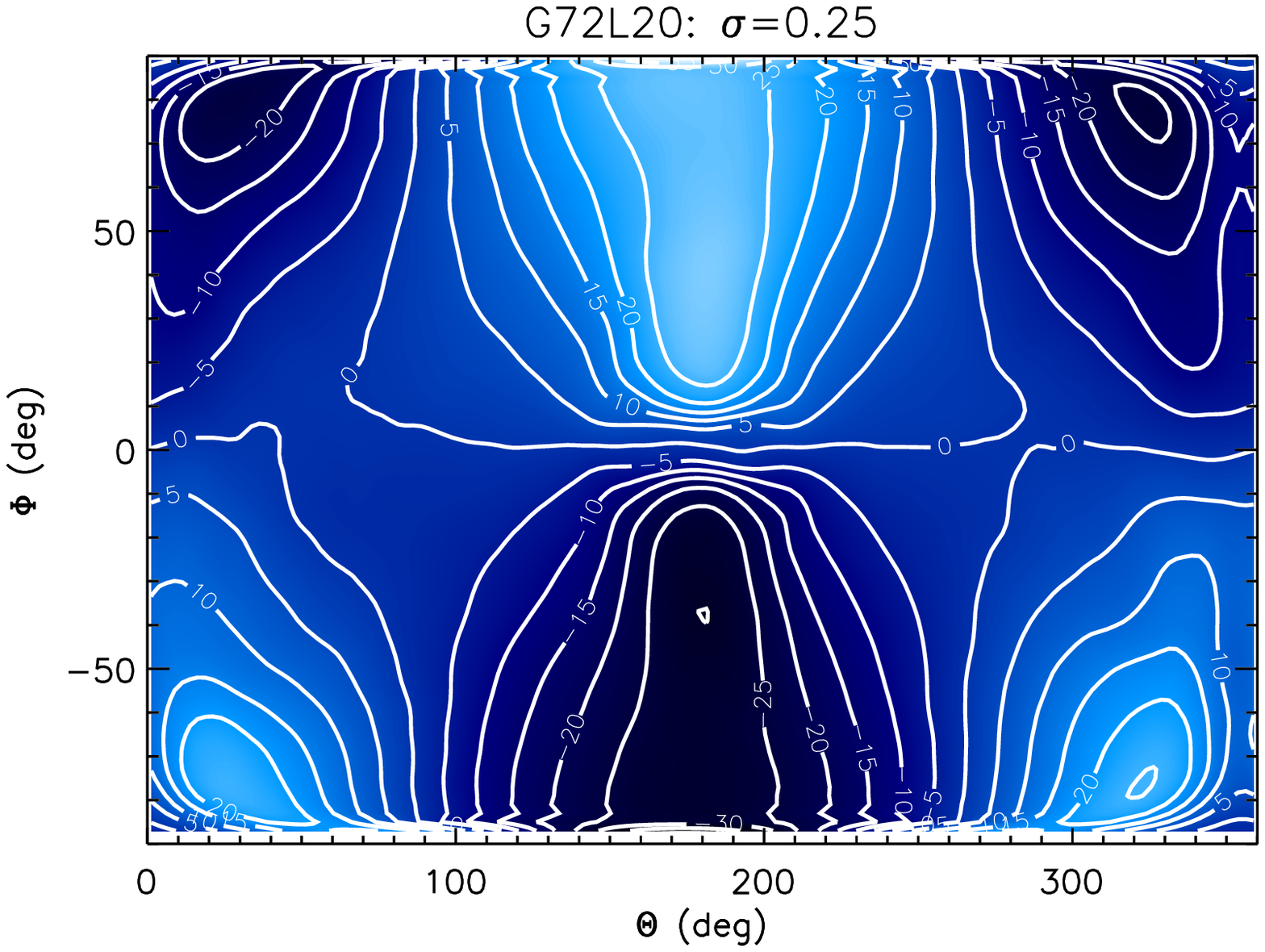}
\includegraphics[width=0.45\columnwidth]{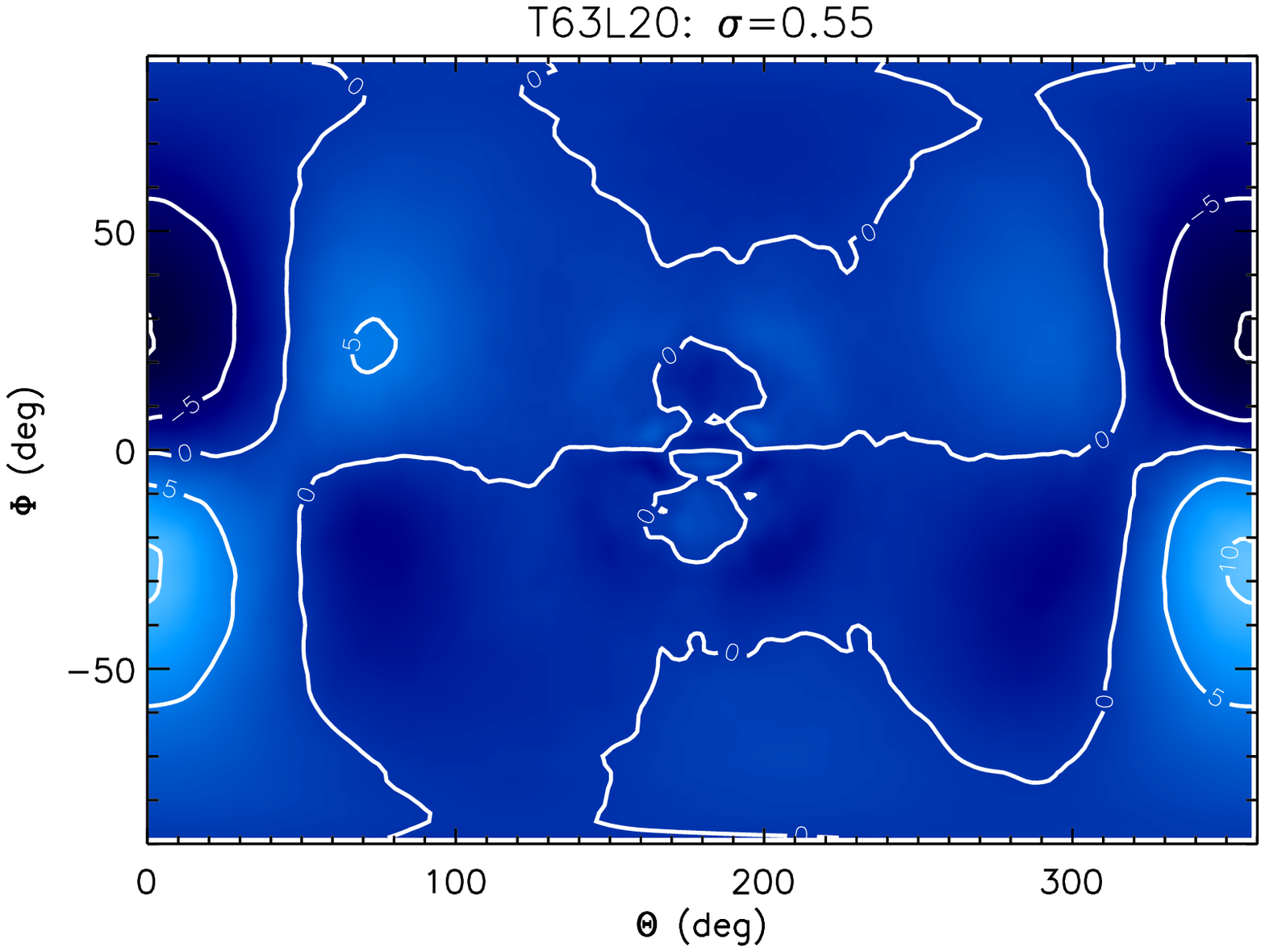}
\includegraphics[width=0.45\columnwidth]{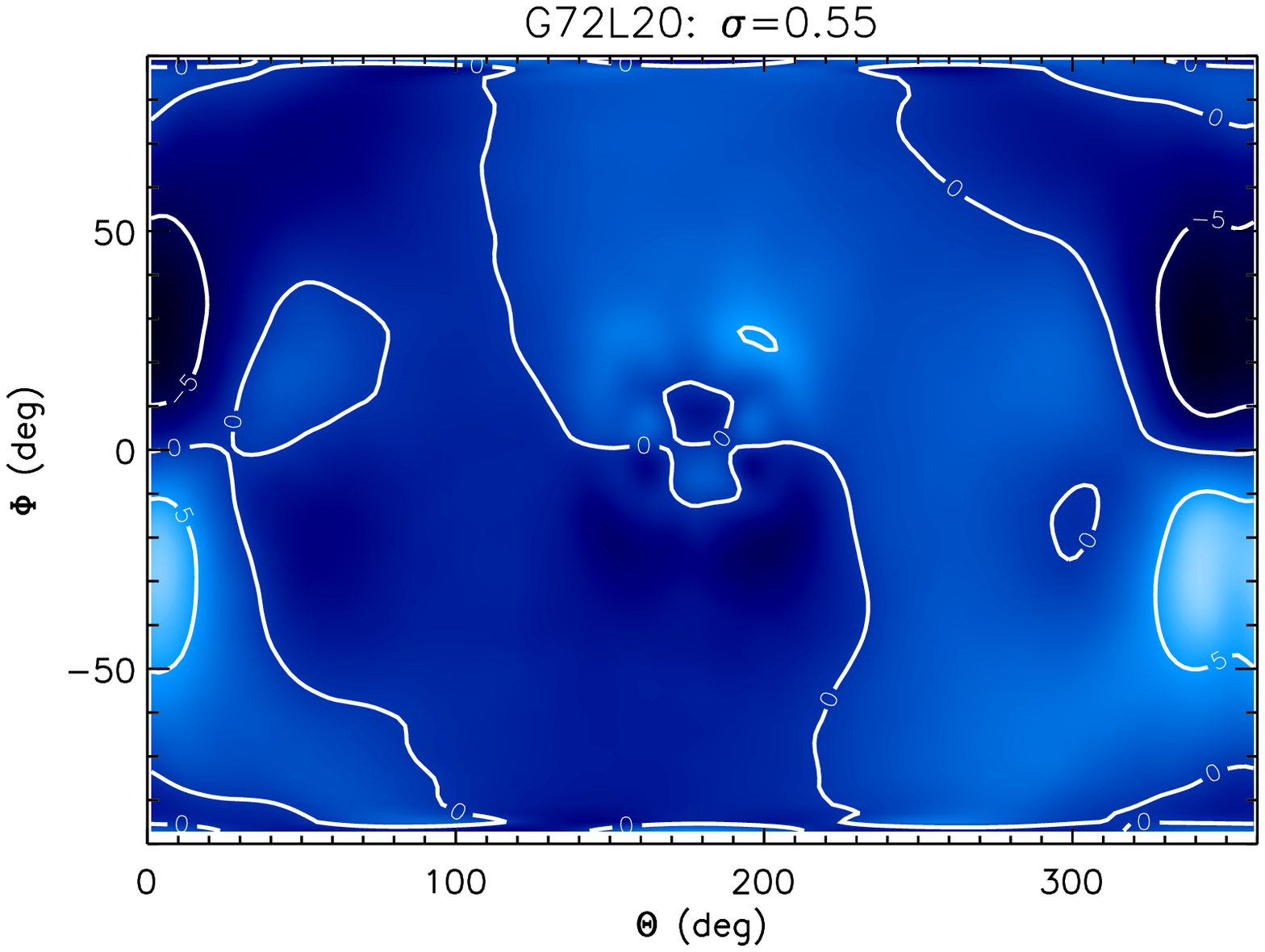}
\includegraphics[width=0.45\columnwidth]{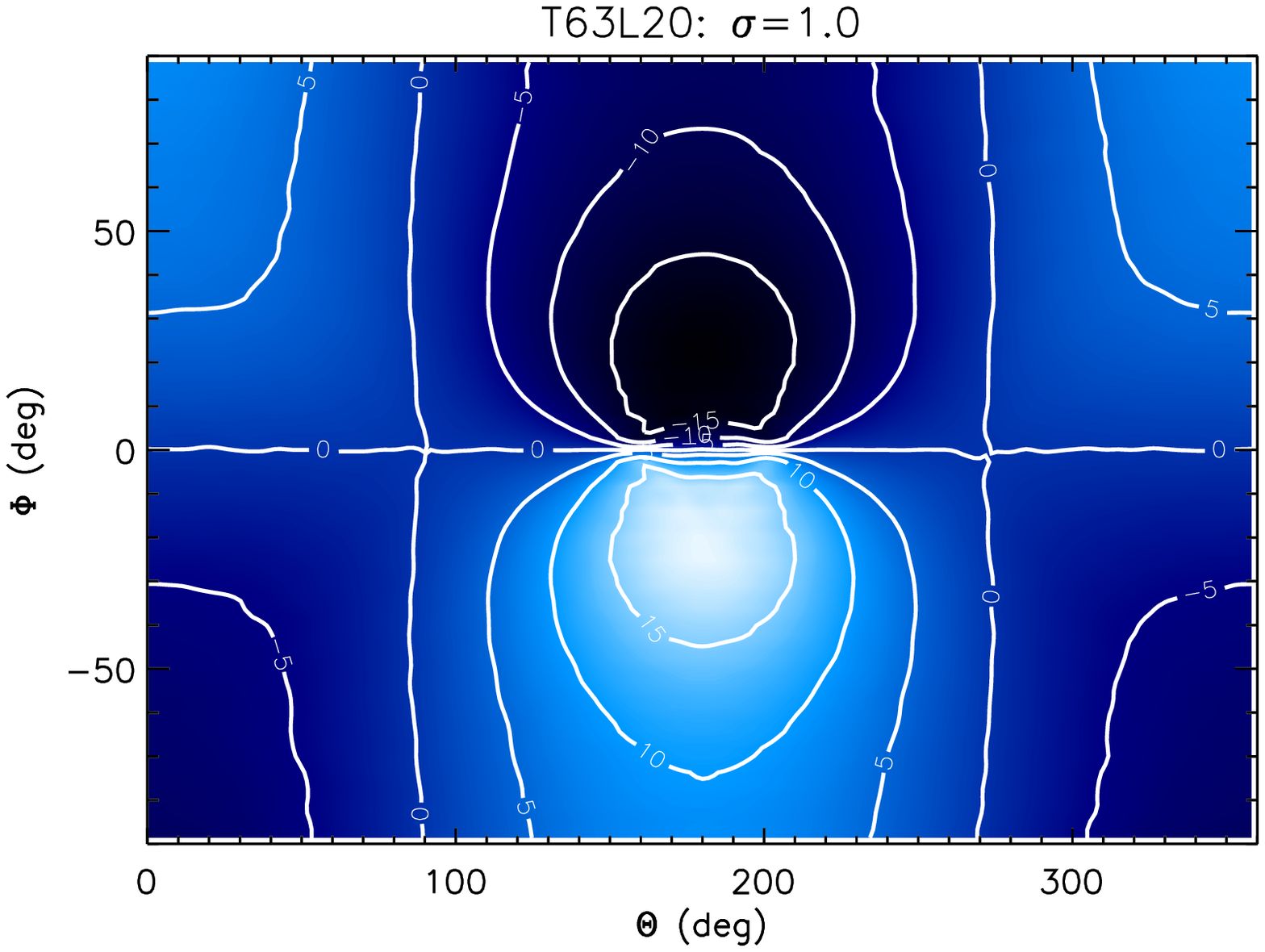}
\includegraphics[width=0.45\columnwidth]{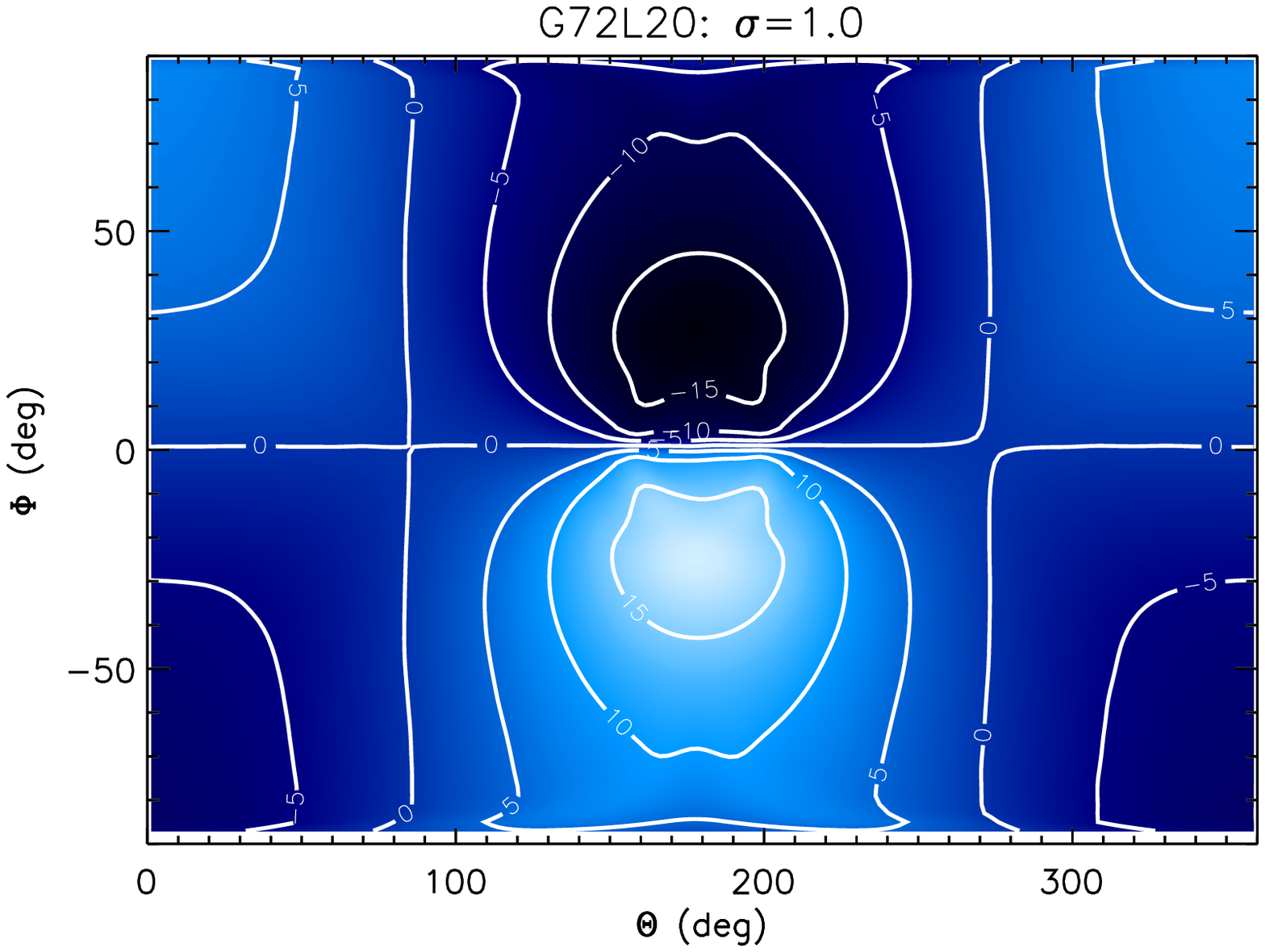}
\end{center}
\vspace{-0.2in}
\caption{Same as the second, third and fourth rows of Figure \ref{fig:tidal_zonal}, but for the temporally averaged meridional wind profiles.}
\label{fig:tidal_meri}
\end{figure}

Thermal forcing for a tidally-locked (exo)planet can be mimicked by replacing the $-\sin^2\Phi$ term in equation (\ref{eq:ths}) with a term that is proportional to $+\cos(\Theta-180^\circ) \cos\Phi$ \citep[see, e.g.,][]{cs05,mr09,ms10}.  Additionally, for a hypothetical tidally-locked Earth the rotation rate has to be reduced,
\begin{equation}
\Omega_p \rightarrow \Omega_p/365,
\end{equation}
such that one planetary day is equal to one planetary year.  As a
prelude to simulating the atmospheric circulation on (tidally-locked)
hot Jupiters, we first examine the case of such a tidally-locked Earth
at 1 AU.  Such a case study was conducted by \cite{ms10}, who
considered more sophisticated physics than is the case for our
dynamical core simulations, including an active hydrological cycle, a
gray radiative transfer scheme with a pressure-dependent opacity, and
an explicit formulation for atmosphere-surface exchanges on an
aquaplanet.

We implement the following thermal forcing
\begin{equation}
\begin{split}
&T_{\rm eq} = \mbox{max}\left\{ T_{\rm stra}, T_{\rm HS} \right\},\\
&T_{\rm HS} \equiv \left[ T_{\rm surf} + \Delta T_{\rm EP} \cos\left(\Theta-180^\circ \right) \cos\Phi - \Delta T_z \ln{\left(\frac{P}{P_0} \right)} \cos^2\Phi \right] \left( \frac{P}{P_0} \right)^\kappa,\\
\end{split}
\label{eq:thermal_tidal}
\end{equation}
placing the substellar point is at $(\Theta=180^\circ, \Phi=0)$.  We adopt the same set of parameters as described in Table \ref{tab:params} for the Held-Suarez benchmark, including for the implementation of Newtonian relaxation and Rayleigh friction.  The time step used is $\Delta t = 600$ s.

The first row of Figure \ref{fig:tidal_zonal} shows snapshots of the temperature field at Day 1200 and $\sigma=1.0$.  The temperature field, which should be compared to Figure 1 of \cite{ms10}, shows an atmospheric temperature structure dominated by radiative forcing rather than advection.  The second, third and fourth rows of Figure \ref{fig:tidal_zonal} show the temporally averaged (over 1000 days) zonal wind profiles, as functions of longitude and latitude, at $\sigma=0.25$, 0.55 and 1.0, respectively.  These values of $\sigma$ were chosen to match as closely as possible the $\sigma=0.28$, 0.54 and 1.0 values adopted by \cite{ms10} in the left column of their Figure 4.  Despite our much simpler setup, our results are in qualitative agreement with those of \cite{ms10}, showing the presence of a large, direct circulation cell centered on the substellar point.  Furthermore, our spectral and finite difference simulations are in general agreement with a clear indication that discrepancies start cropping up near the poles, as may be expected because of the difficulties in treating the poles in the finite difference core.

Figure \ref{fig:tidal_meri} shows the temporally averaged meridional wind profiles at $\sigma=0.25$, 0.55 and 1.0, and should be compared to the right column of Figure 4 of \cite{ms10}.  We again attain qualitative agreement with the results of \cite{ms10}, capturing the large circulation cell centered on the substellar point, which exhibits poleward and equatorward motions in the dayside and nightside hemispheres, respectively, i.e., the atmosphere flows from the day to the night side.

Our results in this subsection provide a useful prelude to the study of hot Jupiter atmospheres, because the simulation of a tidally-locked Earth is less computationally demanding.  The benchmark tests described in this sub-section are thus an efficient way of checking one's code before moving on to the hot Jupiter benchmarks.  For variations on a theme of Earth-like models, please refer to \cite{hv10}.

\subsection{Benchmark Tests for Earth and ``Shallow'' Hot Jupiter}
\label{subsect:mr09}

\cite{mr09} simulated the atmospheric circulation on Earth and hot Jupiters using the 3D \texttt{IGCM} spectral code with equal spacing in $\sigma$ ($N_v=15$).  In the hot Jupiter case, this model is considered to be ``shallow'' because of the limited depth of the 3D atmosphere modelled (down to only 1 bar).  Such a 3D model should not be confused with the ``shallow water'' \citep{menou03} or ``equivalent barotropic'' \citep{cho03,cho08} models, both of which are essentially 2D.  (See also \citealt{lh68}, \citealt{kundu04} and \citealt{hs09}.)

\subsubsection{Earth-like}

\begin{figure}
\begin{center}
\includegraphics[width=0.45\columnwidth]{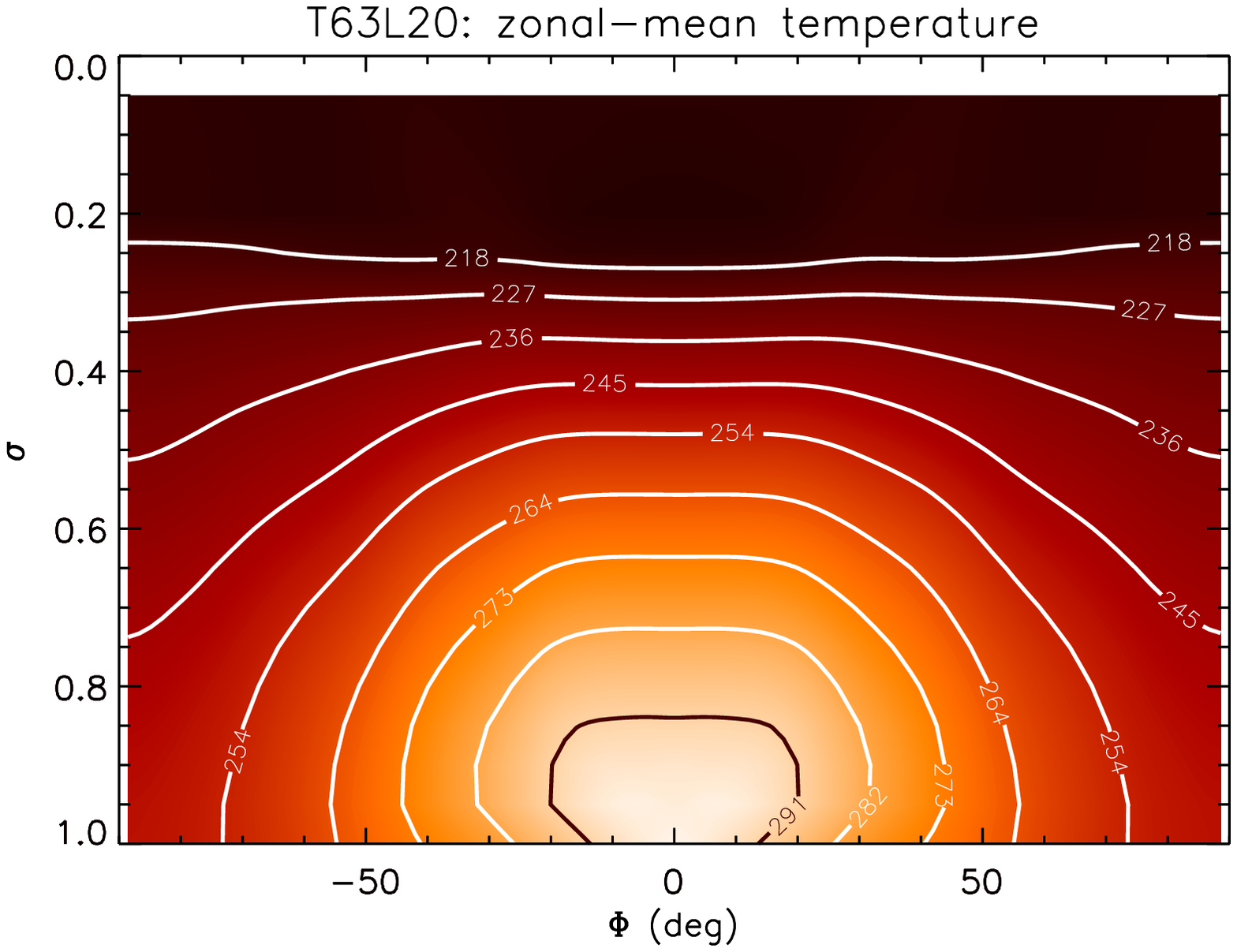}
\includegraphics[width=0.45\columnwidth]{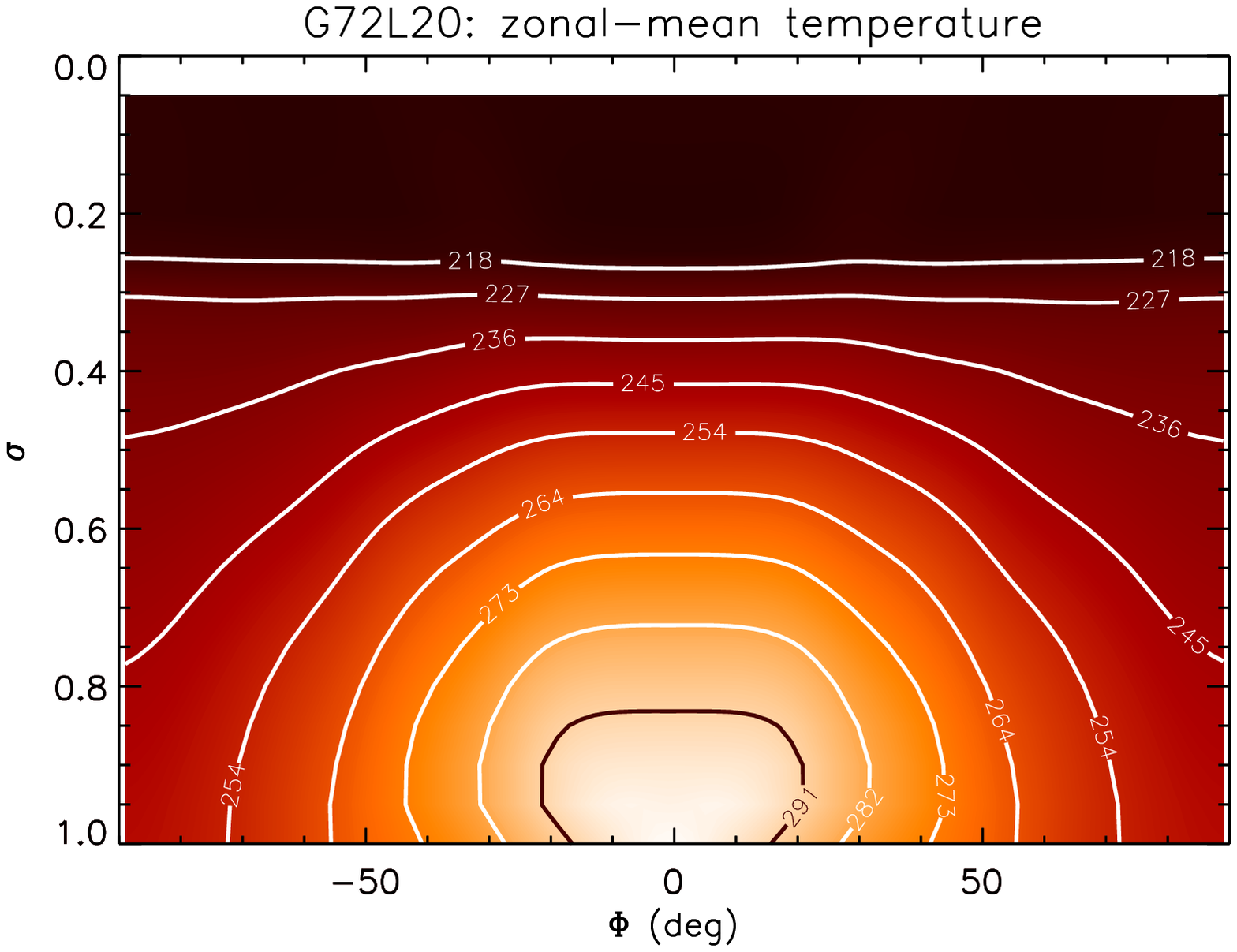}
\includegraphics[width=0.45\columnwidth]{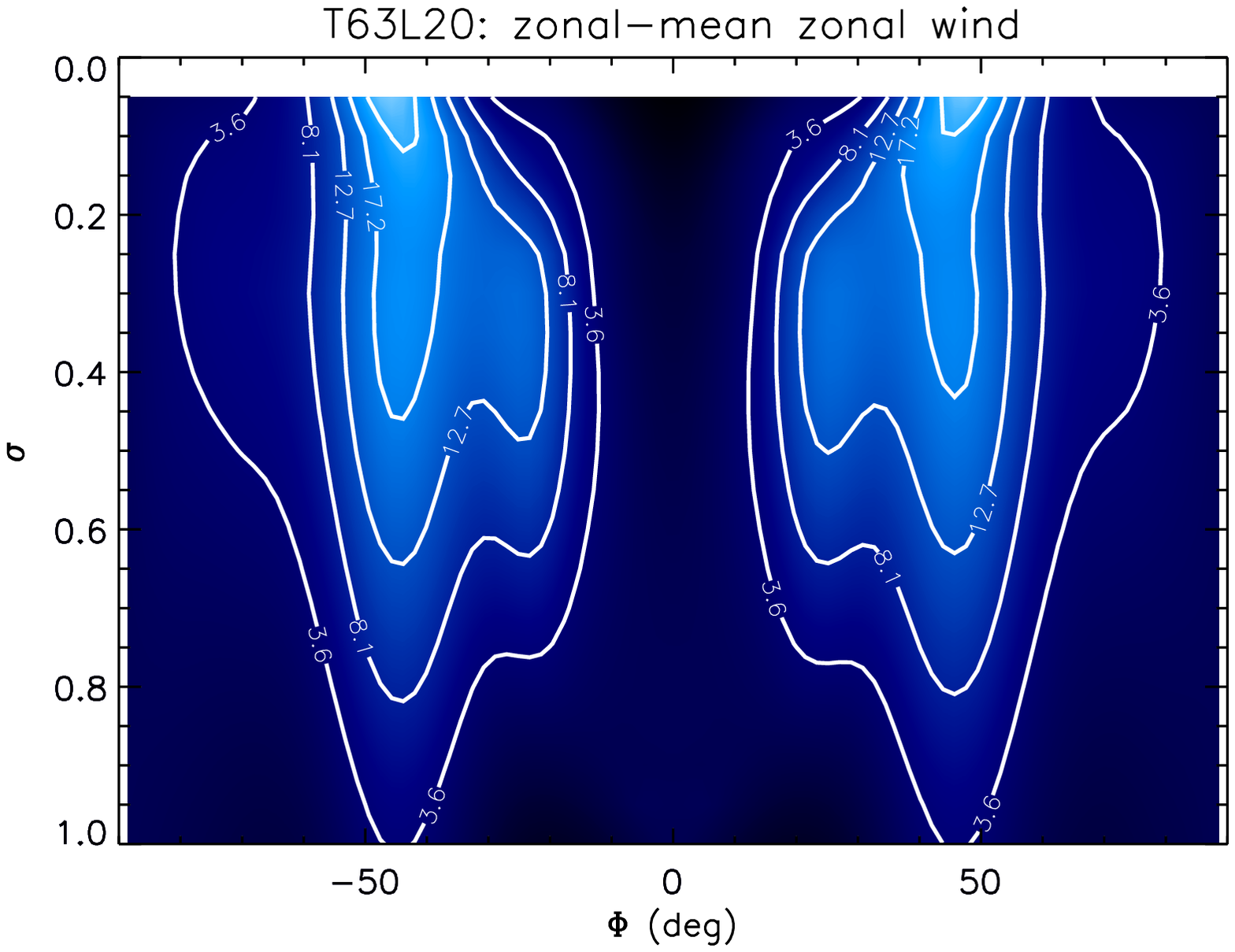}
\includegraphics[width=0.45\columnwidth]{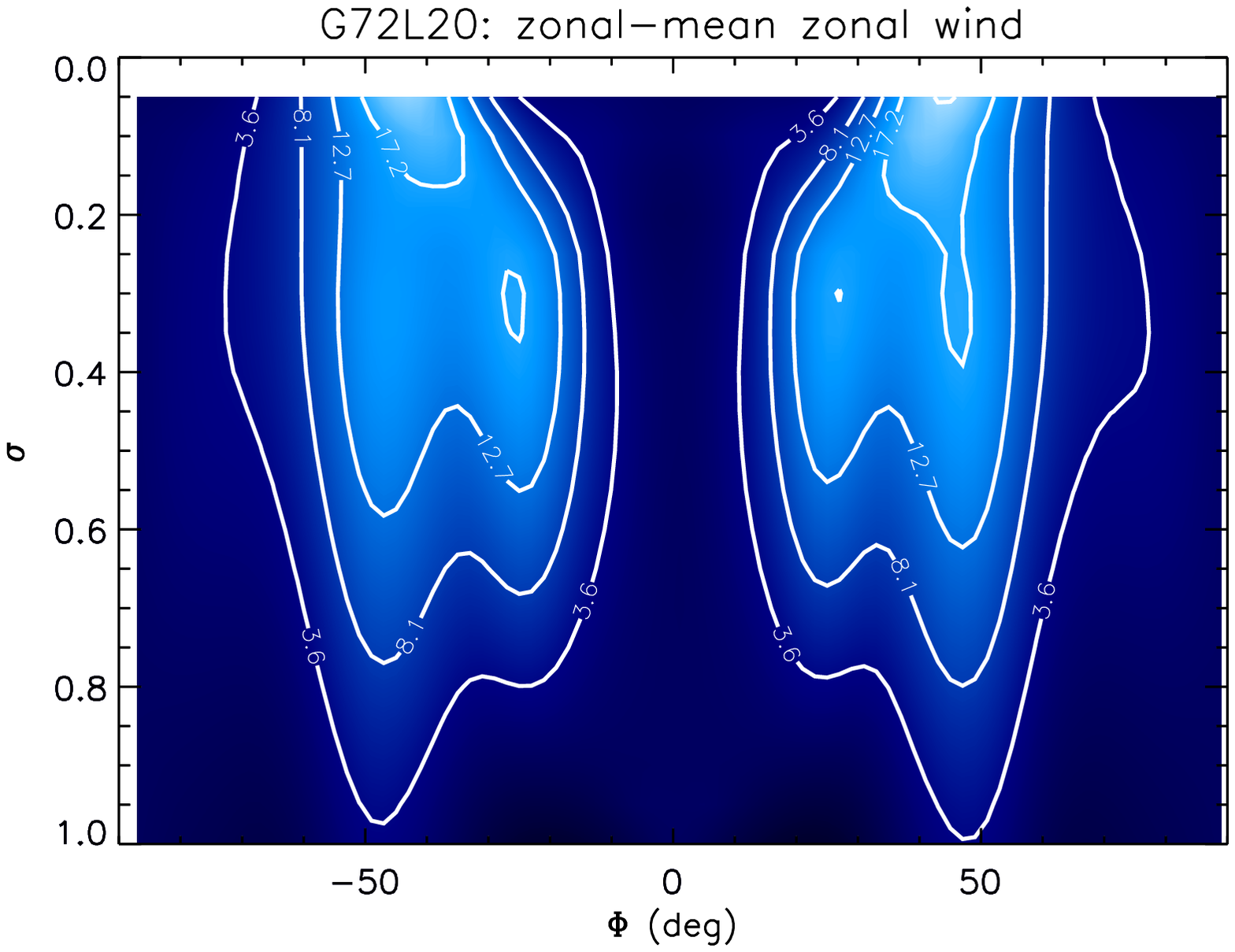}
\end{center}
\vspace{-0.2in}
\caption{Held-Suarez statistics for the Menou-Rauscher thermal forcing of an Earth-like atmosphere, analogous to Figures \ref{fig:temperature} and \ref{fig:zonal_wind}.  Left: T63L20 spectral model.  Right: G72L20 finite difference model.  Temperatures are in K and wind speeds are in m s$^{-1}$.}
\label{fig:mr09_earth}
\end{figure}

For an Earth-like setup, instead of equation (\ref{eq:hs_forcing}), \cite{mr09} employed
\begin{equation}
T_{\rm eq} = T_{\rm vert} + \beta_{\rm trop} ~\Delta T_{\rm EP} \left( \frac{1}{3} - \sin^2\Phi \right)
\label{eq:mr_forcing_earth}
\end{equation}
for the thermal forcing, where
\begin{equation}
T_{\rm vert} \equiv 
\begin{cases}
T_{\rm surf} - \Gamma_{\rm trop} \left( z_{\rm stra}  + \frac{z - z_{\rm stra}}{2} \right) + \left\{ \left[ \frac{\Gamma_{\rm trop}  \left( z - z_{\rm stra} \right)}{2} \right]^2 + \Delta T_{\rm strat}^2 \right\}^{1/2}, & z \le z_{\rm stra}, \\
T_{\rm surf} - \Gamma_{\rm trop} z_{\rm stra} + \Delta T_{\rm stra}, & z > z_{\rm stra}, \\
\end{cases} 
\end{equation}
and
\begin{equation}
\beta_{\rm trop} \equiv
\begin{cases}
\sin\left[\frac{\pi \left( \sigma - \sigma_{\rm stra} \right)}{2 \left(1 - \sigma_{\rm stra} \right)} \right], & z \le z_{\rm stra} \mbox{ or } \sigma \ge \sigma_{\rm stra}, \\
0, & z > z_{\rm stra} \mbox{ or } \sigma < \sigma_{\rm stra}. \\
\end{cases}
\end{equation}
A single value of the Newtonian relaxation time is considered ($\tau_{\rm rad}=15$ days).  We reiterate that $\sigma_{\rm stra}$ is the location of the tropopause in $\sigma$-coordinates.  In \cite{mr09}, the initial temperature is not explicitly specified, so we choose $T_{\rm init}=264$ K following \cite{hs94}.  Also, \cite{mr09} apply Rayleigh friction only to the bottom-most layer of their T42L15 simulation whereas we choose $\sigma_b=0.7$ for our Rayleigh friction scheme.  An important difference between the two schemes is that \cite{mr09} apply Rayleigh friction to the vorticity and divergence fields, while Rayleigh friction is implemented in the \texttt{FMS} as applying directly to the velocity field.  \cite{mr09} consider this test to be a simplified version of the \cite{hs94} benchmark. 

Figure \ref{fig:mr09_earth} shows the Held-Suarez statistics for this benchmark test.  The labelled contours are different from those in Figures \ref{fig:temperature} and \ref{fig:zonal_wind} so as to facilitate direct comparison with Figure 2 of \cite{mr09}.  Our results are temporally averaged over 1000 days, from day 200-1200, while \cite{mr09} present results for day 150.  The temperature profiles computed in our spectral and finite difference simulations are essentially identical; they also match the temperature profile presented in the bottom panel of Figure 2 of \cite{mr09}.  There are some noticeable differences between the zonal-mean zonal winds computed by our spectral and finite difference simulations, yet to a good degree they are mutually consistent and also agree with the top panel of Figure 2 of \cite{mr09}.

\subsubsection{Hot Jupiter}

The shallow hot Jupiter model considers evenly spaced $\sigma$ levels where $P_s=1$ bar.  The thermal forcing implemented by \cite{mr09} is
\begin{equation}
T_{\rm eq} = T_{\rm vert} + \beta_{\rm trop} ~\Delta T_{\rm EP} ~\cos\left(\Theta-180^\circ\right) ~\cos\Phi,
\label{eq:mr_forcing_jup}
\end{equation}
placing the substellar point at $(\Theta=180^\circ, \Phi=0)$.  No Rayleigh friction/drag is implemented, following \cite{mr09}.  Since \cite{mr09} do not specify their choice of the initial temperature, we (arbitrarily) adopt $T_{\rm init}=1800$ K.  The Newtonian relaxation time is half a hot Jupiter day, $\pi/\Omega_p \approx 1.731$ Earth days, which is short enough that the results appear insensitive to the choice of $T_{\rm init}$.  Results produced with $T_{\rm init} = 264$ K and 1470 K are very similar to those reported below.

\begin{figure}
\begin{center}
\includegraphics[width=0.45\columnwidth]{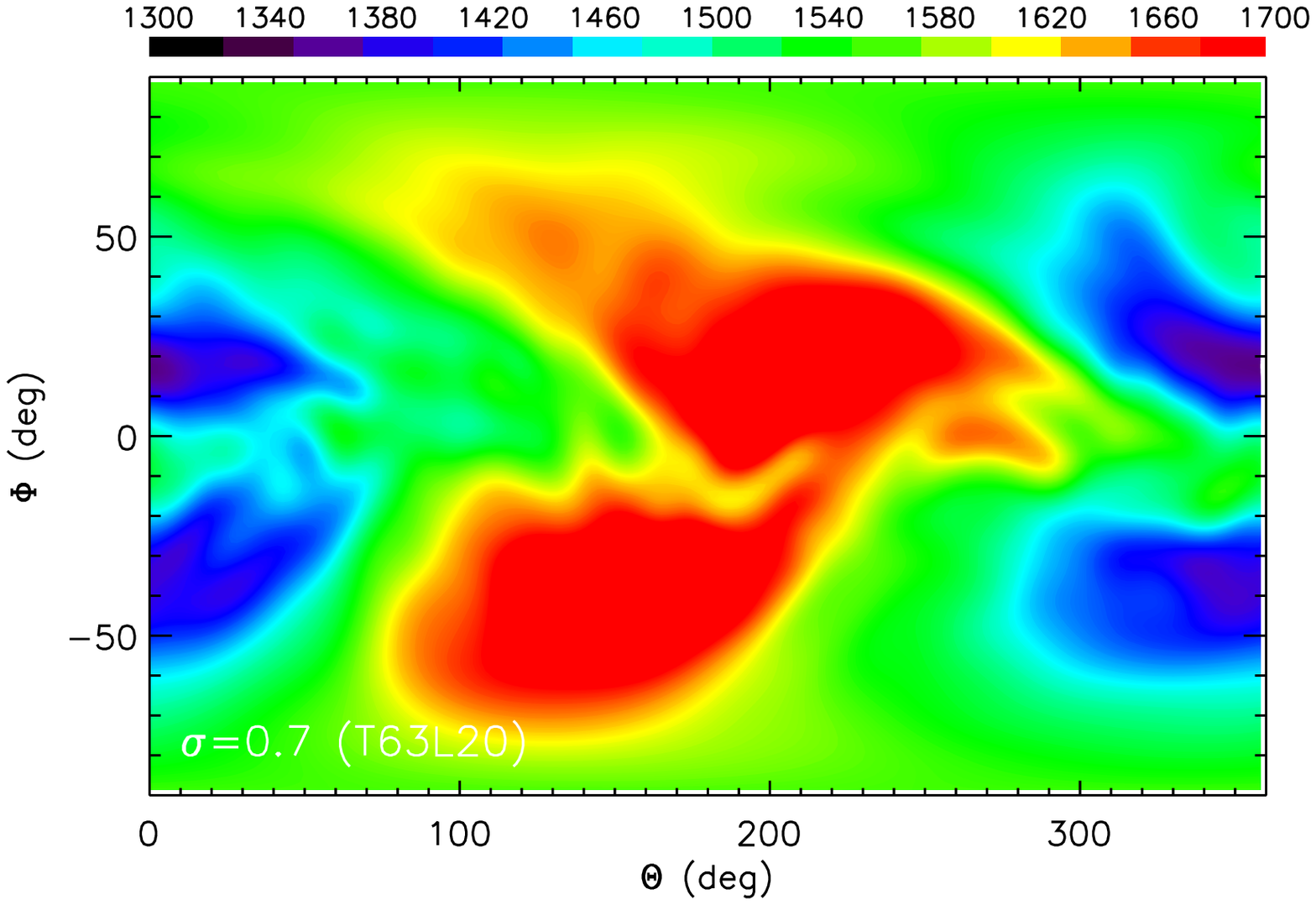}
\includegraphics[width=0.45\columnwidth]{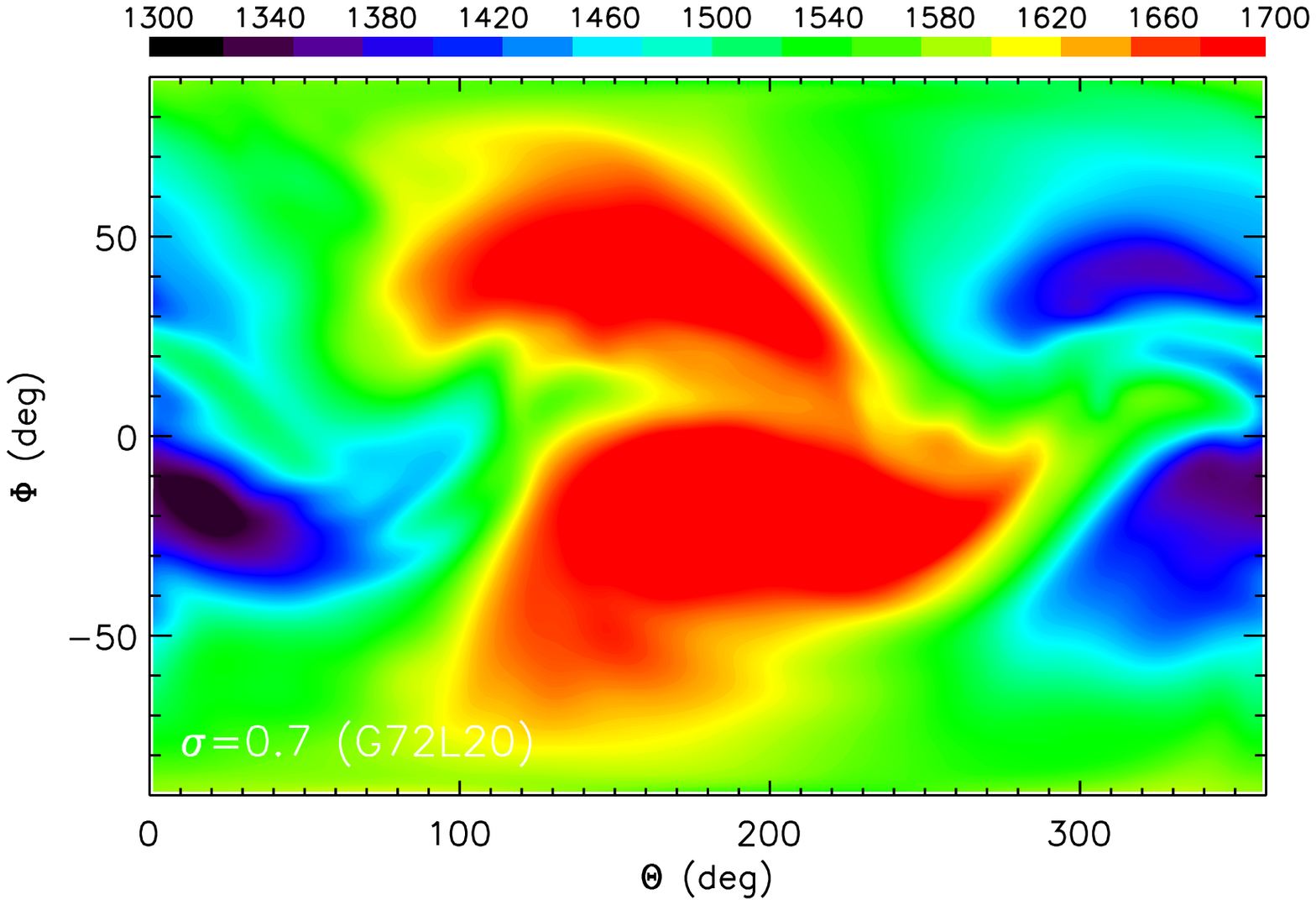}
\includegraphics[width=0.45\columnwidth]{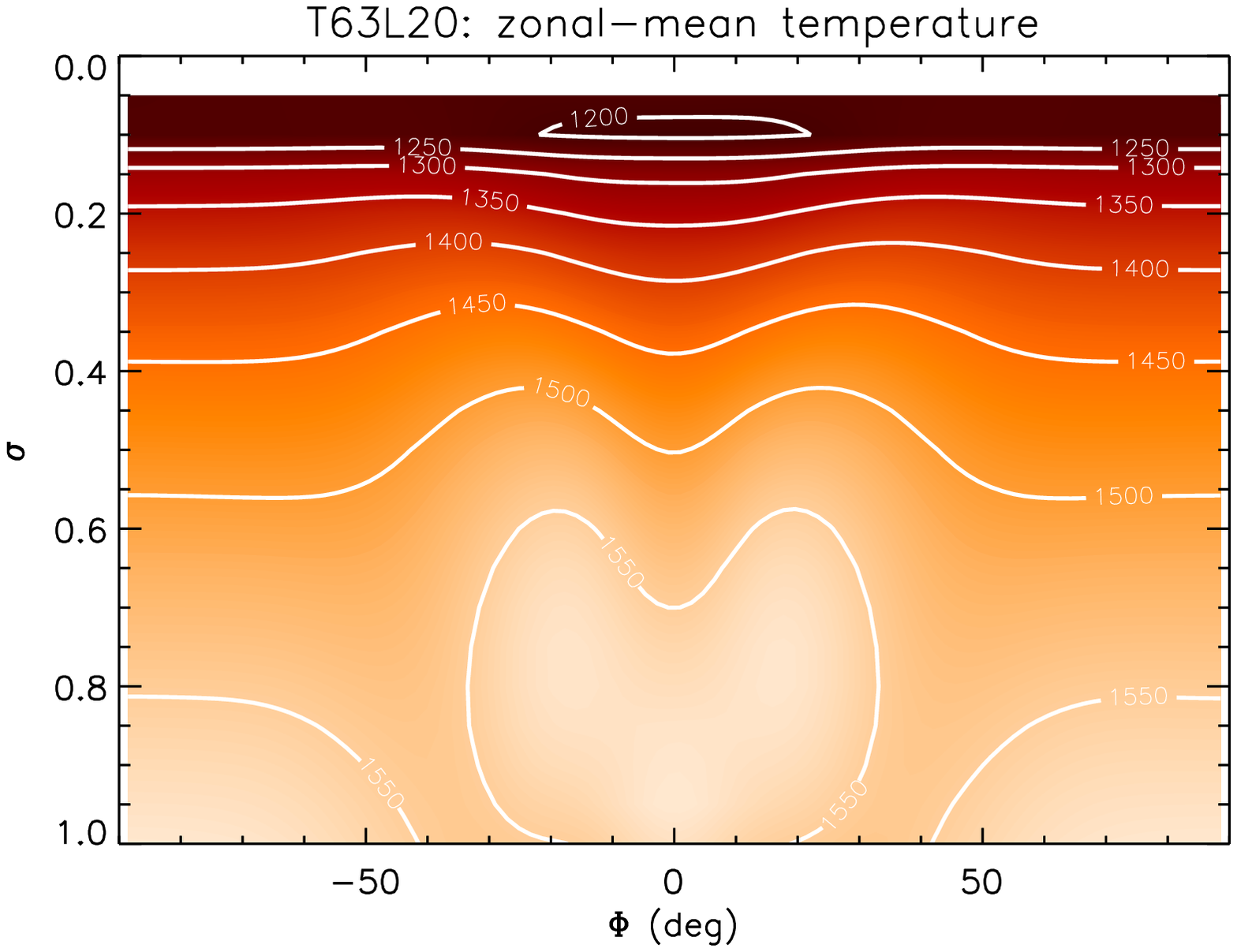}
\includegraphics[width=0.45\columnwidth]{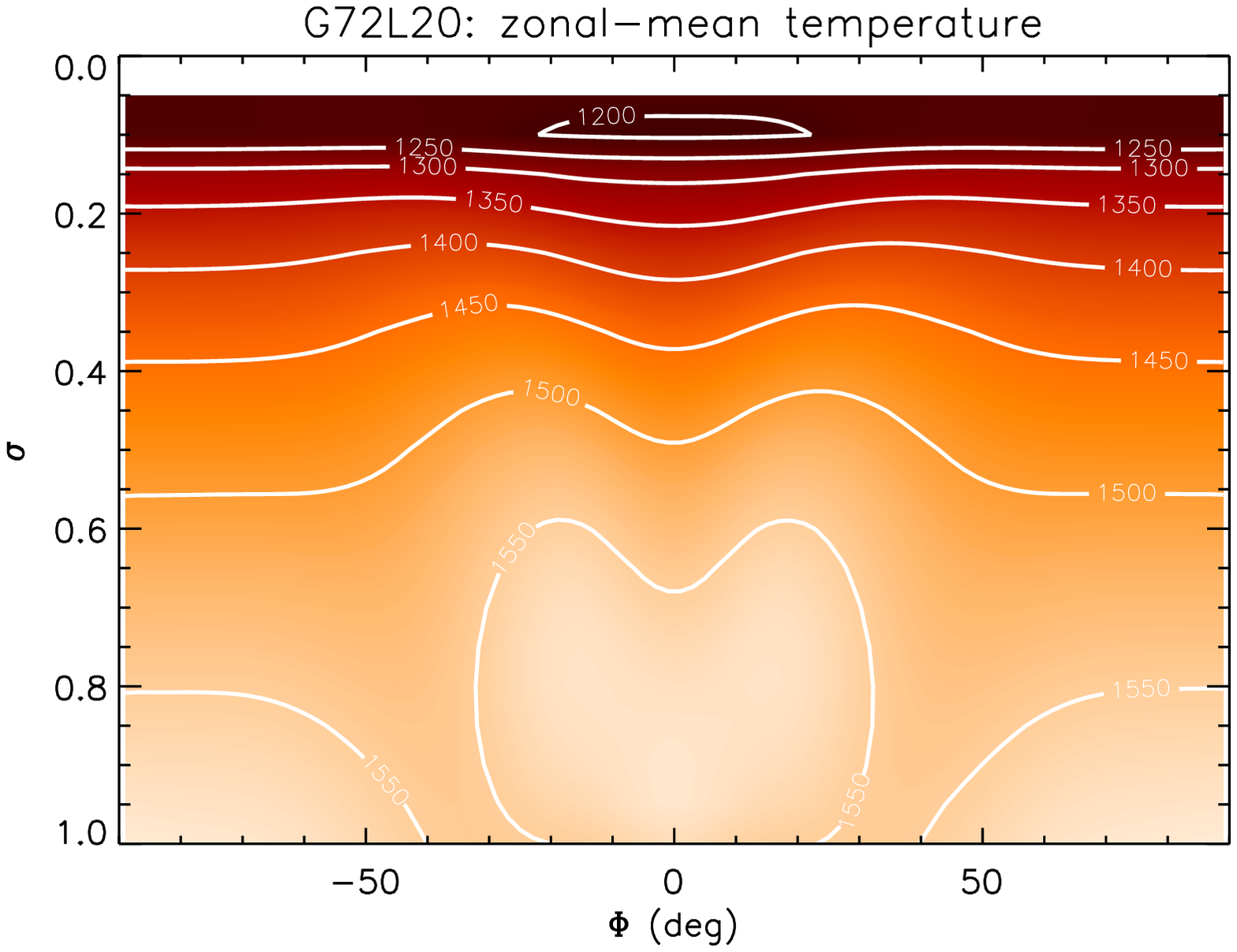}
\includegraphics[width=0.45\columnwidth]{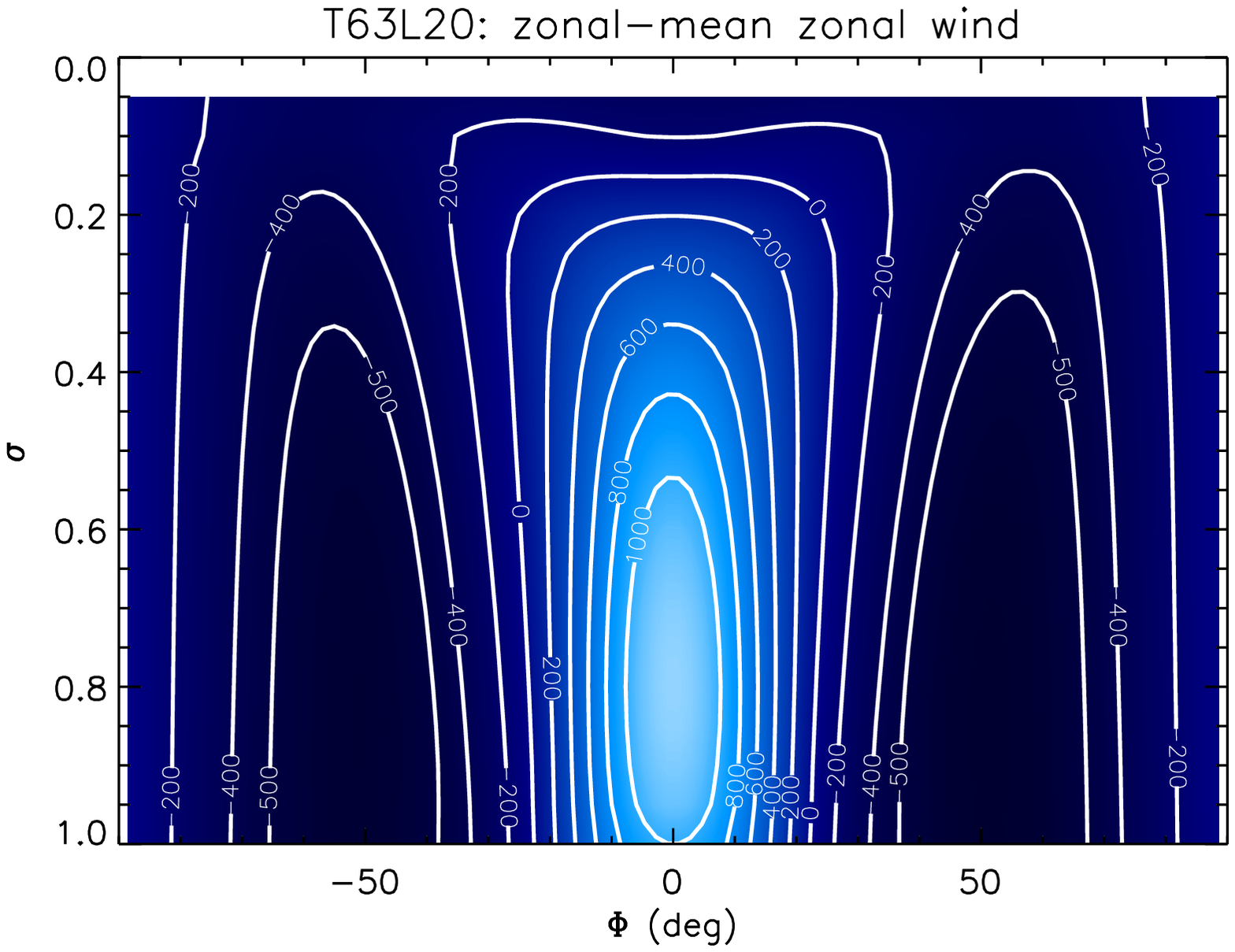}
\includegraphics[width=0.45\columnwidth]{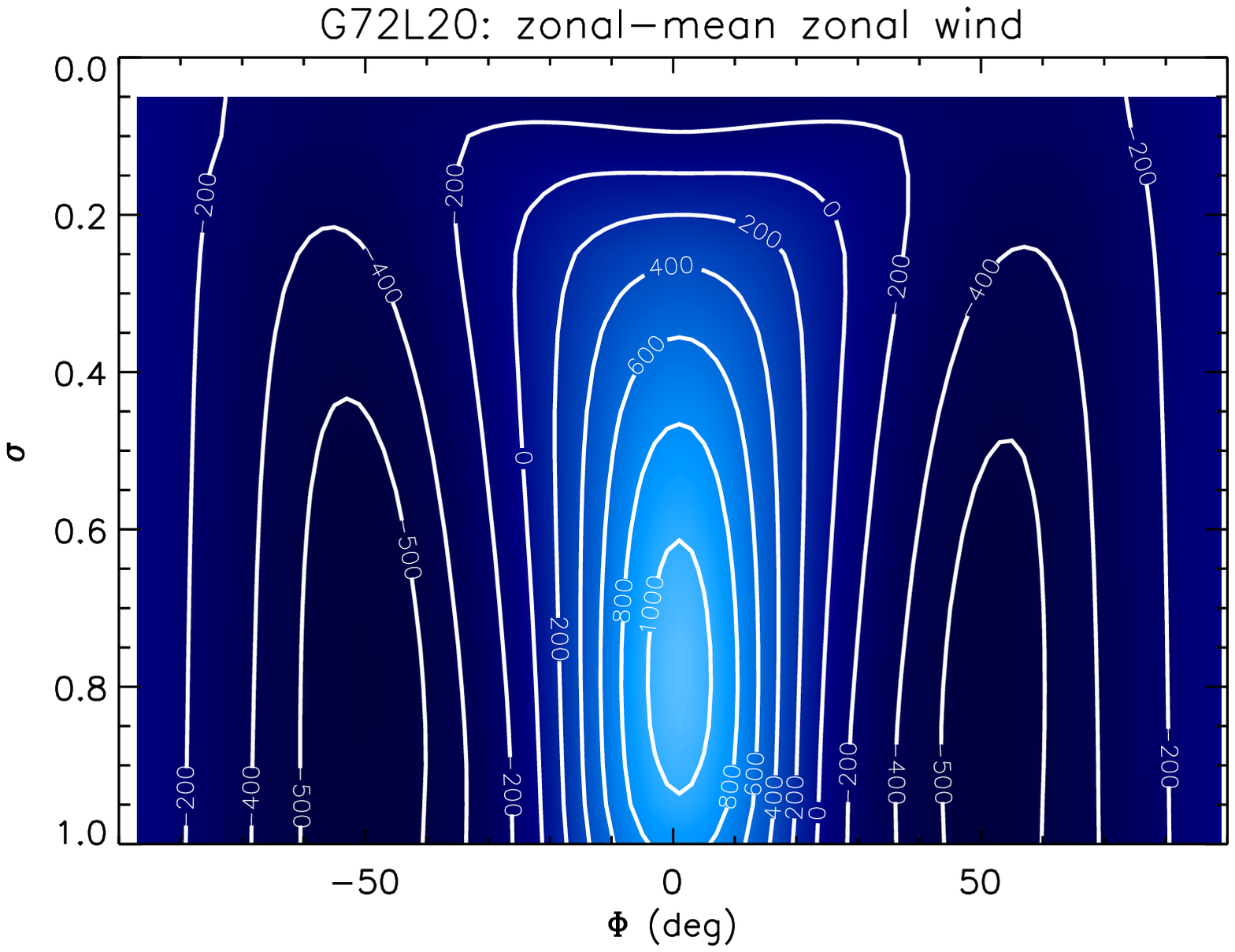}
\end{center}
\vspace{-0.2in}
\caption{Results of spectral (left column; T63L20) and finite difference (right column; G72L20) simulations for the shallow model of hot Jupiters.  Top row: temperature fields at 100 hot Jupiter days ($\approx 346$ Earth days) and $\sigma=0.7$.  Middle row: zonal-mean temperature, temporally averaged over 1000 Earth days.  Bottom row: zonal-mean zonal wind, temporally averaged over 1000 Earth days.  Temperatures are in K and wind speeds are in m s$^{-1}$.}
\label{fig:mr09_jup}
\end{figure}

Both the spectral and finite difference simulations of hot Jupiters were performed using a time step of $\Delta t = 120$ s, a factor of 10 smaller than for the (non-tidally-locked) Earth-like simulations.  As before, we use $t_{\rm total}=1200$ and $t_{\rm discard}=200$ Earth days, both of which are sufficiently long for the simulation to reach quasi-equilibrium; we will retain these values for $t_{\rm total}$ and $t_{\rm discard}$ for simulations involving hot Jupiters throughout the paper.

Figure \ref{fig:mr09_jup} shows the usual Held-Suarez statistics, as well as snapshots of the temperature field at 100 hot Jupiter days ($\approx 346$ Earth days) and $\sigma=0.7$.  These snapshots are meant to be compared to the top panel of Figure 3 of \cite{mr09} --- we can see that there is general agreement with the qualitative features of the flow and the range of temperatures produced.  It is unsurprising that the snapshots are not identical as the time required to reach quasi-equilibrium is slightly different for each simulation --- therefore, one cannot attribute differences in the snapshots (top row of Figure \ref{fig:mr09_jup}) to the different methods of solution (spectral versus finite difference).  A more meaningful comparison is between the Held-Suarez statistics produced, both with Figure 2 of \cite{mr09} and between our pair of simulations (middle and bottom rows of Figure \ref{fig:mr09_jup}), where we witness general agreement.  In general, this point should be kept in mind when examining simulation snapshots produced by different methods of solutions.

Slight quantitative differences between the Held-Suarez statistics
produced by the spectral and finite difference simulations exist.  For
example, the temporally averaged, zonal-mean zonal wind speed ranges
from -630 m s$^{-1}$ to 1239 m s$^{-1}$ for the spectral simulation,
but is -554 m s$^{-1}$ to 1096 m s$^{-1}$ in the finite difference
simulation.  We will not pursue the cause of these differences for the
shallow hot Jupiter model, which are probably due to a combination of
resolution and choice of the magnitude of the horizontal dissipation.
Instead, we feel that the ``deep'' model of HD 209458b provides a more
meaningful exploration of these issues, which we will examine in
\S\ref{subsect:hd209458b}.

\subsection{Deep Benchmark Test for Hot Jupiter: HD 209458b}
\label{subsect:hd209458b}

\cite{rm10} modelled atmospheric circulation on the hot Jupiter HD
209458b from $P=1$ mbar down to $P=220$ bar with $N_v=33$ unevenly
spaced vertical levels.  They find an upper atmosphere dominated by
radiative forcing, due to the short radiative time scales, and an
advection-dominated lower atmosphere with a low level of variability.
General features of the flow (Figure 1 of \citealt{rm10}) agree
qualitatively with Figure 1 of \cite{cs05}, who employed the finite
difference (instead of the spectral) method.  \cite{cs05,cs06} also
modelled the atmosphere of HD 209458b down to deeper levels: $P=3$
kbar instead of 220 bar.

Key differences in the results of \cite{cs05,cs06} and \cite{rm10} are
of interest because they may indicate limitations in the different
methods of solutions for the hot Jupiter
regime.  We summarize the differences:
\begin{enumerate}

\item The super-rotating equatorial jet descends down to only about 7 bar in the simulations of \cite{rm10}, but can be found down to 50 bar in those of \cite{cs05,cs06};

\item The simulations of \cite{cs06} show predominantly super-rotating and counter-rotating flows in the upper and lower atmosphere, respectively.  By contrast, \cite{rm10} find flows in both directions throughout the active layers of the atmosphere.  \cite{rm10} interpret this difference as being due to the deeper reservoir of inert ($\tau_{\rm rad} = \infty$) atmospheric layers in the models of \cite{cs06}, which counter-balances the angular momentum of the super-rotating wind higher up in the atmosphere.

\end{enumerate}

Using both the spectral and finite difference cores of \texttt{FMS}, we will see that the discrepancies described above vanish, implying that they probably arise from differences in initial/boundary conditions as well as setup.  However, our initial simulations reveal other quantitative differences, which we will discuss.

\subsubsection{Setup}

To perform simulations spanning several orders of magnitude in $\sigma$ or $P$, one needs to implement uneven vertical spacing, the technical details of which are described in Appendix \ref{append:levels}.  Figure \ref{eq:uneven_sigma} illustrates the setup needed for a simulation with $N_v=33$ to cover $1 \mbox{ mbar} \lesssim P \le 220 \mbox{ bar}$, similar to the simulation of HD 209458b by \cite{rm10}.

\begin{figure}
\begin{center}
\includegraphics[width=0.45\columnwidth]{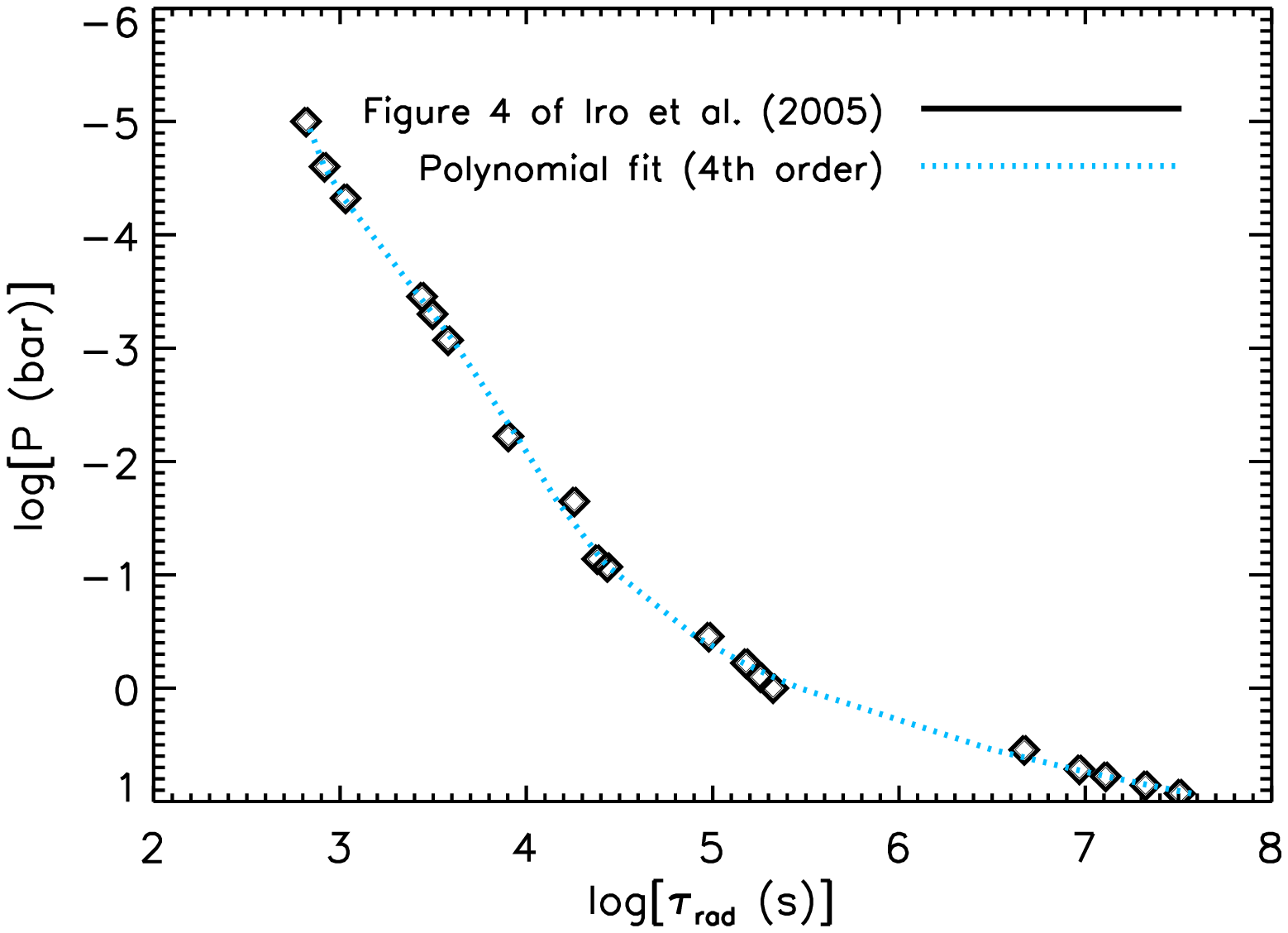}
\includegraphics[width=0.45\columnwidth]{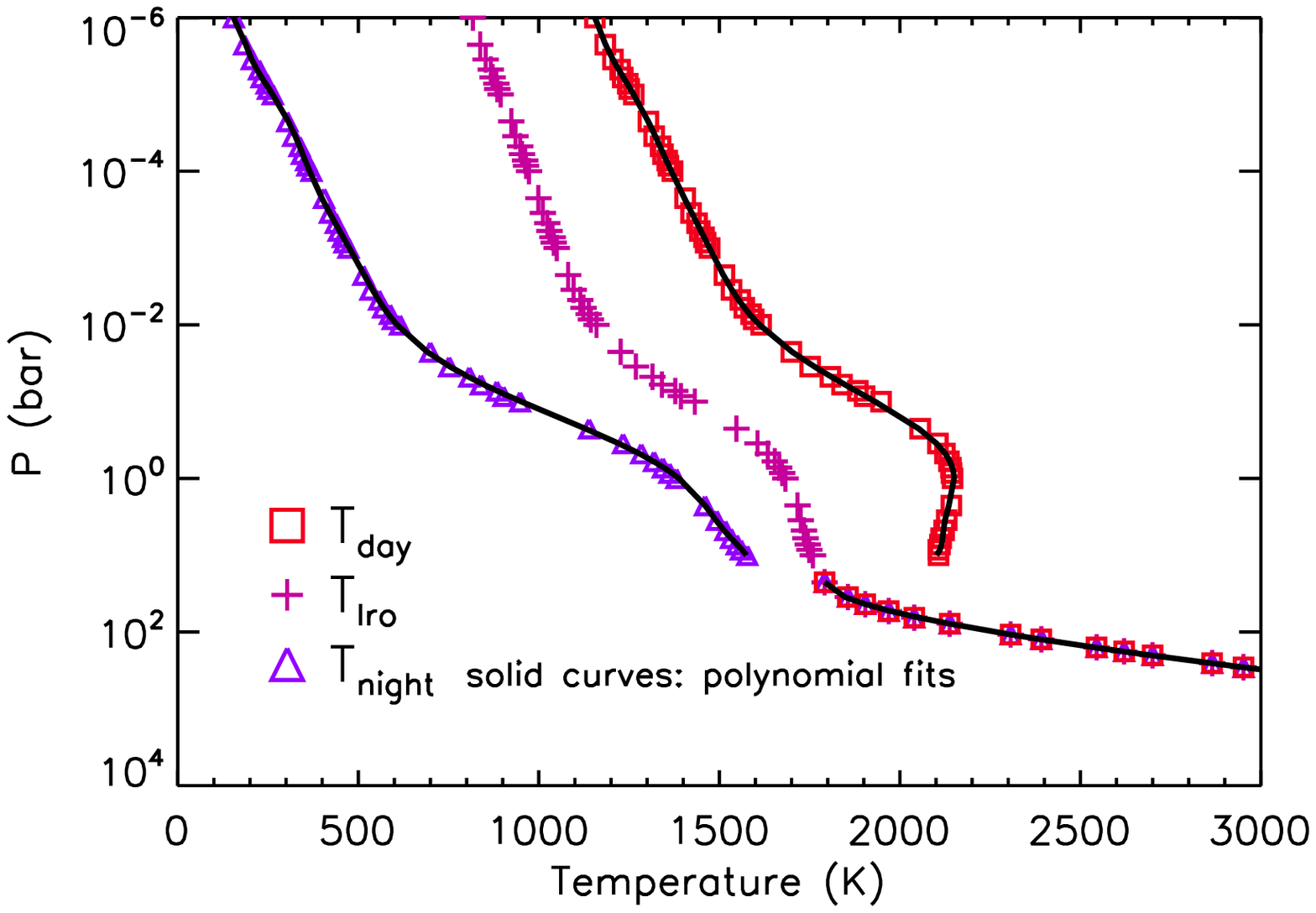}
\end{center}
\vspace{-0.2in}
\caption{Newtonian relaxation time and day/night side temperature profiles as computed for HD 209458b.  Left: diamonds represent the calculations from Figure 4 of Iro et al. (2005), while the dotted curve is our 4th order polynomial fit.  Right: day ($T_{\rm day}$) and night ($T_{\rm night}$) side temperature profiles computed from the globally-averaged profile of Iro et al. (2005; $T_{\rm Iro}$).}
\label{fig:iro05}
\end{figure}

The other ingredients needed are the functional forms of $\tau_{\rm rad}$ and $T_{\rm eq}$.  In the case of HD 209458b, \cite{iro05} have computed the Newtonian relaxation time in their Figure 4 using 1D, time-dependent, radiative transfer models.  \cite{rm10} apply this calculation of $\tau_{\rm rad}$ when $P < 10$ bar.  Appendix \ref{append:fits} contains a polynomial fit to $\tau_{\rm rad}=\tau_{\rm rad}(P)$.  The specification of the radiative relaxation time in turn specifies the ``active'' ($\tau^{-1}_{\rm rad} > 0$) and ``inert'' ($\tau^{-1}_{\rm rad} = 0$) layers of the atmosphere.

The temperature profile for the thermal forcing of HD 209458b is given by equation (2) of \cite{cs05},
\begin{equation}
T_{\rm eq} = 
\begin{cases}
\left[ T^4_{\rm night} + \left( T^4_{\rm day} - T^4_{\rm night} \right)~\cos\left(\Theta-180^\circ\right) ~\cos\Phi  \right]^{1/4}, & 90^\circ \le \Phi \le 270^\circ, \\
T_{\rm night}, & \mbox{ otherwise}, \\
\end{cases}
\label{eq:forcing_hd209458b}
\end{equation}
where $T_{\rm night}$ and $T_{\rm day}$ are the temperature profiles as functions of pressure on the night and day sides, respectively.  \cite{iro05} have computed the globally-averaged (between night and day) temperature profile ($T_{\rm Iro}$; see Appendix \ref{append:fits}), which \cite{rm10} have used to calculate $T_{\rm night}$ and $T_{\rm day}$; see also Figure 1 of \cite{cs06}.  We recompute these profiles using the calculations of \cite{iro05} by solving the transcendental equation for $T_{\rm night}$ (equation [22] of \citealt{cs06}),
\begin{equation}
4 T^4_{\rm Iro} = 3 T^4_{\rm night} + \left(T_{\rm night} + \Delta T_{\rm eq} \right)^4,
\end{equation}
at each value of $P$.  The temperature difference between the night and day sides, $\Delta T_{\rm eq}$, is set equal to 1000 K for $P\le 1$ mbar and 530 K at $P=10$ bar.  In between, $\Delta T_{\rm eq}$ is equally spaced at uniform intervals in $\log{P}$.  Our polynomial fits for $T_{\rm night}$ and $T_{\rm day}$ are given in Appendix \ref{append:fits}.

Figure \ref{fig:iro05} shows the radiative relaxation time and thermal forcing function used in our simulations of HD 209458b.  The initial temperature is set to $T_{\rm Iro}(P=10\mbox{ bar})=1759$ K, which is the value of $T_{\rm eq}$ at $P=10$ bar.  Our use of a constant initial temperature is simpler than what \cite{rm10} implement, which is $T_{\rm init}=T_{\rm night}$ for $P<10$ bar and $T_{\rm init}=T_{\rm Iro}$ otherwise.  In their T31L33 simulation, \cite{mr09} used $N_{\rm lat}=48$ such that small-scale numerical noise is dissipated on a time scale $t_\nu \sim 9 \times 10^{-4}$ HD 209458b day.  For the spectral simulations, we set the dissipation rate to be exactly $t^{-1}_\nu = 0.32785918$ s$^{-1}$ ($t_\nu = 10^{-5}$ HD 209458b day) which is consistent with the \cite{rm10} value (to within a factor of a few) if the scaling relation in equation (\ref{eq:tnu_scale}) is considered.  For the finite difference simulations, we initially adopt ${\cal K}=0.35$ for the horizontal mixing coefficient.  We will explore variations in $t_\nu$ and ${\cal K}$ as well as in $N_{\rm h}$ and $N_{\rm v}$.

\subsubsection{Results}

\begin{figure}
\begin{center}
\includegraphics[width=0.48\columnwidth]{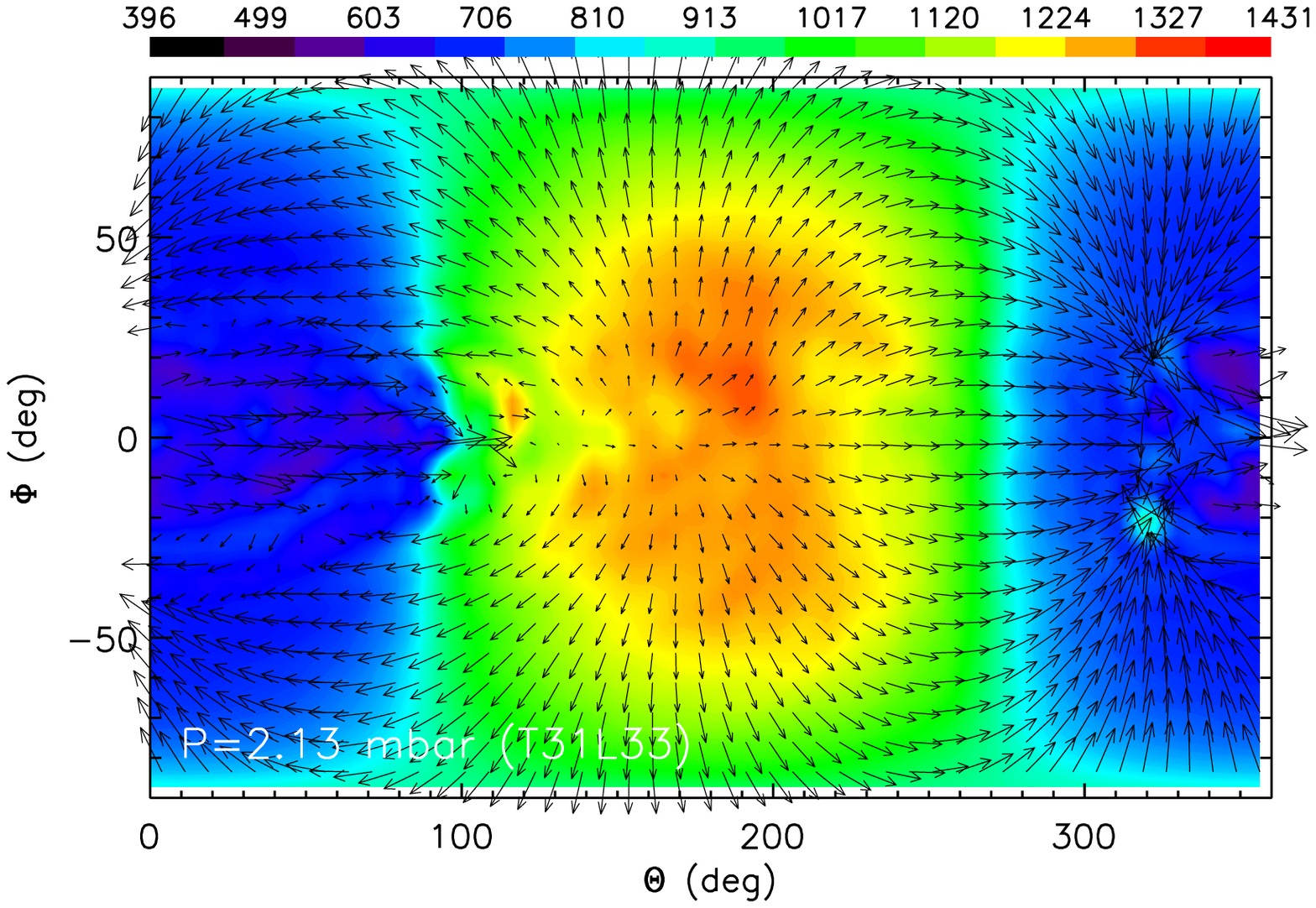}
\includegraphics[width=0.48\columnwidth]{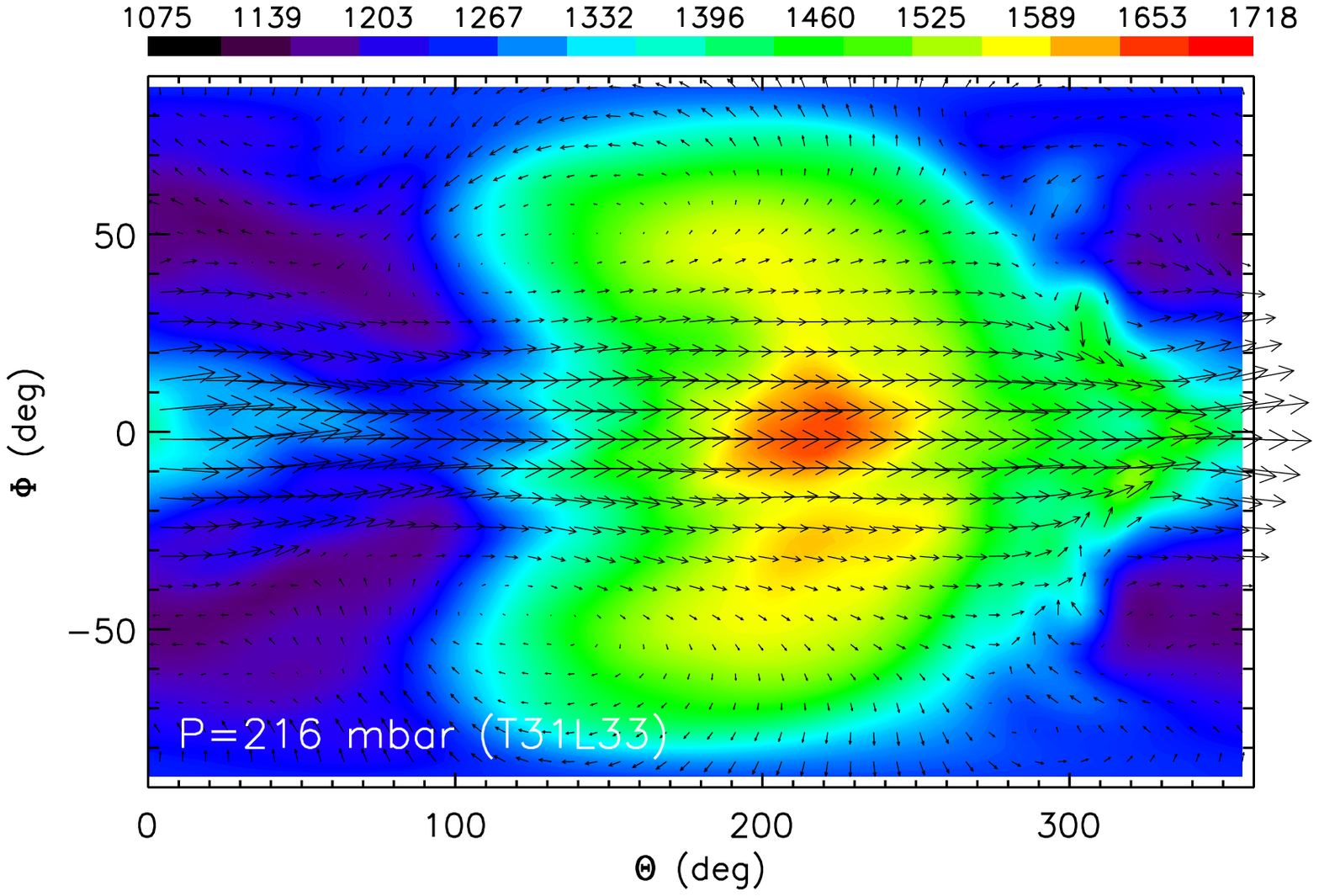}
\includegraphics[width=0.48\columnwidth]{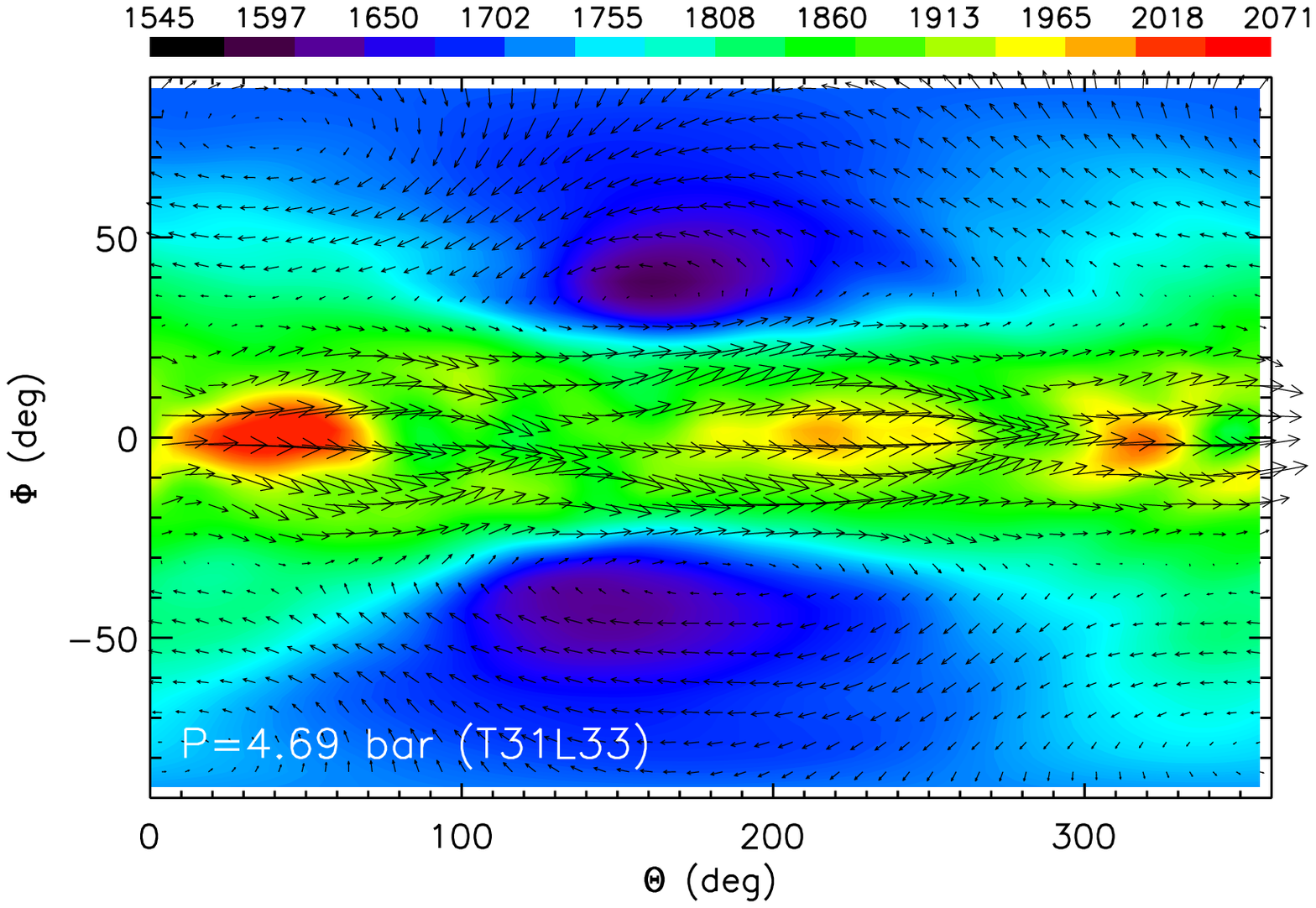}
\includegraphics[width=0.48\columnwidth]{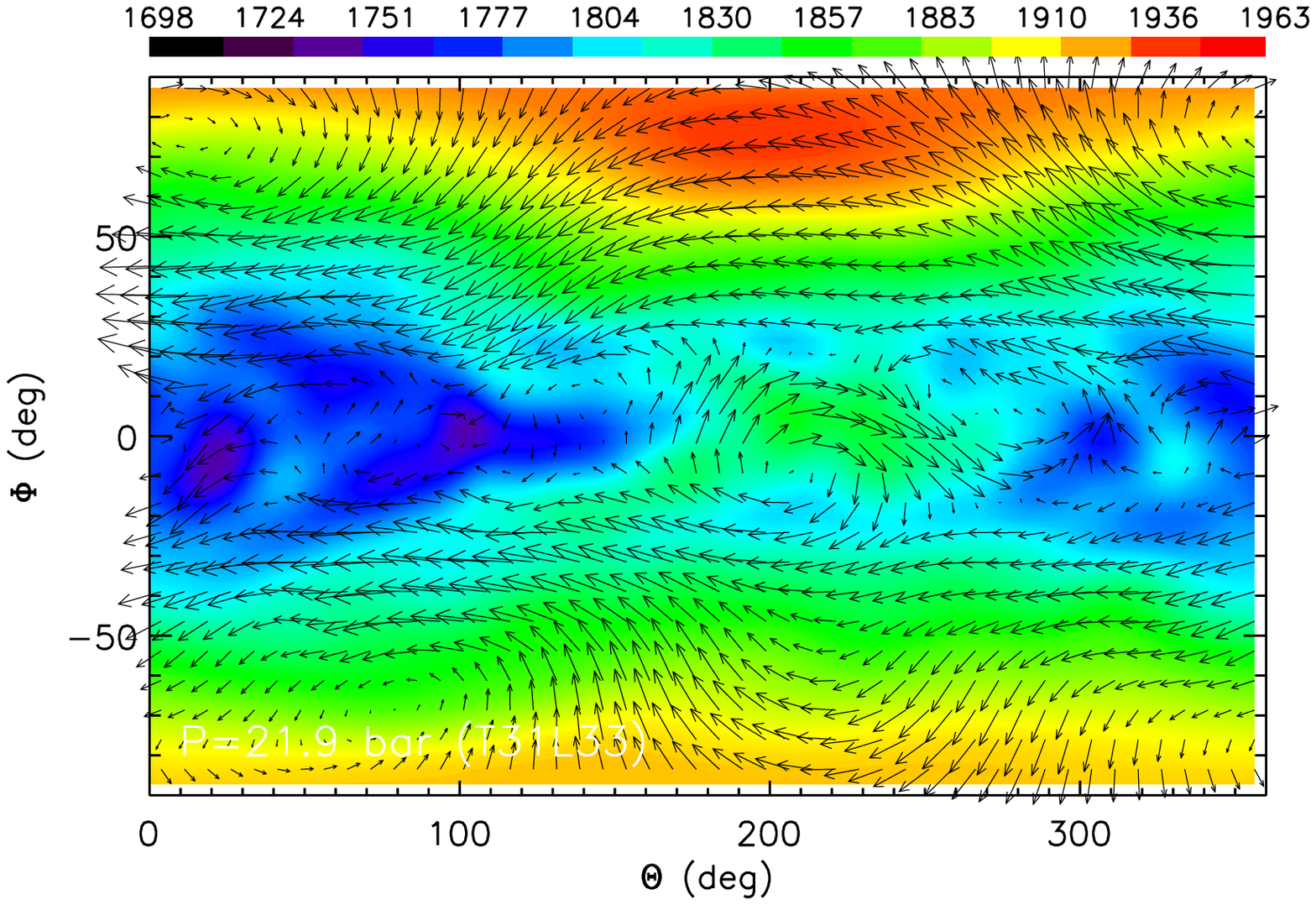}
\end{center}
\vspace{-0.2in}
\caption{Snapshots of the flow field at about 340 HD 209458b days at $P=2.13$ mbar (top left), 216 mbar (top right), 4.69 bar (bottom left) and 21.9 bar (bottom right).  The T31L33 simulation was performed using the \texttt{FMS} spectral core with $t_\nu = 10^{-5}$ HD 209458b day.  Colors indicate temperature in K, while arrows represent the velocity vectors.}
\label{fig:rm10_hd209458b}
\end{figure}

\begin{figure}
\begin{center}
\includegraphics[width=0.48\columnwidth]{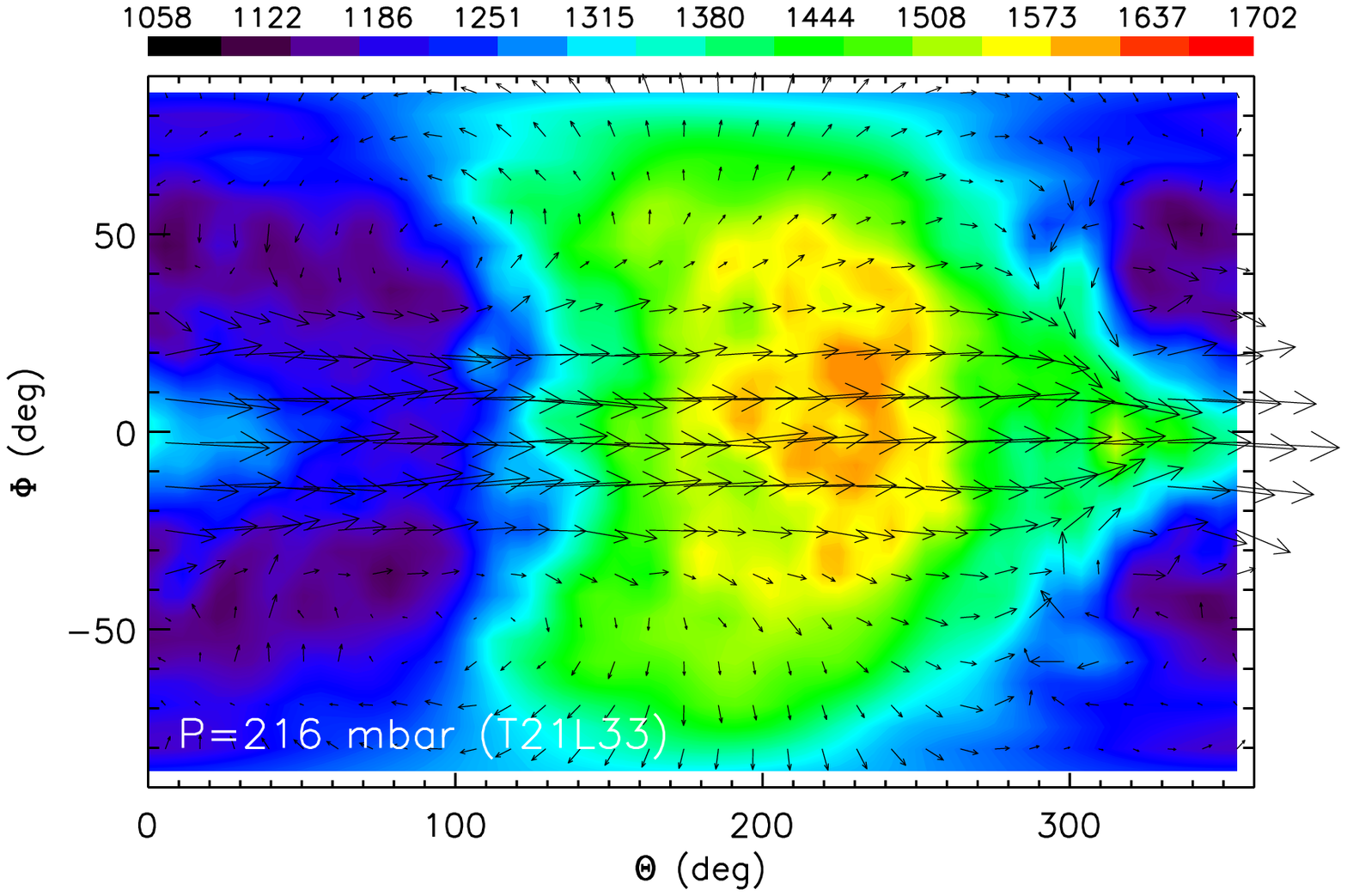}
\includegraphics[width=0.48\columnwidth]{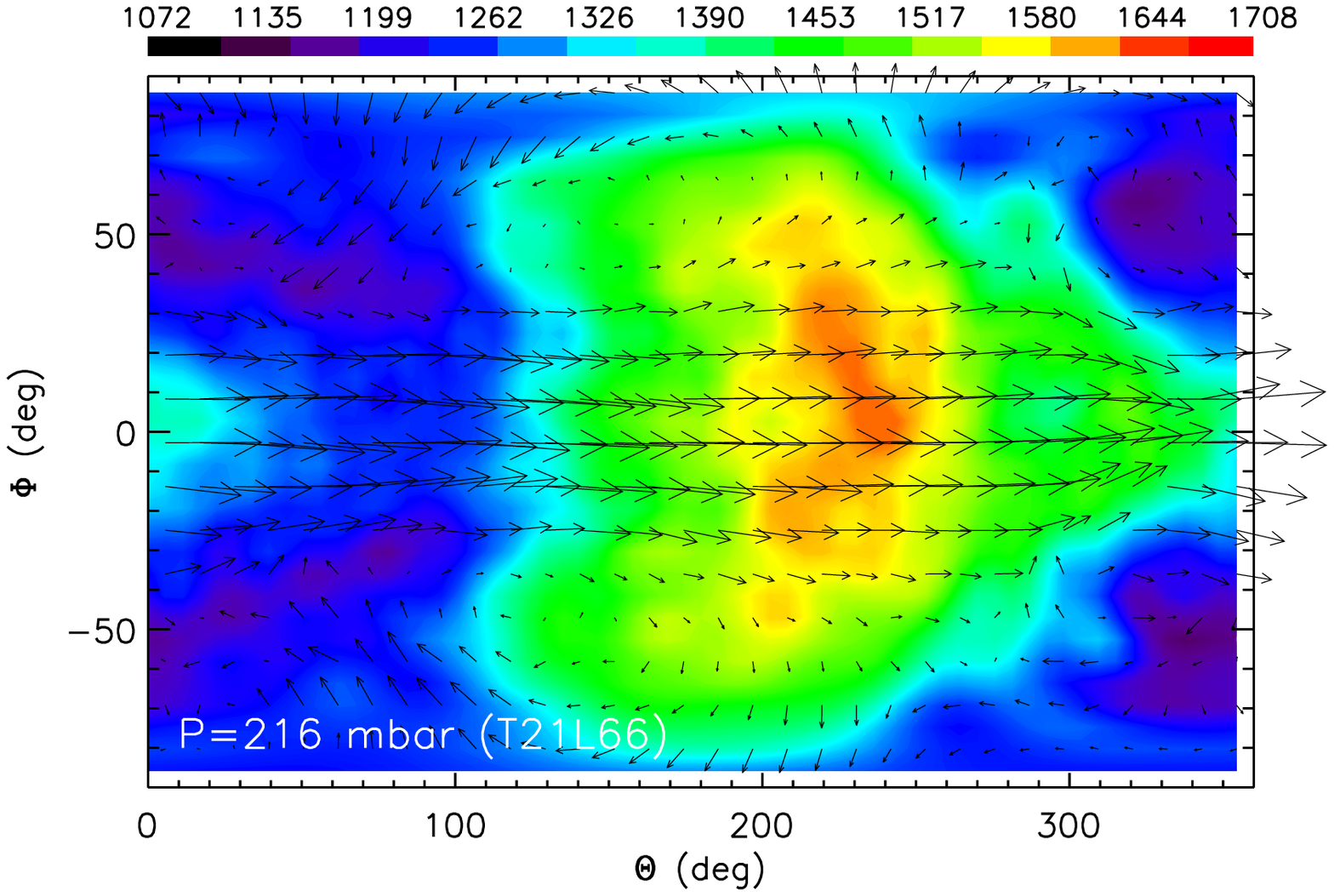}
\includegraphics[width=0.48\columnwidth]{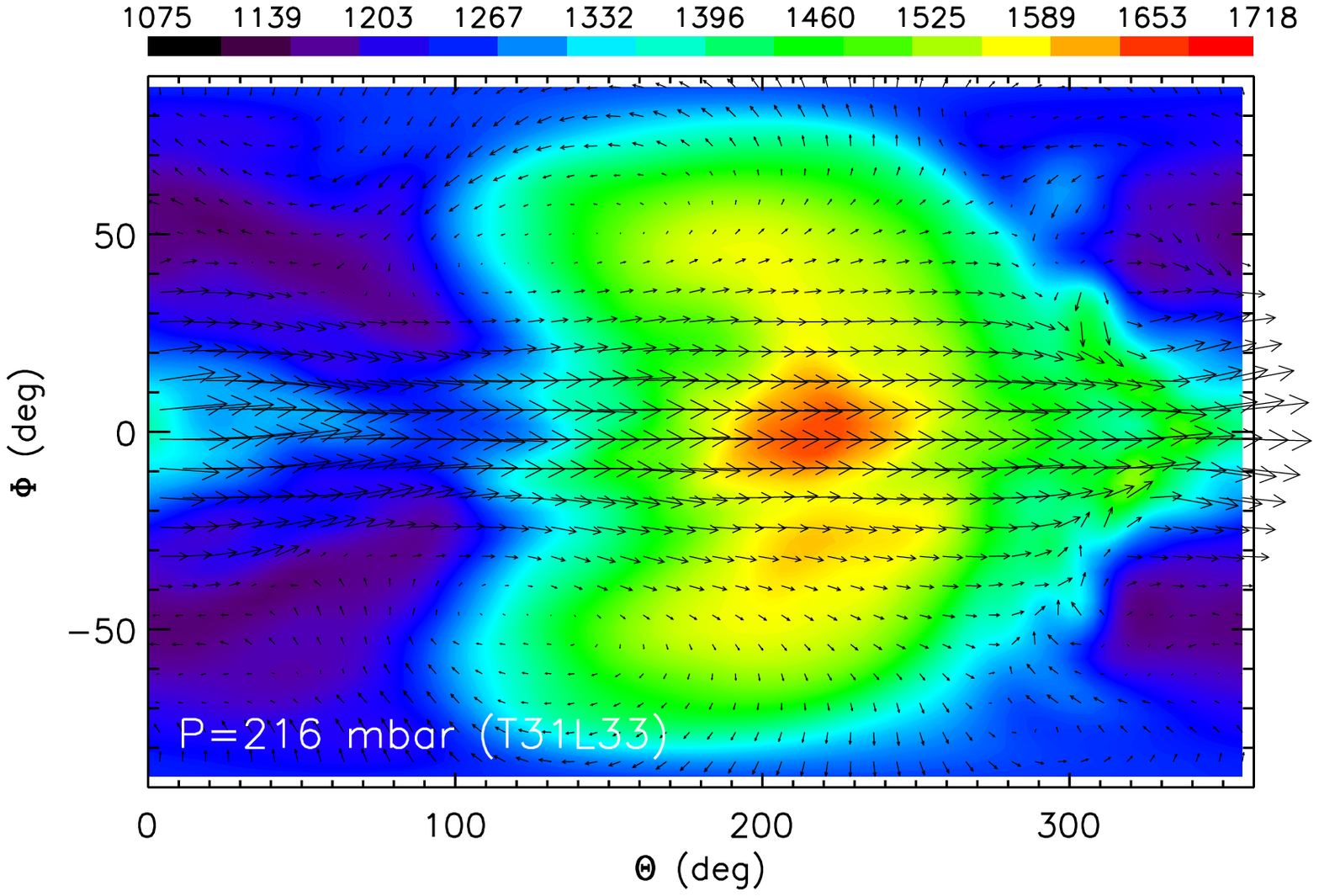}
\includegraphics[width=0.48\columnwidth]{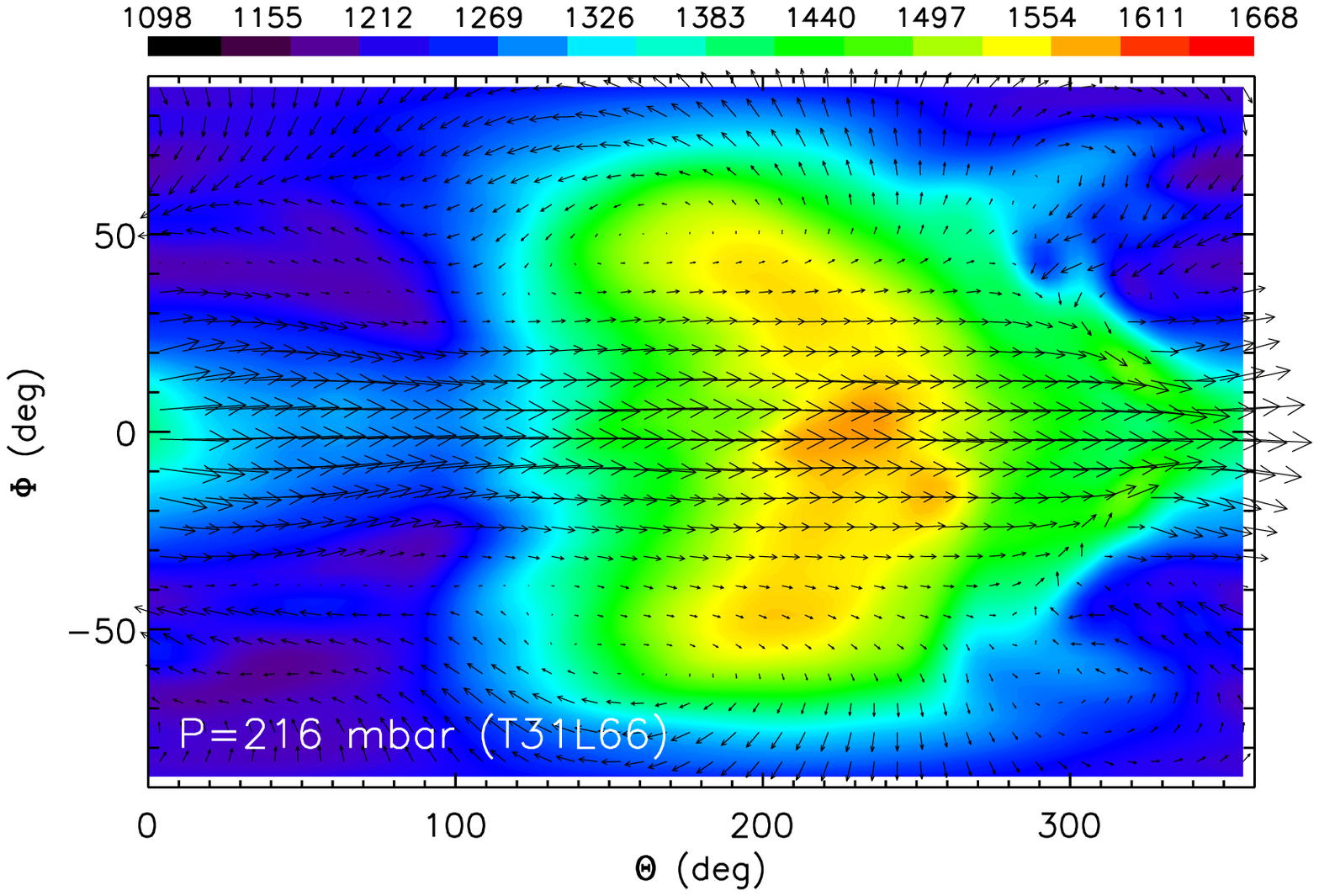}
\includegraphics[width=0.48\columnwidth]{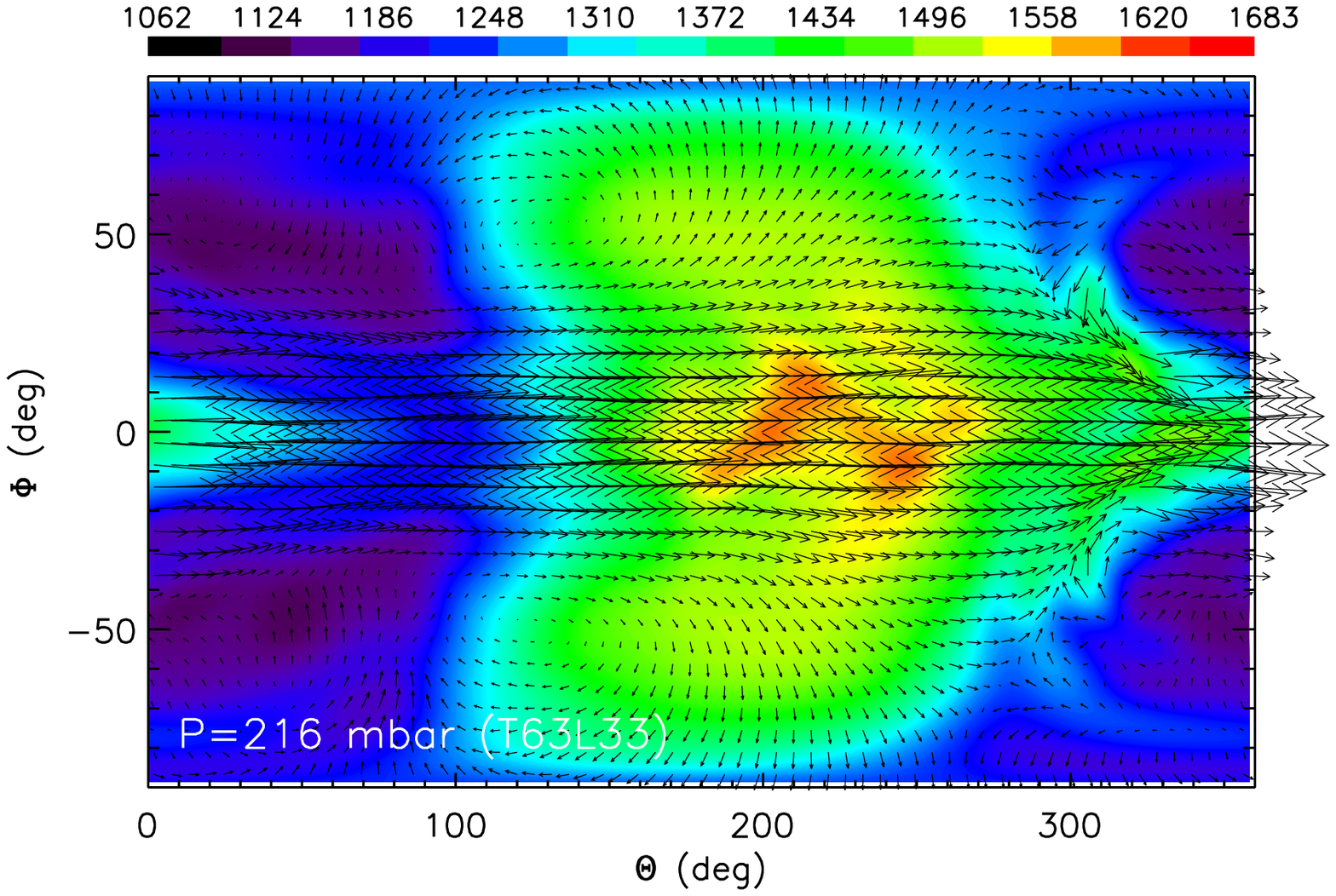}
\includegraphics[width=0.48\columnwidth]{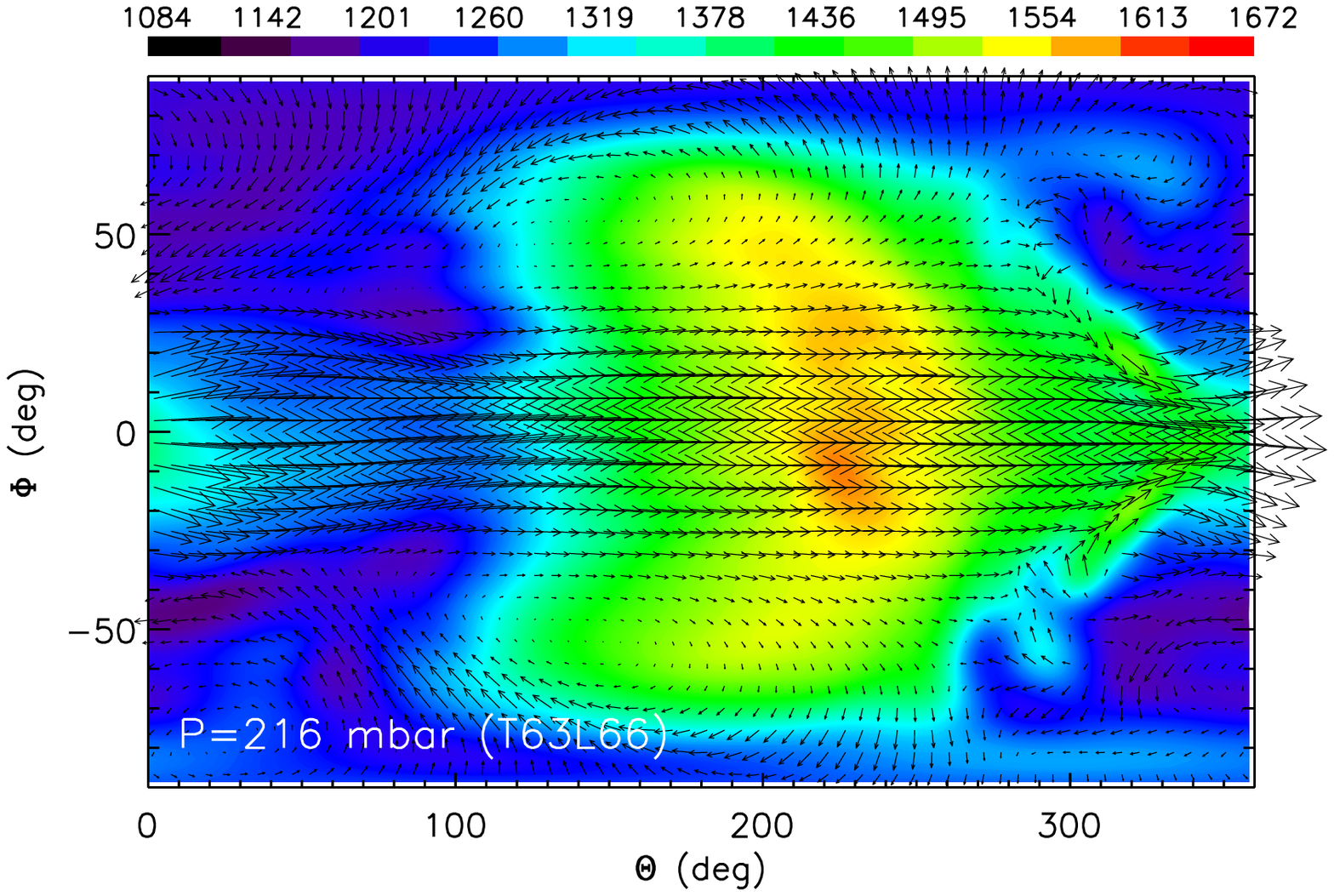}
\end{center}
\vspace{-0.2in}
\caption{Snapshots of the flow field at about 340 HD 209458b days and at $P=216$ mbar.  Shown are results from the T21L33 (top left), T21L66 (top right), T31L33 (middle left), T31L66 (middle right), T63L33 (bottom left) and T63L66 (bottom right) spectral simulations.  The hyperviscosity $\nu$ is kept fixed such that the dissipation time varies and scales up when the resolution coarsens (see equation [\ref{eq:tnu_scale}]).  Colors indicate temperature in K, while arrows represent the velocity vectors.  In these plots, the color bar range is fixed for clarity of comparison.}
\label{fig:rm10_hd209458b_res}
\end{figure}

\begin{figure}
\begin{center}
\includegraphics[width=0.48\columnwidth]{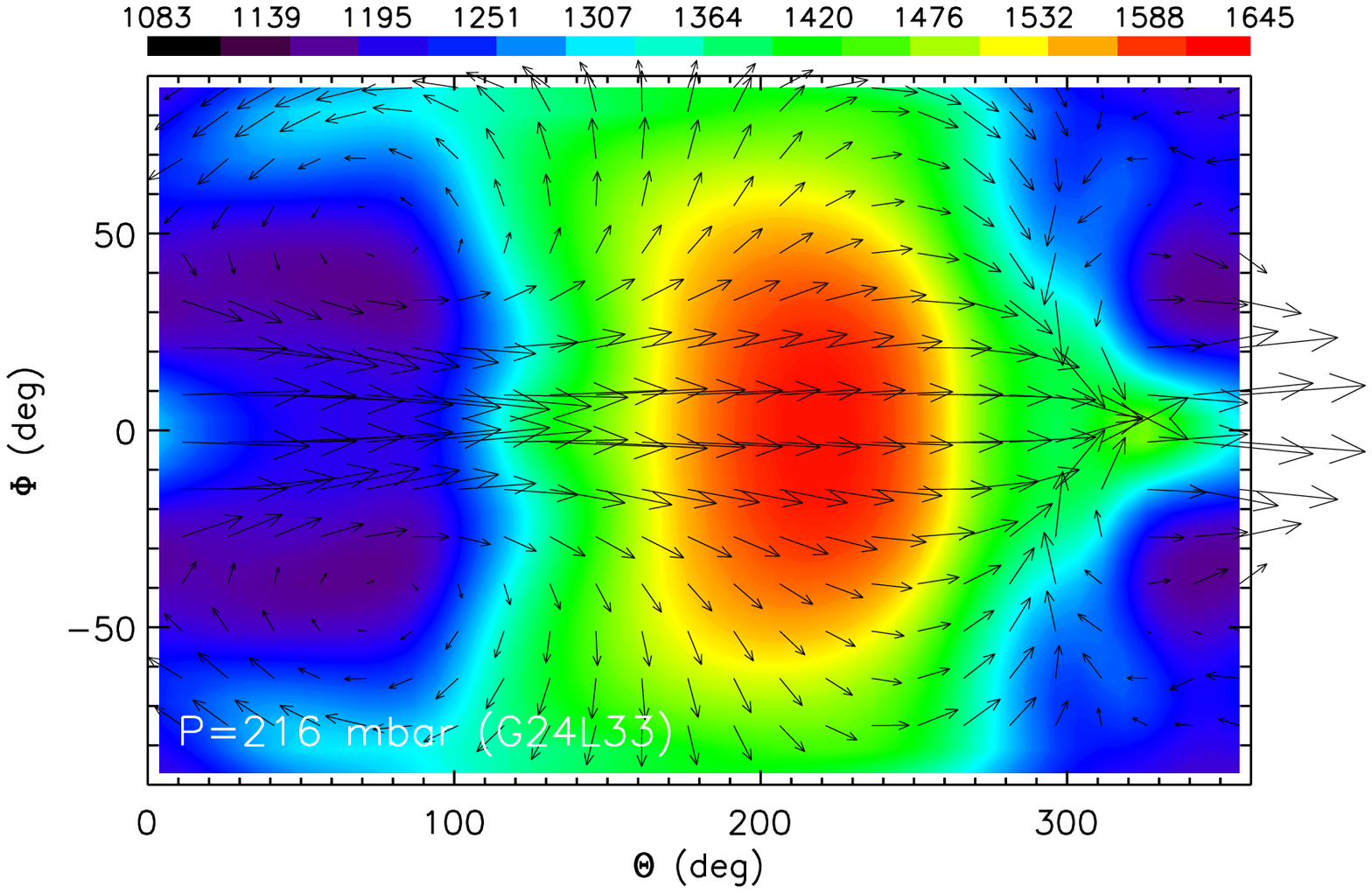}
\includegraphics[width=0.48\columnwidth]{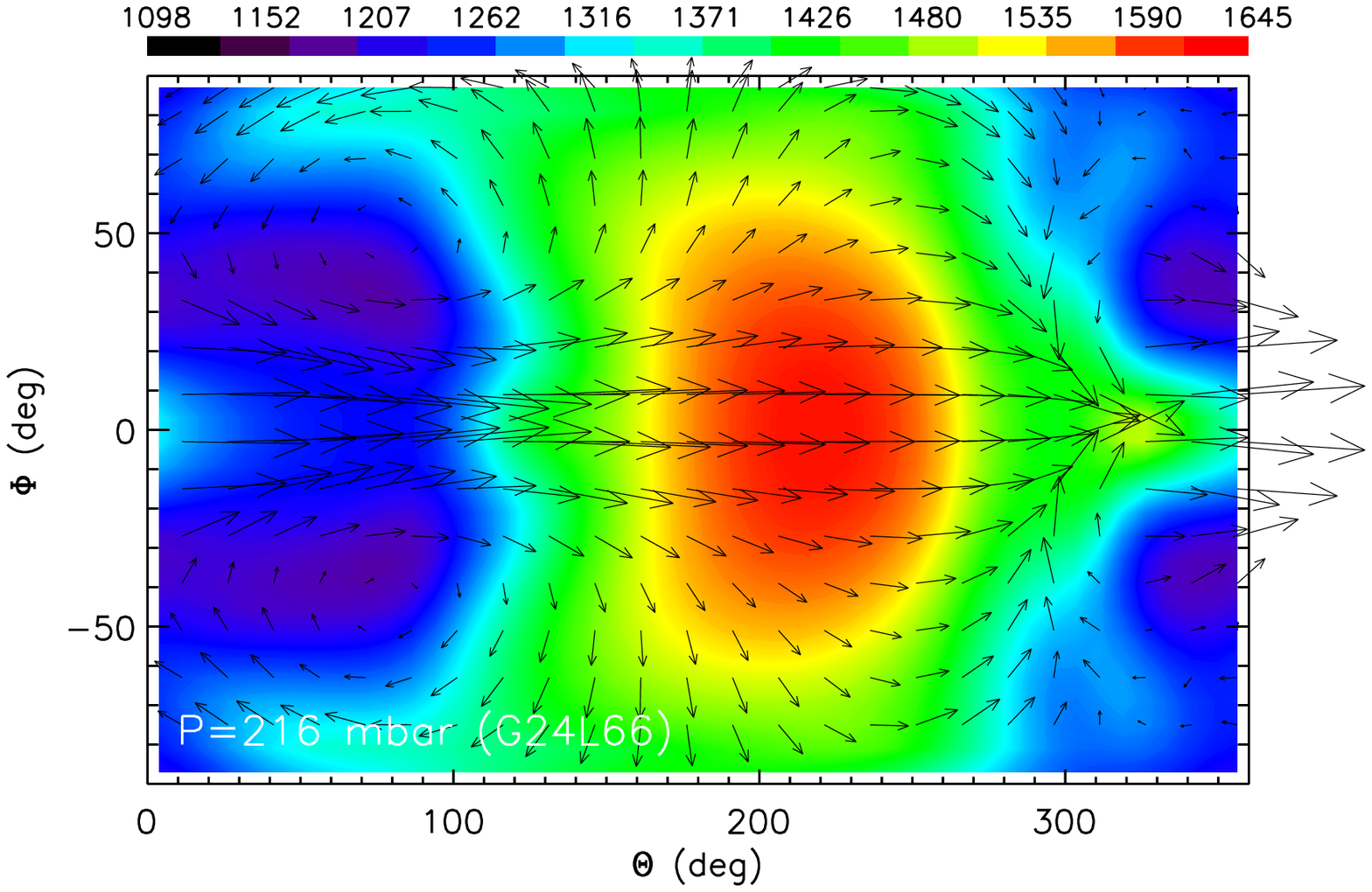}
\includegraphics[width=0.48\columnwidth]{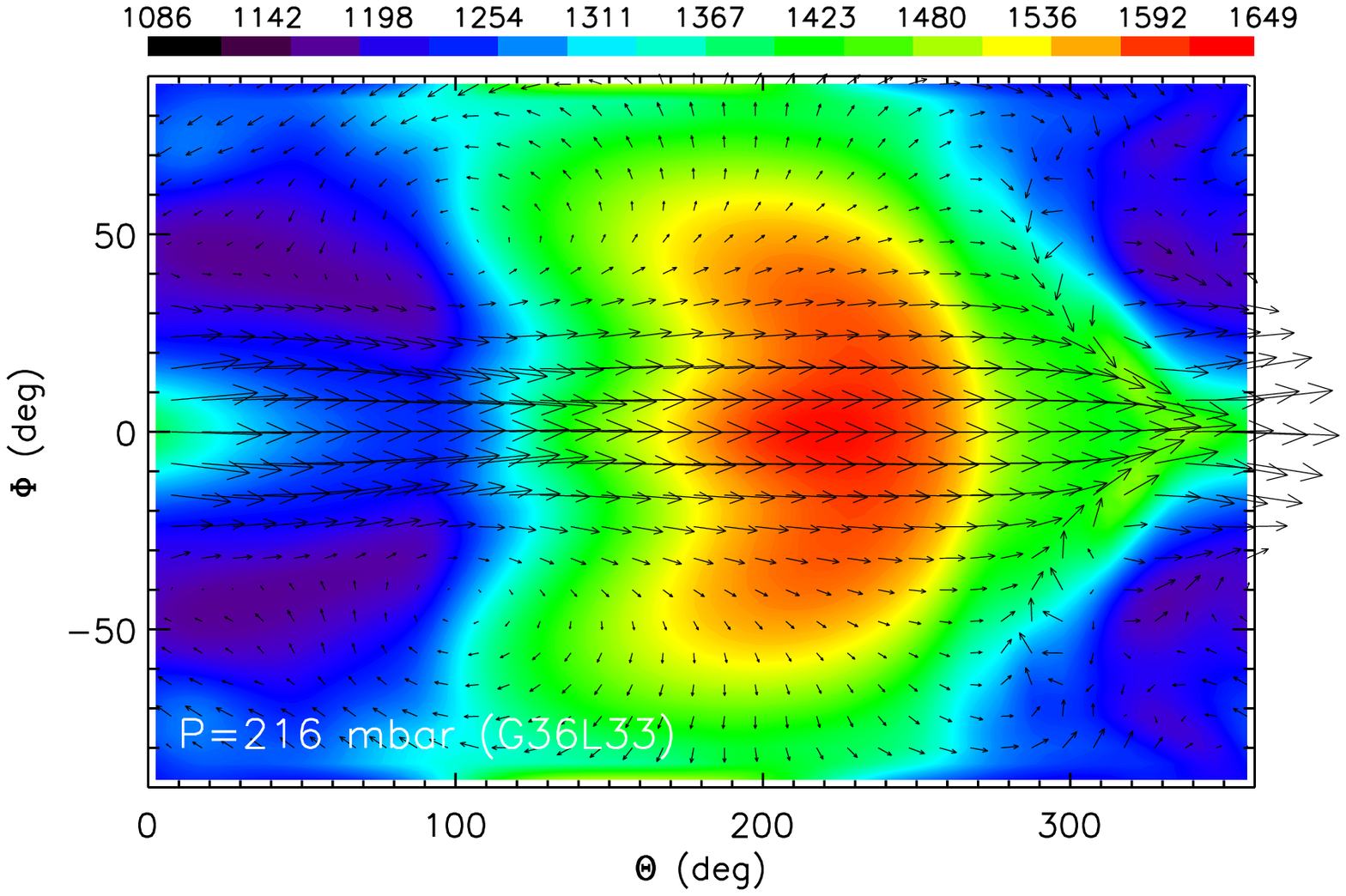}
\includegraphics[width=0.48\columnwidth]{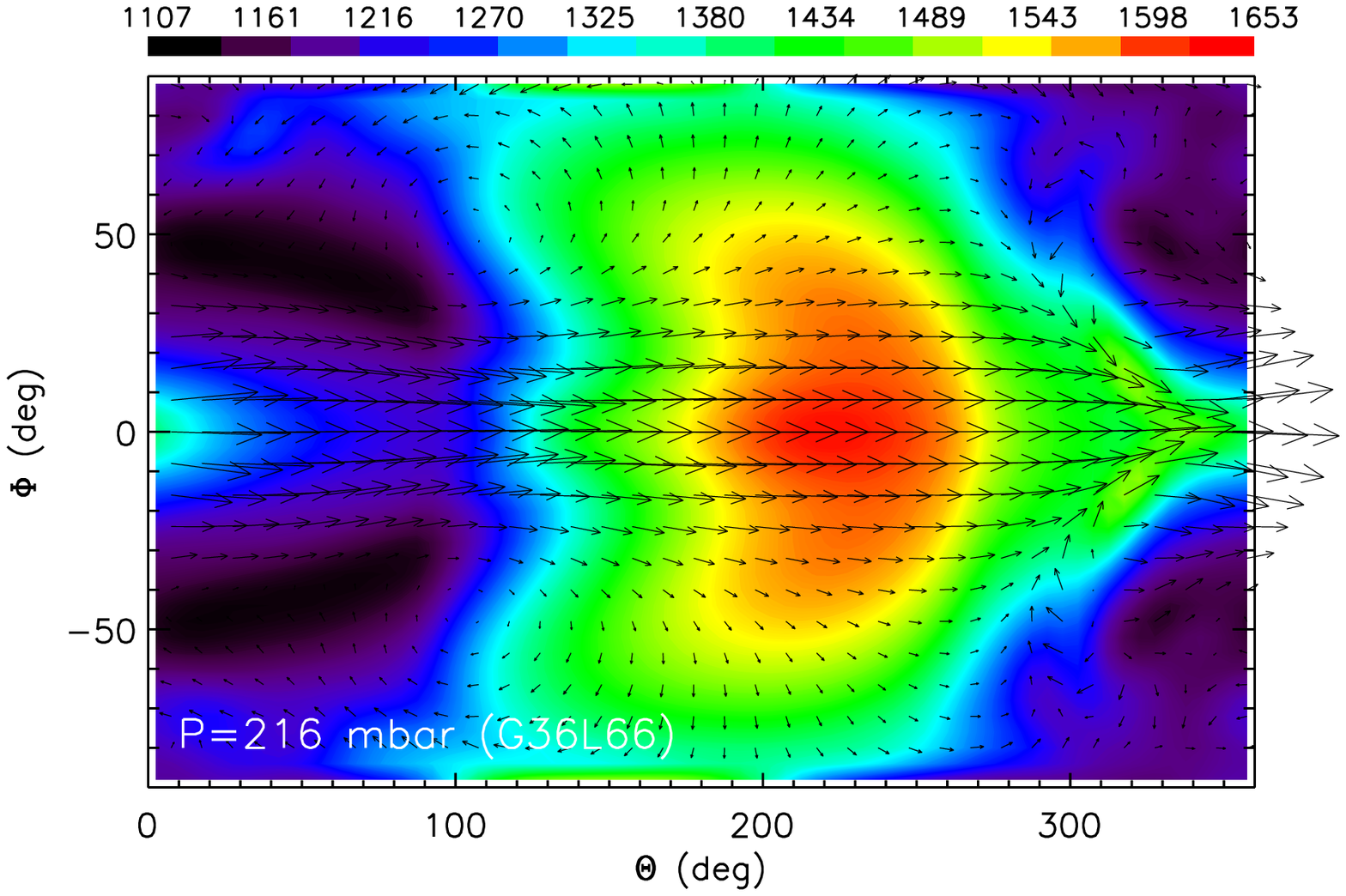}
\includegraphics[width=0.48\columnwidth]{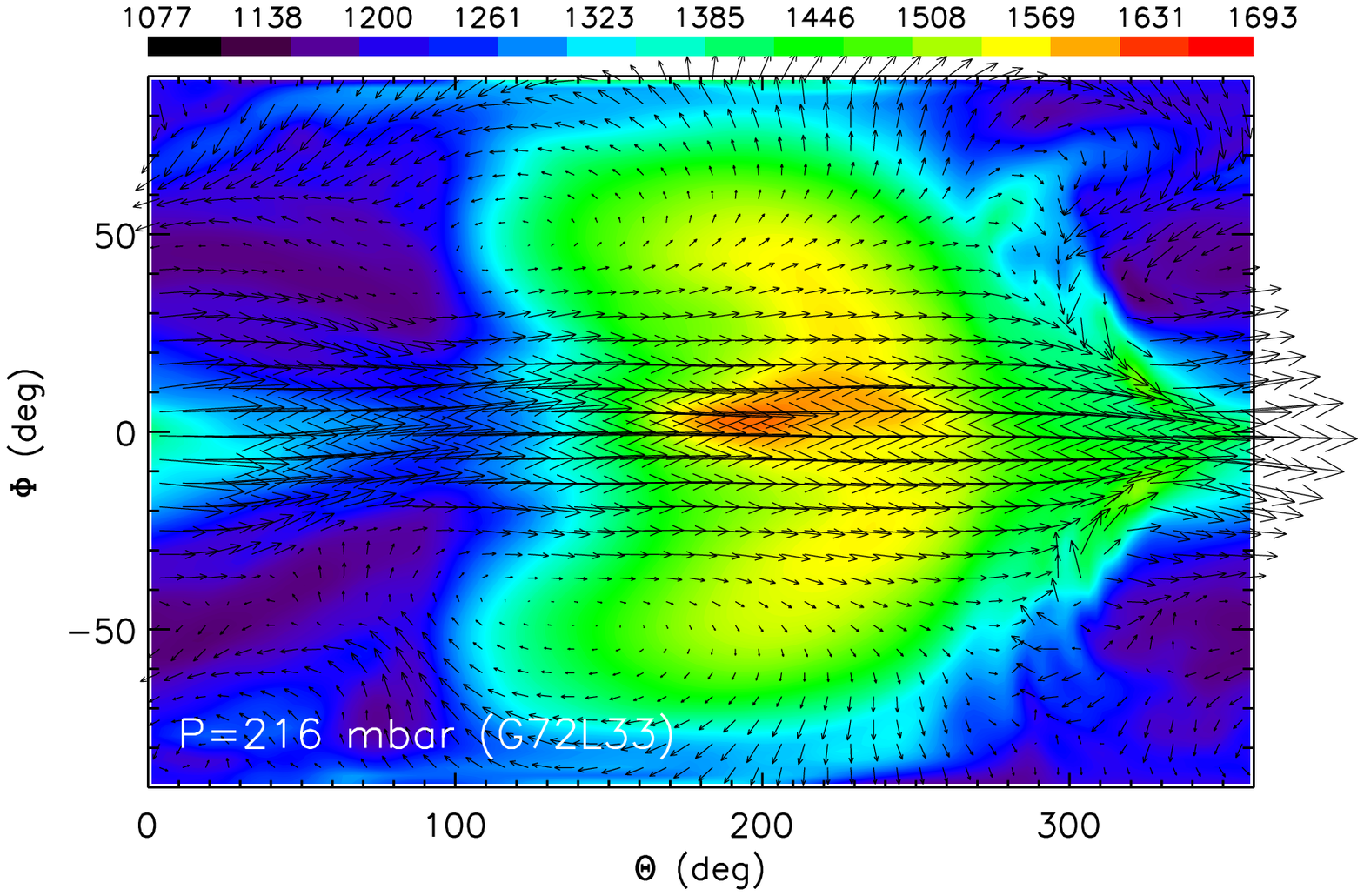}
\includegraphics[width=0.48\columnwidth]{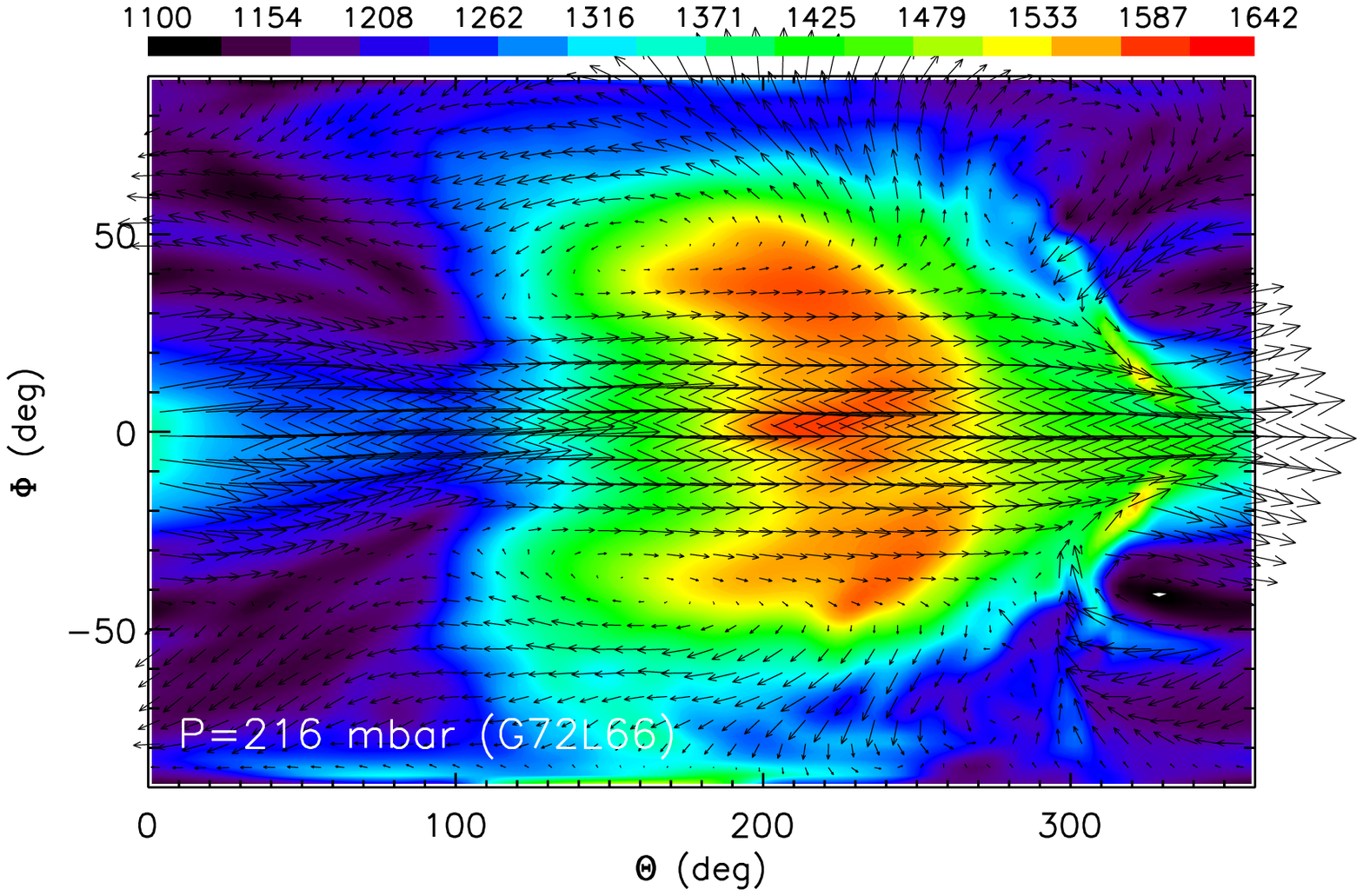}
\end{center}
\vspace{-0.2in}
\caption{Same as Figure \ref{fig:rm10_hd209458b_res}, but for the finite difference core with ${\cal K}=0.35$.  Shown are the G24L33 (top left), G24L66 (top right), G36L33 (middle left), G36L66 (middle right), G72L33 (bottom left) and G72L66 (bottom right) simulations.}
\label{fig:rm10_hd209458b_res2}
\end{figure}

\begin{figure}
\begin{center}
\includegraphics[width=0.55\columnwidth]{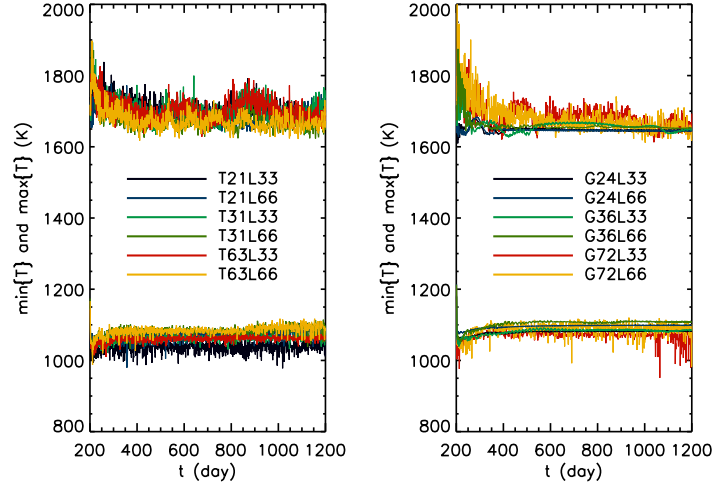}
\end{center}
\vspace{-0.2in}
\caption{Minimum and maximum temperatures at $P=216$ mbar, for both spectral (left panel) and finite difference (right panel) simulations, as functions of time in Earth days and as computed from the simulation snapshots taken in Figures \ref{fig:rm10_hd209458b_res} and \ref{fig:rm10_hd209458b_res2}.}
\label{fig:temp_var}
\end{figure}

\begin{figure}
\begin{center}
\includegraphics[width=0.45\columnwidth]{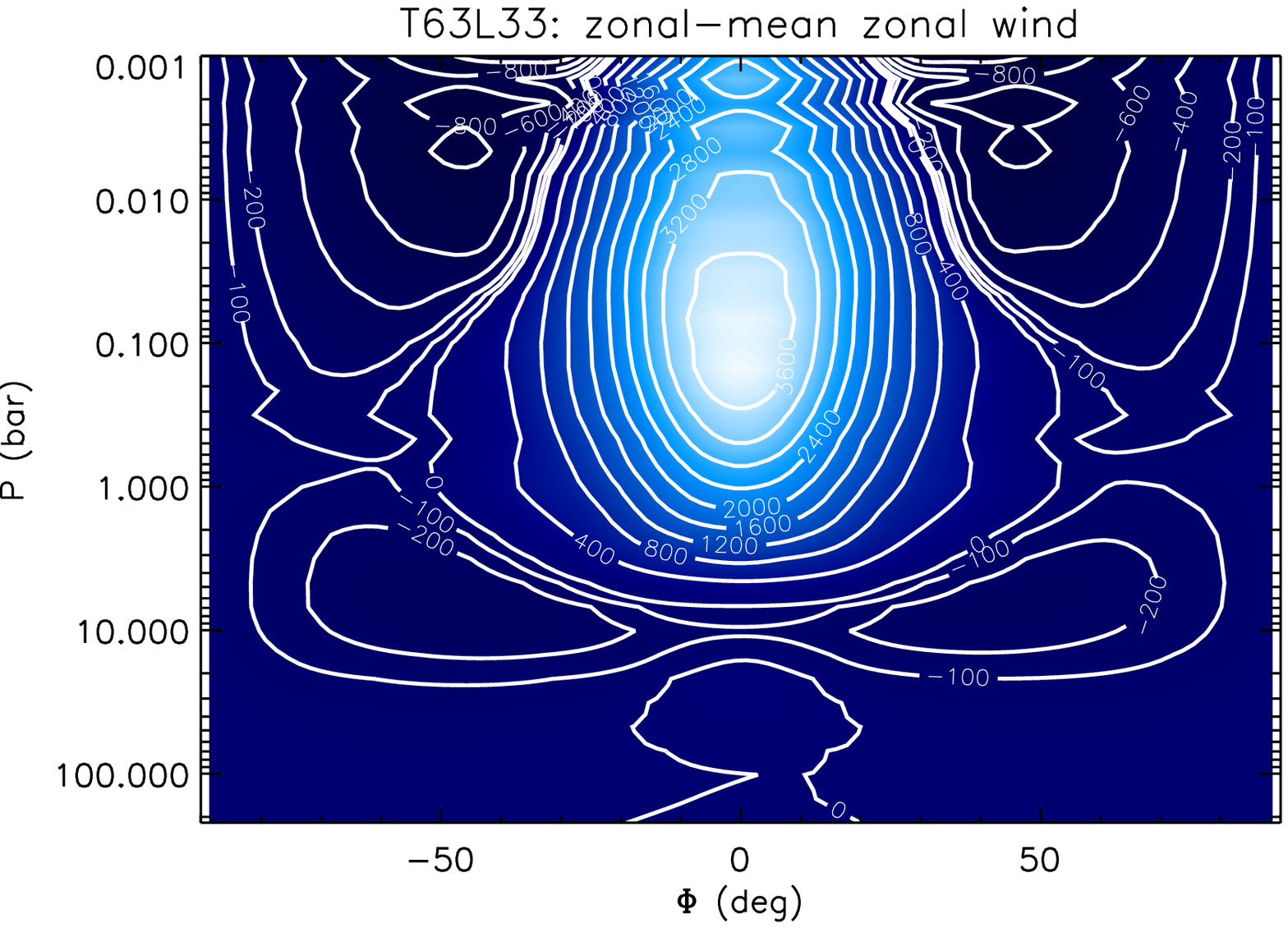}
\includegraphics[width=0.45\columnwidth]{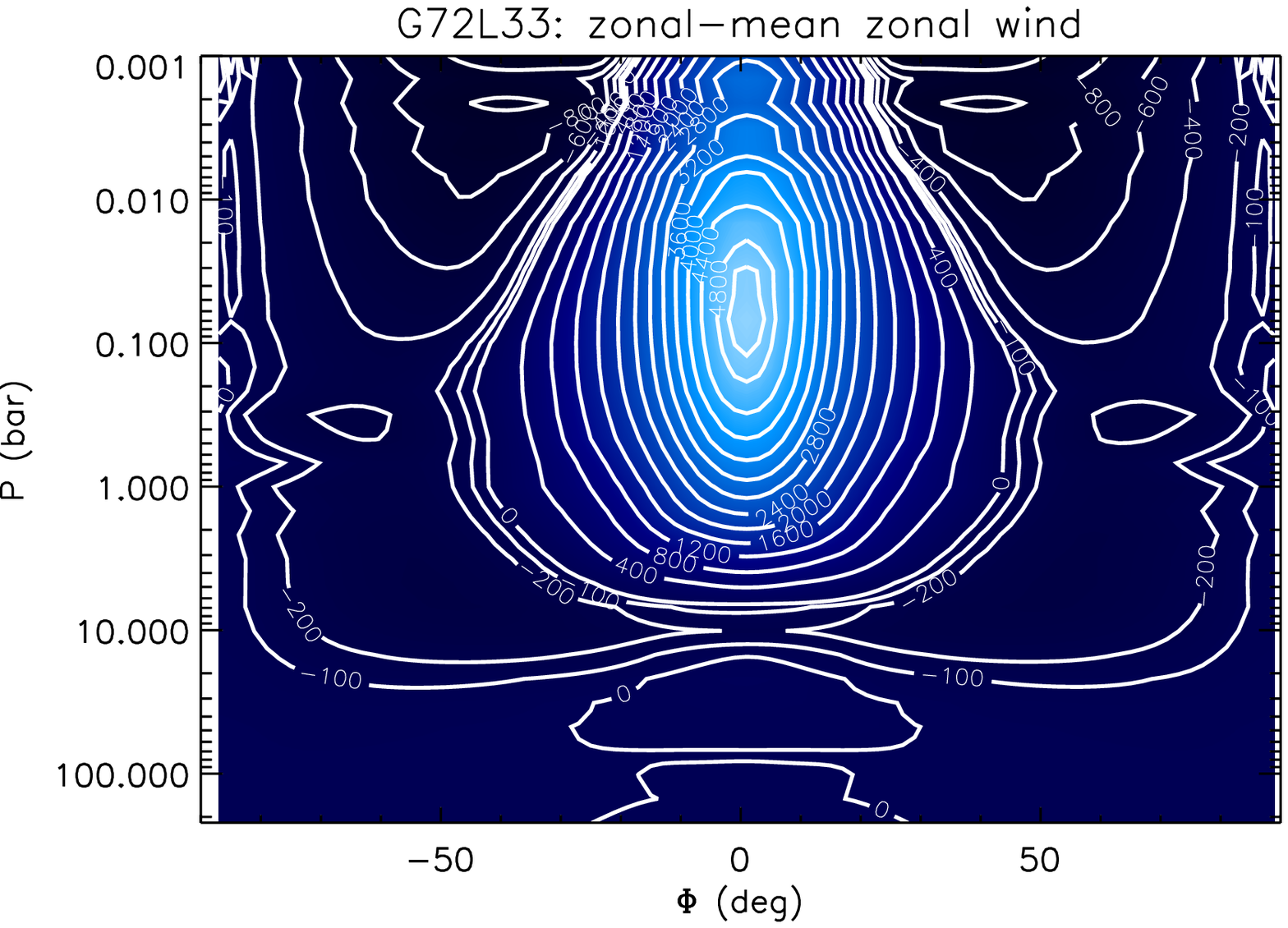}
\includegraphics[width=0.45\columnwidth]{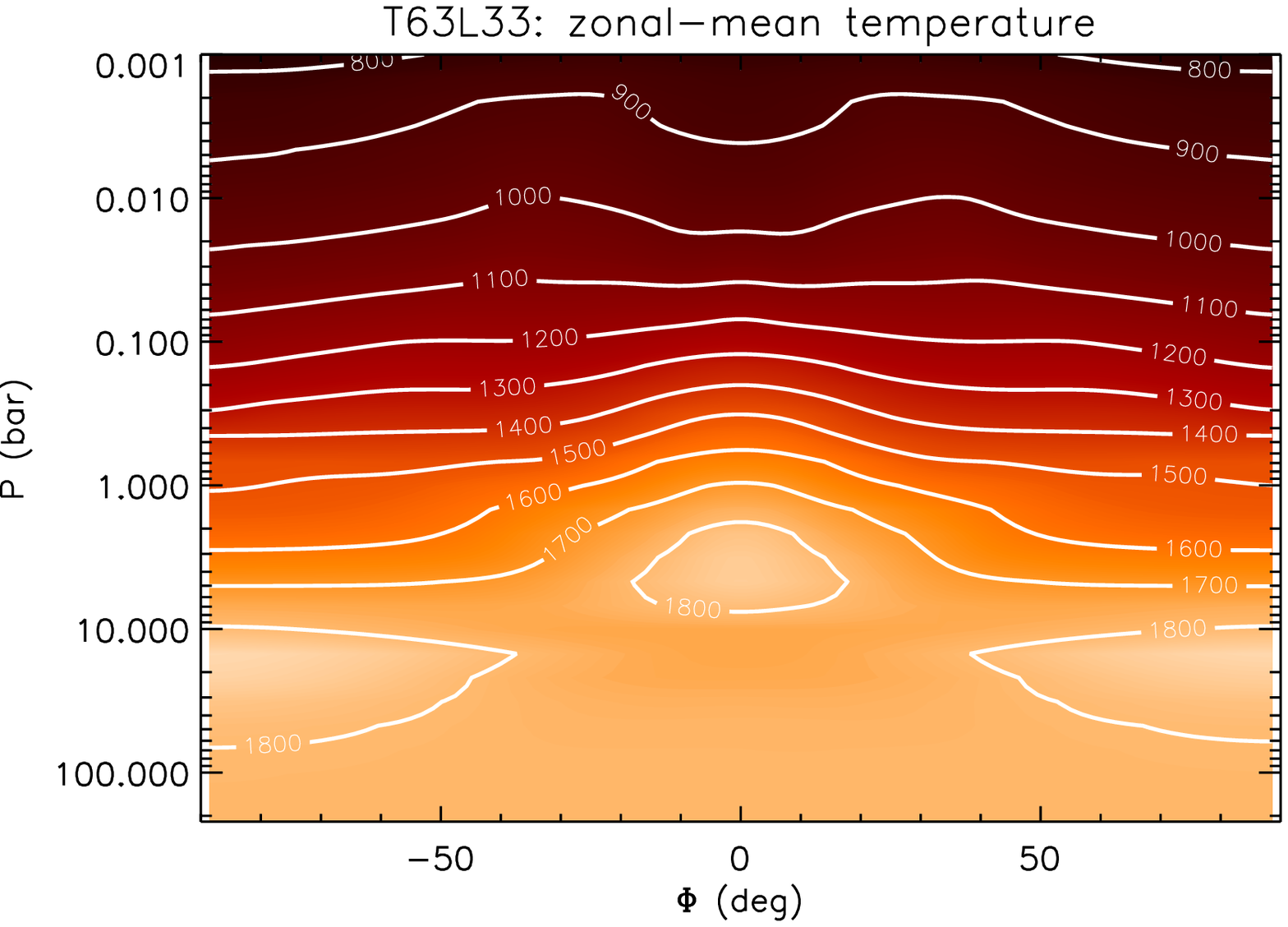}
\includegraphics[width=0.45\columnwidth]{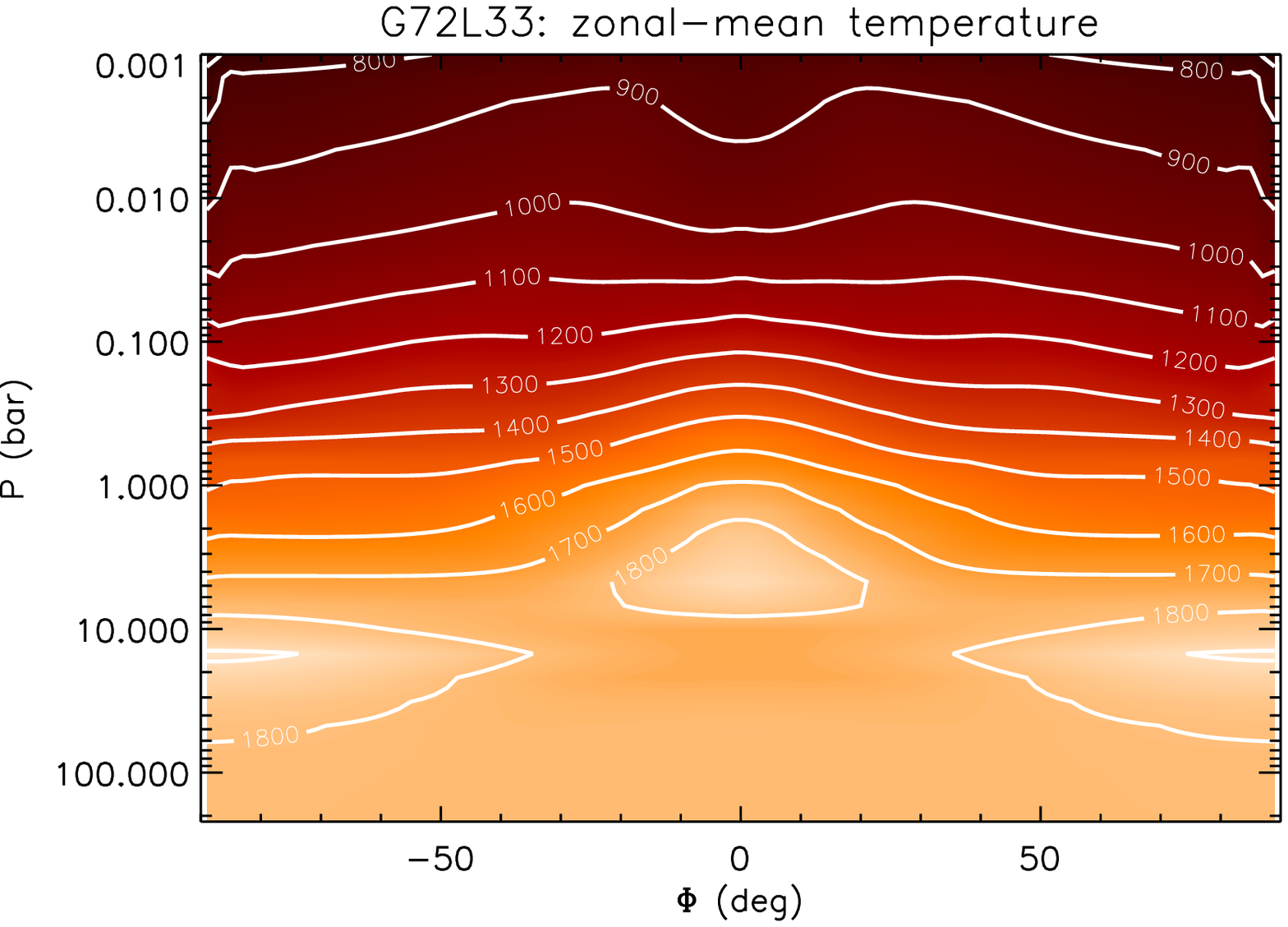}
\end{center}
\vspace{-0.2in}
\caption{Temporally averaged, zonal-mean zonal wind (top row) and
  temperature (bottom row) profiles simulated for HD 209458b.  Results
  from the spectral (left column; T63L33) and finite difference (right
  column; G72L33) simulations are shown.  The horizontal dissipation
  parameters take their fiducial values of $t_\nu = 10^{-5}$ HD
  209458b day and ${\cal K}=0.35$.  Temperatures are in K and wind
  speeds are in m s$^{-1}$.}
\label{fig:rm10_hd209458b_hs}
\end{figure}

Figure \ref{fig:rm10_hd209458b} shows snapshots of the temperature and velocity field at 1200 Earth days after the simulation, where the first 200 Earth days were disregarded.  Thus, the snapshots are of the exoplanet at about 340 HD 209458b days.  The four figure panels show the flow at different pressures and are chosen to approximately match Figures 1 and 2 of \cite{rm10}, who presented similar plots at $P=2.5$ mbar, 220 mbar, 4.4 bar and 20 bar at 1450 HD 209458b days.  Our plot for $P=2.13$ mbar (top left panel) shows an upper atmosphere dominated by radiative forcing, similar to the top panel of Figure 1 of \cite{rm10}.  Farther down in the atmosphere at $P=216$ mbar, a chevron-shaped feature is displaced eastwards of the substellar point and should be compared to the bottom panel of Figure 1 of \cite{rm10}.  At $P=4.69$ bar (bottom left panel), the advective time scales start to become shorter than $\tau_{\rm rad}$, resulting in the longitudinal homogenization of temperature.  At $P=21.9$ bar (bottom right panel), the equatorial wind becomes more counter-rotating than super-rotating, partly as a result of the conservation of total angular momentum (which is set by starting the simulation from a windless initial state, in the absence of drag).  Overall, there is a good degree of qualitative and quantitative agreement between our computed flow fields and those presented in Figures 1 and 2 of \cite{rm10}, despite the snapshots being taken at different times.  However, some discrepancies remain: the velocity features in our simulations are stronger at the various vertical heights; the temperature ranges are discrepant from those shown in \cite{rm10}, especially at $P=2.13$ and 216 mbar.

Figure \ref{fig:rm10_hd209458b_res} focuses on the $P=216$ mbar
snapshot at about 340 HD 209458b days, but for six different
simulation resolutions: T21L33, T21L66, T31L33, T31L66, T63L33 and
T63L66.  We note that $P \sim 0.1$ bar is the pressure/height at which
the infrared emission emerges and where the stratosphere (if any)
begins.  Details concerning the numerical resolution are given in Table
\ref{tab:resolution}.  In general, the chevron-shaped flow feature is
seen at all six resolutions --- its substructure shows up in all of
the simulations and is (expectedly) most clearly visible at
T63.  There are hints that the details of the flow, such as zonal
wind speed, depend on the simulation resolution.  The same conclusions
may be drawn from the finite difference simulations presented in Figure
\ref{fig:rm10_hd209458b_res2}, but we note that comparing B-grid simulations at different resolutions, with the same value of ${\cal K}$, may not constitute a fair exercise (see Appendix \ref{append:efold}).  In other words, Figure \ref{fig:rm10_hd209458b_res} displays results from spectral simulations which are equally dissipative (same value of $\nu$) and at progressively finer resolutions, whereas Figure \ref{fig:rm10_hd209458b_res2} shows results from finite difference simulations which are more dissipative (numerically viscous) at lower resolutions.

Other points deserve to be emphasized.
Firstly, the spectral simulations manifestly capture the details of
fine flow features better than the finite difference simulations,
which is likely because the latter (with ${\cal K}=0.35$) are more
dissipative than the former (with $t_\nu=10^{-5}$ HD 209458b day).  Secondly, there are
clear qualitative differences between the snapshots from the spectral
and finite difference simulations, which show $\sim 10\%$ variations
in the temperature field.  The third point concerns model variations
and the level of variability in the temperature field at $P=216$ mbar,
which we illustrate in Figure \ref{fig:temp_var}.  For the spectral simulations (left panel of Figure \ref{fig:temp_var}), it is apparent that there are $\sim 10\%$ variations in the temperature field as a function of time and the magnitudes of the variations are roughly equal across resolution (T21--T63).  For the finite difference simulations (right panel of Figure \ref{fig:temp_var}), the G72 simulations also show $\sim 10\%$ variations in the temperature field at $P=216$ mbar whereas the G24 and G36 simulations show less variation, consistent with our earlier statement that the latter simulations are more dissipative (compared to their G72 counterpart).  It is worth noting that numerical noise may contribute to these variations and thus complicate the comparisons.  The level of variability present and its dependence on the magnitude of the horizontal dissipation applied will have implications for the study of variability in these model atmospheres for hot Jupiters.  The discrepancies in the predicted temperature fields may be due to the different temporal evolutions within each simulation --- even a pair of identical models seeded with slightly different (and random) initial perturbations may lead to different temporal evolutions.  Furthermore, the time taken to reach quasi-equilibrium is different for each method of solution. 

\begin{figure}
\begin{center}
\includegraphics[width=0.55\columnwidth]{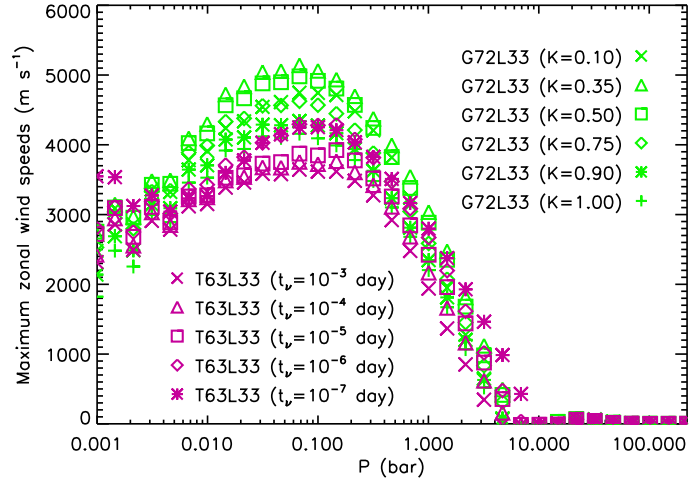}
\end{center}
\vspace{-0.2in}
\caption{Maximum values of the temporally averaged, zonal-mean zonal
  wind speeds, from different simulations, as functions of the
  vertical pressure, for various magnitudes of the horizontal
  dissipation.  Here, ``day" refers to one HD 209458b day which is about 3.5 Earth days.}
\label{fig:winds}
\end{figure}

\begin{figure}
\begin{center}
\includegraphics[width=0.45\columnwidth]{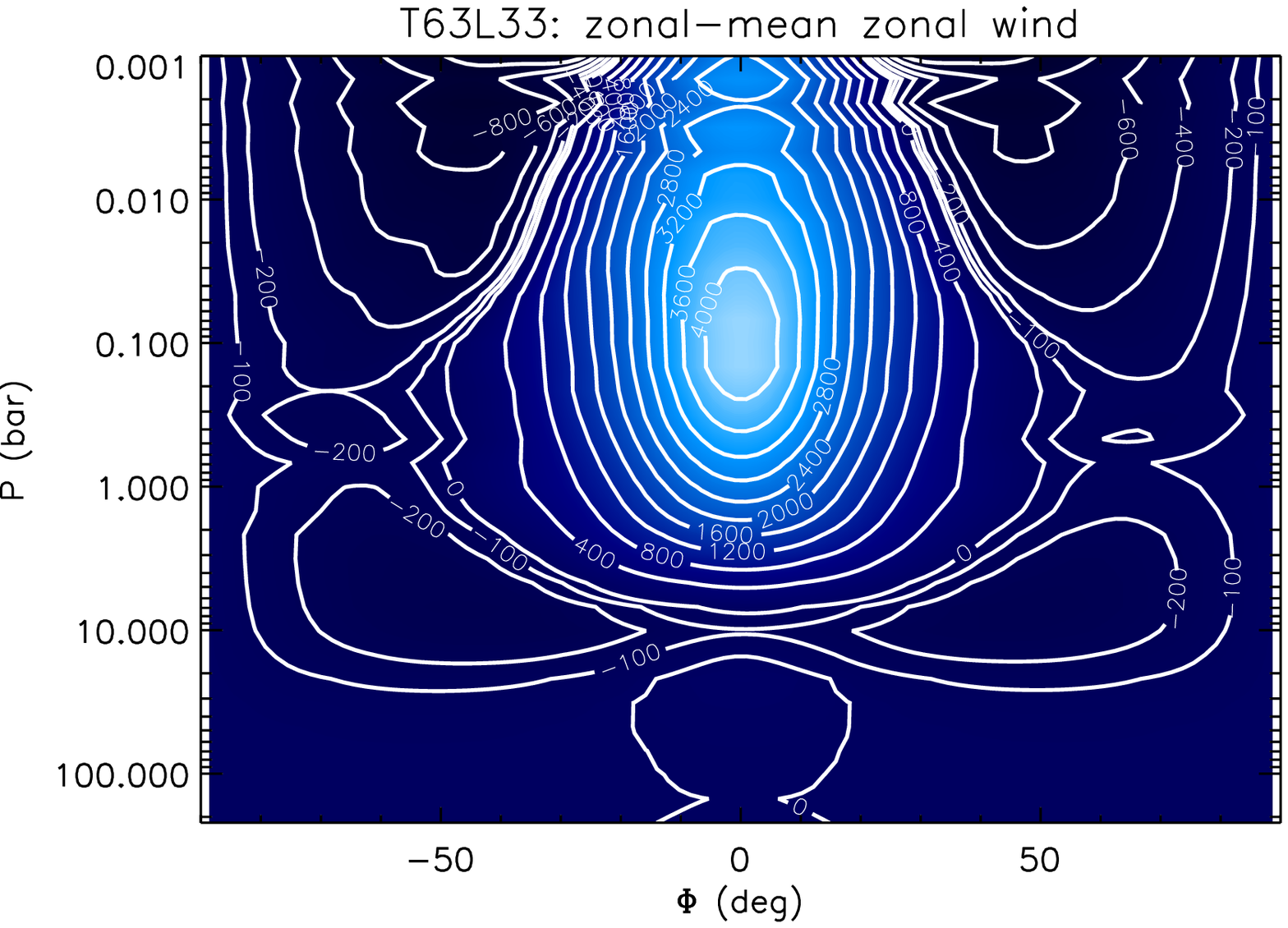}
\includegraphics[width=0.45\columnwidth]{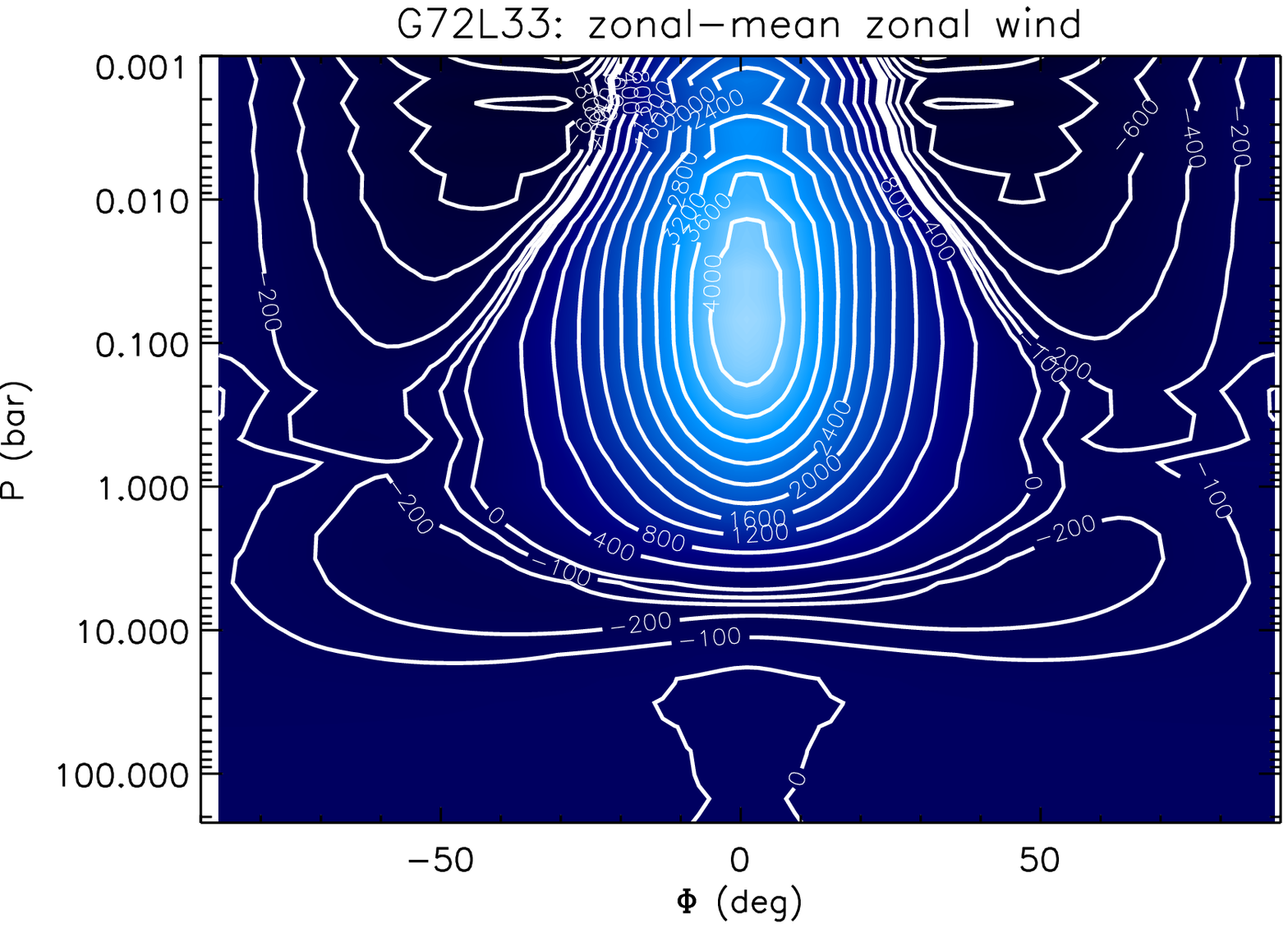}
\includegraphics[width=0.45\columnwidth]{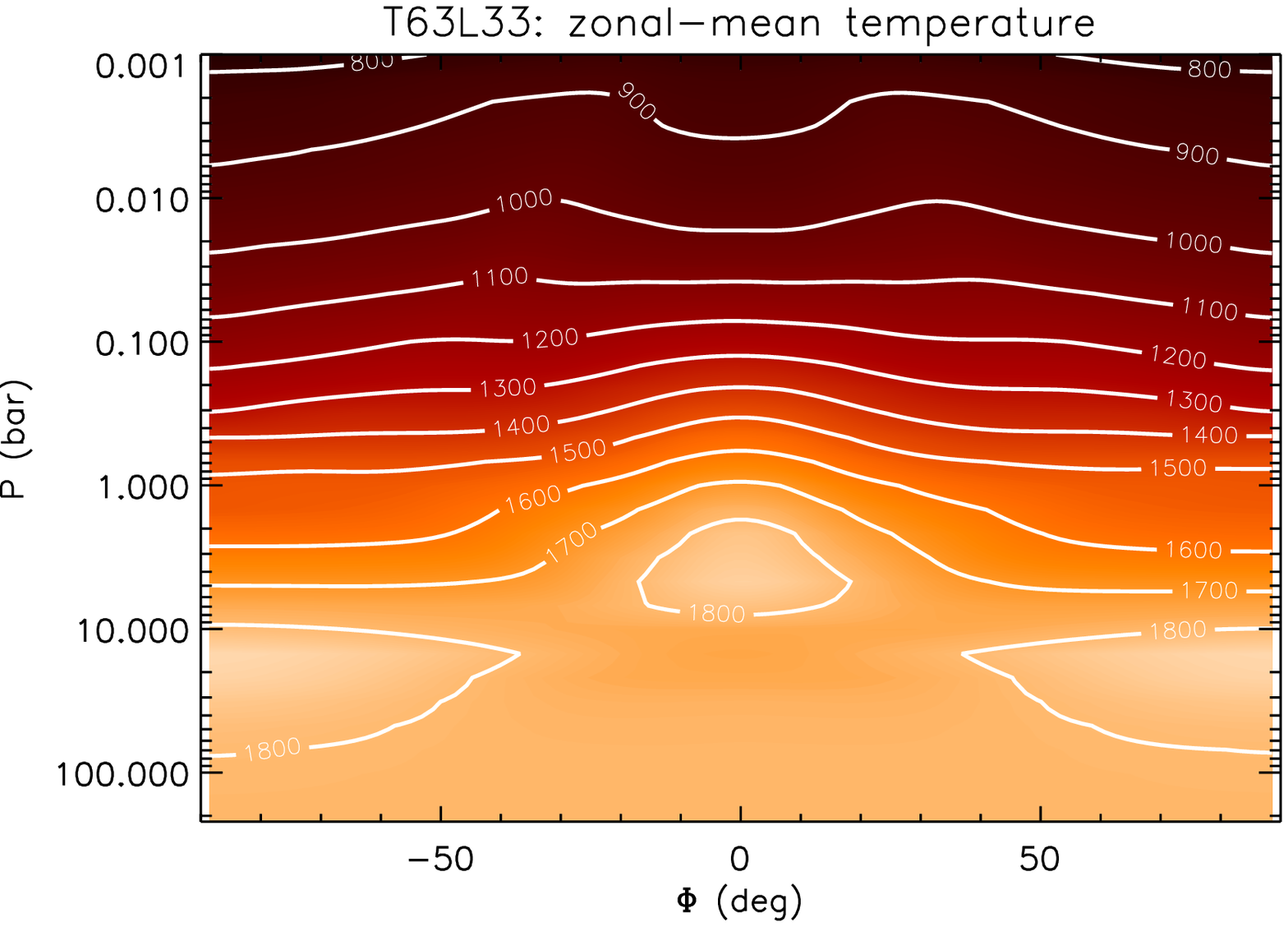}
\includegraphics[width=0.45\columnwidth]{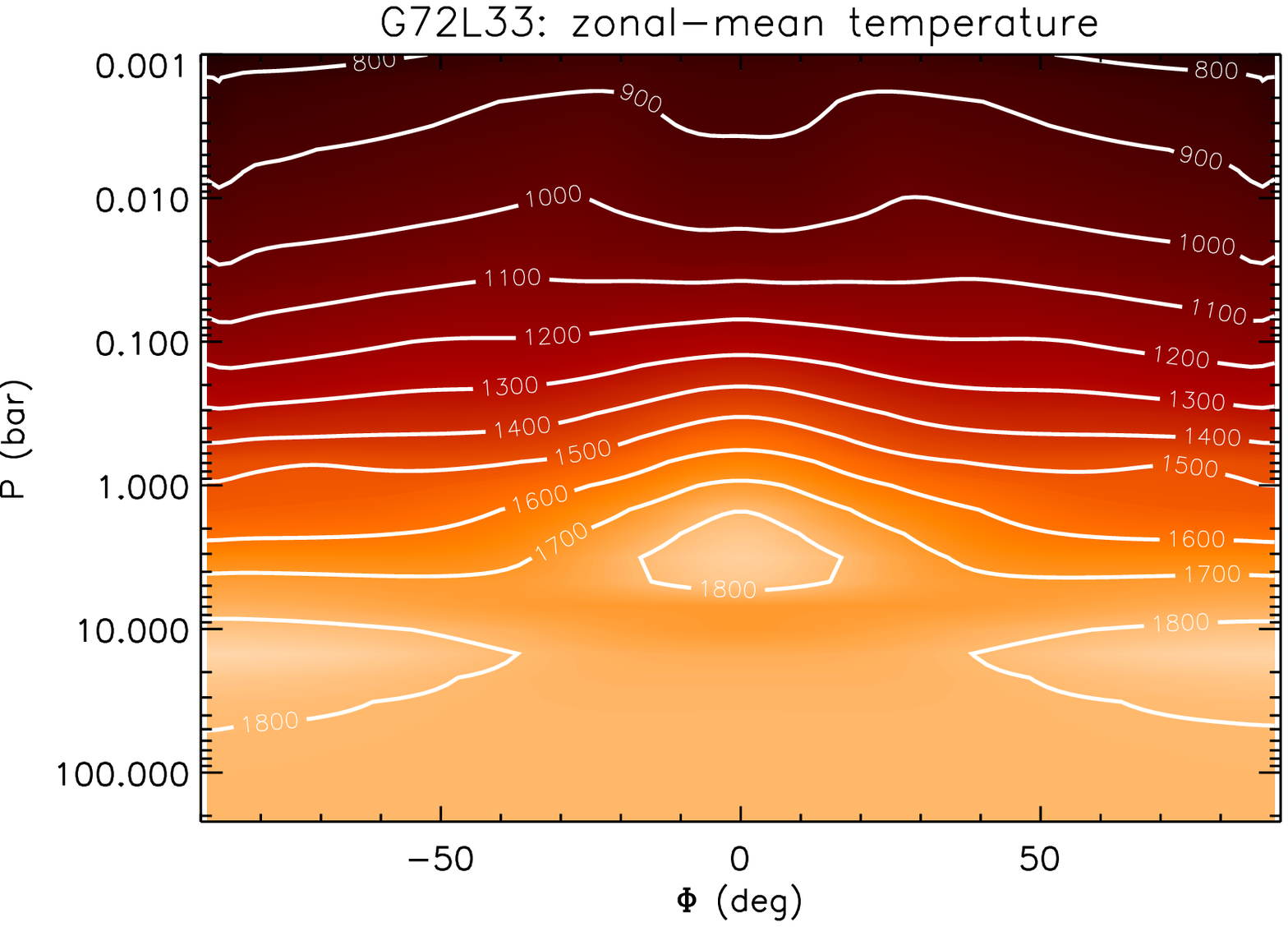}
\end{center}
\vspace{-0.2in}
\caption{Same as Figure \ref{fig:rm10_hd209458b_hs}, but with the
  values of the horizontal dissipation parameters adjusted ($t_\nu =
  10^{-6}$ HD 209458b day and ${\cal K}=0.90$) such that the zonal
  wind profiles (as shown in Figure \ref{fig:winds}) approximately
  match.  Temperatures are in K and wind speeds are in m s$^{-1}$.}
\label{fig:rm10_hd209458b_hs_adjusted}
\end{figure}

Figure \ref{fig:rm10_hd209458b_hs} shows the usual Held-Suarez
statistics for HD 209458b.  The temporally-averaged, zonal-mean zonal
wind profiles depict an equatorial, super-rotating wind down to $\sim
10$ bar, flanked by counter-rotating jets at mid-latitude.  Wind
speeds have typical magnitudes $\sim 1$ km s$^{-1}$.  The top row of
Figure \ref{fig:rm10_hd209458b_hs} should be compared to Figure 3 of
\cite{rm10}.  While the temporally averaged and zonal-mean
temperatures are not presented in \cite{rm10}, we still present these
figures (bottom row) for comparison between the spectral and finite
difference simulations.  The temperature profiles between our pair of
simulations are in good agreement.  The similarity of the temperature
profiles indicates that we have chosen $t_{\rm discard}$ (=200 Earth
days) to be sufficiently large, such that differences due to
initialization have been erased.  However, noticeable differences
exist particularly with respect to the wind field --- the maximum
speed of the equatorial, super-rotating wind is about 5 km s$^{-1}$ in
the finite difference simulation, but is only about 3.6 km s$^{-1}$ in
the spectral simulation.

The discrepancies in the predictions for the zonal wind profile and
the maximum speed of the super-rotating jet motivate us to explore the
issue further by varying the values of $t_\nu$ and ${\cal K}$ --- the magnitude of horizontal dissipation (see \S\ref{subsect:damping}) --- and subsequently varying the
simulation resolution.  Figure \ref{fig:winds} shows T63L33 and G72L33
simulations with various values of $t_\nu = 10^{-7}$--$10^{-3}$ HD
209458b day and ${\cal K}=0.1$--1, respectively.  We note that ${\cal
  K}=1$ is not used in normal circumstances, except near the poles to
prevent the numerical problems previously described.\footnote{We also examined a simulation with no horizontal mixing applied (${\cal K}=0$), which did not complete successfully (produced multiple output values of \texttt{NaN}s).}  \emph{The key point is that there are $\gtrsim 10\%$ uncertainties associated with
  the predictions for the zonal wind profiles within each method of
  solution and also between them.}  The wind profile from the finite difference
simulation with ${\cal K}=0.9$ appears to approximately match that
from the spectral simulation with $t_\nu=10^{-6}$ HD 209458b day.
Figure \ref{fig:rm10_hd209458b_hs_adjusted} shows the Held-Suarez
statistics with these adjusted values of the horizontal dissipation
parameters --- it is clear that there is now closer agreement between
the pairs of simulations, both qualitatively and quantitatively.
Overall, Figures \ref{fig:winds} and
\ref{fig:rm10_hd209458b_hs_adjusted} demonstrate that \emph{one can,
  by trial and error, obtain consistent results for arbitrarily
  adjusted values of $t_\nu$ and ${\cal K}$, but there is still no
  rigorous way to choose these dissipation times/rates.}

\begin{figure}
\begin{center}
\includegraphics[width=0.45\columnwidth]{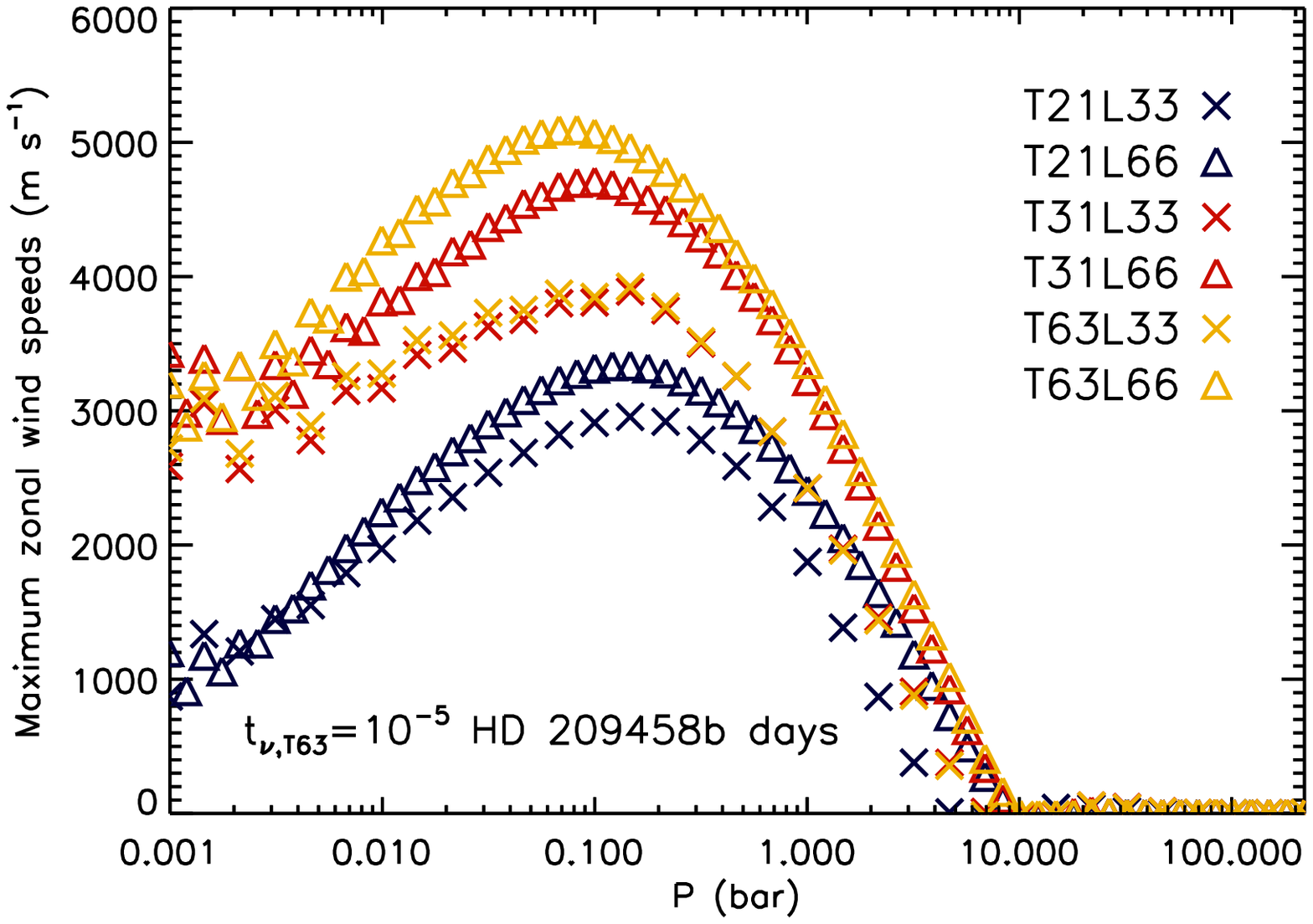}
\includegraphics[width=0.45\columnwidth]{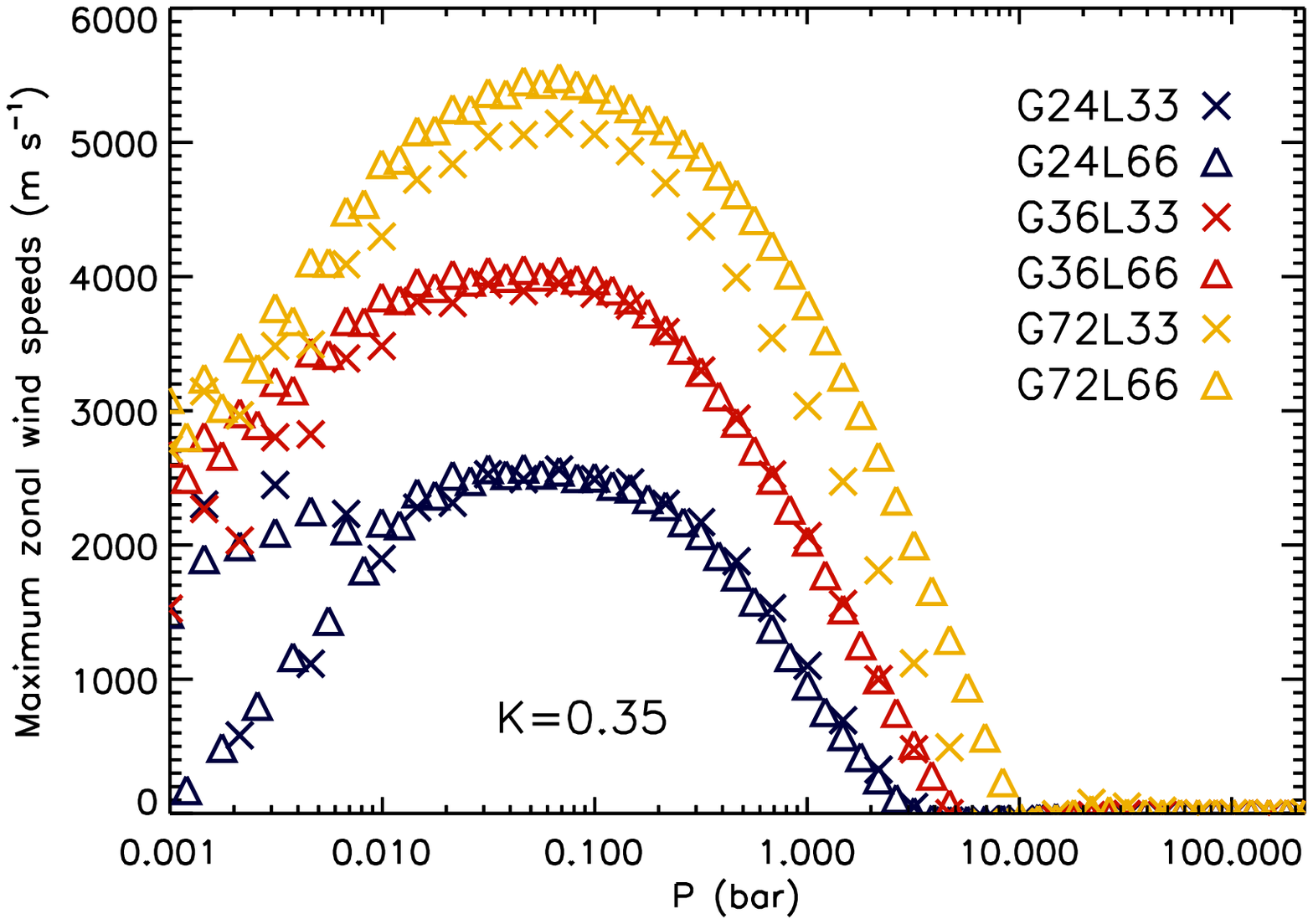}
\end{center}
\vspace{-0.2in}
\caption{Maximum values of the temporally averaged, zonal-mean zonal
  wind speeds, from different simulations, as functions of the
  vertical pressure.  Left: spectral simulations with different
  resolutions.  Right: finite difference simulations with different
  resolutions.  The fiducial value of the horizontal dissipation
  parameter used is ${\cal K}=0.35$ in the finite difference simulations, 
  while the hyperviscosity $\nu$ is kept fixed in the spectral simulations.  Inter-comparing the left and right panels does not constitute a fair exercise, because the assumed magnitude of horizontal dissipation is different; even the intra-comparison of results within the right panel may not be straightforward (see text).}
\label{fig:winds2}
\end{figure}

To investigate the effects of varying the numerical resolution, we revert to the fiducial values of ${\cal K}$ and the hyperviscosity $\nu$ for the finite difference and spectral simulations, respectively.  Keeping $\nu$ constant while varying the resolution results in the variation of the numerical dissipation time/rate, thus allowing the spectral simulations to be compared on an equal footing (see equation [\ref{eq:tnu_scale}]).  However, keeping ${\cal K}$ at a fixed value is strictly speaking analogous to varying $\nu$ (see Appendix \ref{append:efold}), so examining finite difference simulations at different resolutions, with the same value of ${\cal K}$, may not constitute a fair comparison.  Nevertheless, we show in Figure \ref{fig:winds2} the results of our resolution studies from both the spectral and finite difference simulations, where we record the maximum speed of the  temporally averaged, zonal-mean zonal wind as a function of $P$.  For the spectral simulations, there is a spread of about 50\% in the predicted wind speeds.  The T31 and T63 simulations agree reasonably well with one another at a given vertical resolution (L33 or L66), whereas the wind speed predictions from the T21 simulations are discrepant with the T31 and T63 ones, suggesting that the T21 simulations are under-resolved.  For the finite difference simulations, the predictions for the wind speeds are also in agreement at a given vertical resolution (L33 versus L66).  Simulations with lower horizontal resolution predict lower wind speeds, consistent with our earlier statement that they are more dissipative (at a fixed value of ${\cal K}$).  Nevertheless, the depth at which the super-rotating wind is the fastest is a robust feature of the simulations, occurring at about 0.05 bar in the finite difference simulations (versus about 0.08 in the corresponding set of spectral simulations).

We note that inter-comparison of the results between each panel of Figure \ref{fig:winds2} does not constitute a fair exercise, because the assumed magnitude of horizontal dissipation is different for each suite of simulations.  Even the intra-comparison of results within the right panel of Figure \ref{fig:winds2} may not be straightforward, because as we discussed previously keeping ${\cal K}$ fixed while varying the resolution in effect changes the analogue of the hyperviscosity, but we do not know of a clear way of varying ${\cal K}$ in a ``resolution independent" manner (see Appendix \ref{append:efold}).  The best we can conclude from Figure \ref{fig:winds2} is that the predictions for the maximum zonal wind speed, from both the spectral and finite difference simulations, are resolution-dependent.  In tandem with Figure \ref{fig:winds}, one may also conclude that since the specification of horizontal dissipation is a more lucid endeavour within the spectral core, comparing results from the spectral and finite difference simulations should only be performed when Held-Suarez statistics from the latter are calibrated to match those produced by the former.  In this sense, the spectral simulations are more robust.

We conclude that while we have achieved a rather satisfactory level of
agreement between our spectral and finite difference simulations of HD
209458b, discrepancies arise in the quantitative predictions which may
limit our ability to accurately model these extreme atmospheres,
especially in terms of wind speeds.  The main lessons we learn are
that there are $\sim 10\%$ uncertainties associated with the
temperature field and $\gtrsim 10\%$ uncertainties associated with the velocity field, due to the choice of
the magnitude of the horizontal dissipation as well as the resolution
of the simulations.

\section{Discussion}
\label{sect:discussion}

\subsection{Broader Implications}

Using a single and consistent simulation platform, we have performed a
suite of benchmark tests concerning the atmospheric circulation of
Earth and tidally-locked extrasolar planets.  We find that while the
dynamical cores of the \texttt{FMS} produce qualitative and
quantitative agreement for the Earth, tidally-locked Earth and shallow
hot Jupiter tests, the agreement is less than satisfactory for a deep model
of the hot Jupiter HD 209458b.  Further investigation reveals that
closer agreement can be attained by arbitrarily adjusting the values
of the horizontal dissipation parameters in the two dynamical cores,
but there is no rigorous way to pick the magnitude of the horizontal
dissipation in these models.

Our findings suggest that even without dealing with additional
physics such as radiative transfer or atmospheric chemistry,
discrepancies in the temperature and velocity fields, at the level of
$10\%$ and several tens of percent respectively, already exist
for the dynamics alone. In this context, direct measurements of wind
velocity in a hot Jupiter atmosphere, as recently reported by
\citet{snellen10}, are important as they could prove to be particularly
constraining for the models.

In general, weakly-dissipative spectral simulations are expected to be
sensitive to the choice of horizontal dissipation parameter ($t_\nu$),
in the sense that they tend to fail as a result of small-scale noise
accumulation (``spectral blocking'') if the horizontal dissipation is
not chosen to be strong enough.  Therefore, to the extent that
spectral simulations with the largest possible value of $t_\nu$ (i.e.,
the weakest possible horizontal dissipation) are more trustworthy --- as
conventional wisdom would suggest --- our findings could also be
interpreted as indicating that results in the literature based on
finite difference methods may somewhat \emph{over-estimate} the
magnitude of wind speeds in hot Jupiter atmospheres (see Figure
\ref{fig:winds}).  However, until the nature of horizontal dissipation in these
atmospheres is better understood \citep[e.g.,][]{goodman09},
one should probably not interpret these trends as more than
suggestive.

Operationally, our suite of benchmark tests provides a reference for
researchers wishing to adapt their codes to simulate atmospheric
circulation on tidally-locked extrasolar planets, regardless of
whether the codes solve the primitive or full Navier-Stokes equations.

\subsection{Summary}

The salient points of our study are:
\begin{itemize}

\item We have generalized the Held-Suarez dynamical benchmark for
  Earth to include tidally-locked exoplanets using a single simulation
  platform (the \texttt{FMS}).  Our suite of benchmark tests provides
  a reference for researchers wishing to adapt their codes to study
  atmospheric circulation on tidally-locked Earths/Neptunes/Jupiters.

\item We have found that the differences in the HD 209458b simulations
  of \cite{cs05,cs06} and \cite{rm10} are probably due to
  initial/boundary conditions and setup, and not due to the method of
  solution utilized.

\item Qualitative and quantitative agreement between the spectral and
  finite difference simulations of the deep-atmosphere benchmark test
  for the hot Jupiter HD 209458b can be attained if arbitrarily
  adjusted values of the horizontal dissipation parameters are
  adopted.  However, the difficulty remains that the \emph{magnitude}
  of the horizontal dissipation cannot (yet) be specified from first
  principles.  This in turn leads to dynamical uncertainties at the
  level of $\gtrsim 10\%$ which limit our ability to
  accurately model these atmospheres, especially with respect to wind
  velocities.  Direct wind measurements from transit observations of extrasolar planets should thus be particularly constraining for the models.

\end{itemize}

\section*{Acknowledgments}

K.H. acknowledges support from the Zwicky Prize Fellowship at ETH
Z\"{u}rich, the Frank \& Peggy Taplin Membership of the Institute for
Advanced Study (IAS), NASA grant NNX08AH83G and NSF grant AST-0807444, as well as
encouragement from Scott Tremaine.  K.M. was supported in part by the Perimeter Institute for
Theoretical Physics.  We acknowledge useful conversations with Lucio
Mayer, Hans Martin Schmid, Michael Meyer, Isaac Held, Josh Schroeder,
Jonathan Mitchell, Adam Burrows and Dave Spiegel, as well as useful comments from the anonymous referee which greatly improved the quality of the manuscript.  We thank Nicolas
Iro for providing selected results from Iro et al. (2005) in
electronic form.  Multiple suites of simulations were started at the IAS using the
\texttt{aurora} computing cluster (managed by Prentice Bisbal, James
Stephen et al.) and finished on the \texttt{Brutus} computing cluster
at ETH Z\"{u}rich (managed by Olivier Bryde et al.).  The website
\texttt{http://www.dfanning.com} provided useful technical advice for
\texttt{IDL} on many occasions, while animation/visualization was done
using the \texttt{VisIt} platform created by the Lawrence Livermore
National Laboratory.  This work benefited from the collegial environment
at the Institute for Astronomy of ETH Z\"{u}rich.


\appendix

\section{Implementing Uneven Vertical Levels}
\label{append:levels}

\begin{figure}
\begin{center}
\includegraphics[width=0.55\columnwidth]{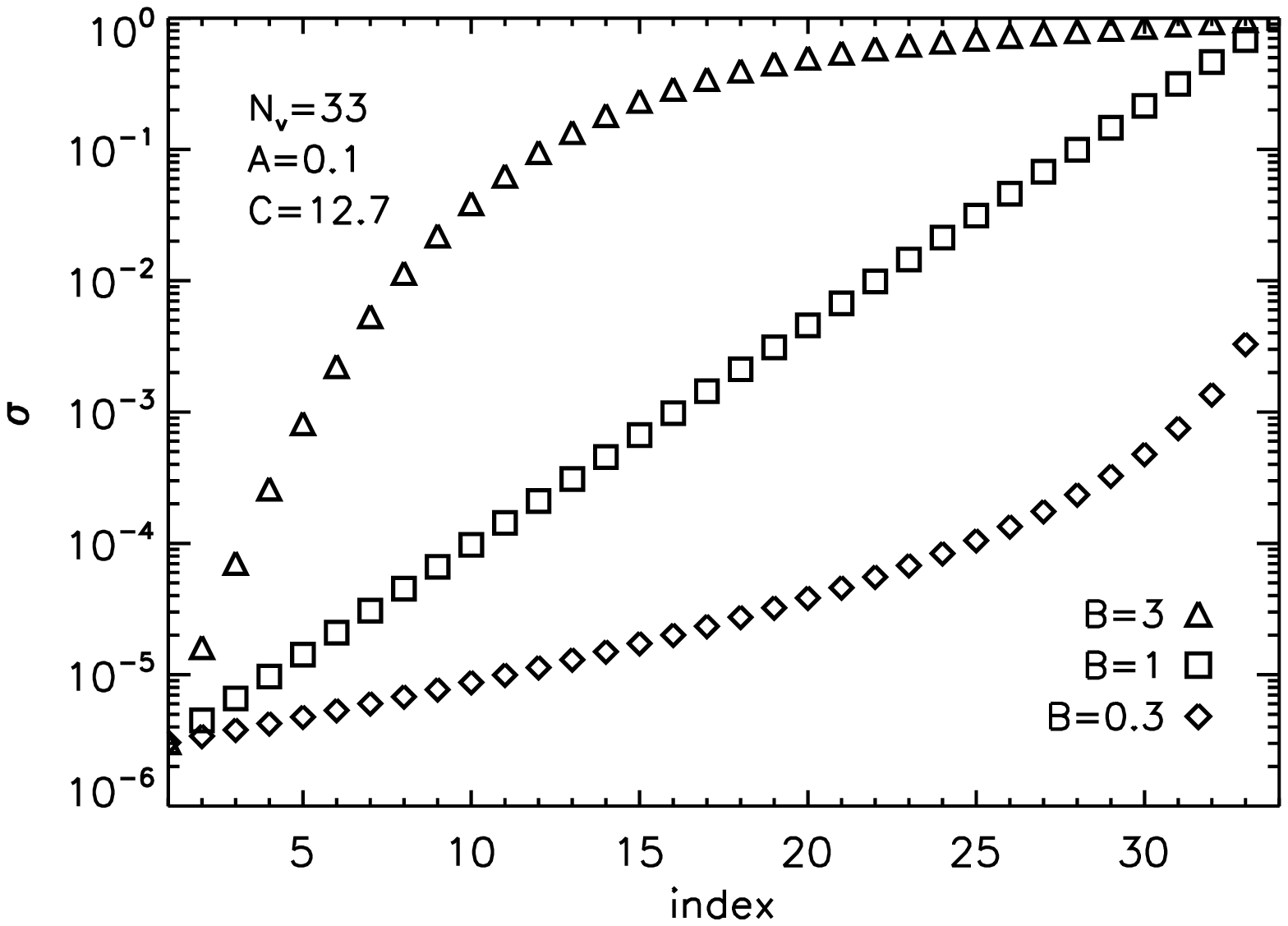}
\end{center}
\vspace{-0.2in}
\caption{Uneven spacing of the $\sigma$ vertical coordinate accomplished using equation (\ref{eq:uneven_sigma}).  Note that $\sigma(i=1)=0$ and is not shown.}
\label{fig:uneven_sigma}
\end{figure}

The ability to perform dynamical core simulations with arbitrary spacing of the vertical levels comes as a default in the \texttt{FMS}, accomplished using the following prescription:
\begin{equation}
\begin{split}
&\zeta = 1 - \left(i-1\right)/N_v, \\
&\tilde{\zeta} = A\zeta + \left(1-A\right)\zeta^B,\\
&\sigma = \exp\left(-\tilde{\zeta} C\right),\\
\end{split}
\label{eq:uneven_sigma}
\end{equation}
where the index $i$ runs from 1 to $N_v+1$.  Varying the parameters $A$, $B$ and $C$ allows one to control the range of $\sigma$ covered and the spacing between the points.  For example, Figure \ref{fig:uneven_sigma} shows three implementations of equation (\ref{eq:uneven_sigma}), where $\log\sigma$ is evenly spaced only for $B=1$.

\section{Polynomial Fits for Thermal Forcing of HD 209458b}
\label{append:fits}

To aid the reader in reproducing our results, we provide convenient polynomial fits to $\tau_{\rm rad}$, $T_{\rm night}$ and $T_{\rm day}$.  The radiative relaxation time is approximated by a 4th order polynomial fit,
\begin{equation}
\log\left(\frac{\tau_{\rm rad}}{\mbox{1 s}}\right) = 
\begin{cases}
5.4659686 + 1.4940124 \tilde{P} + 0.66079196 \tilde{P}^2 + 0.16475329 \tilde{P}^3 + 0.014241552 \tilde{P}^4, & P < \mbox{10 bar}, \\
\infty, & \mbox{otherwise},\\
\end{cases}
\label{eq:trad_hd209458b}
\end{equation}
where $\tilde{P} \equiv \log(P/\mbox{1 bar})$.  The preceding fit is valid for $10 ~\mu\mbox{bar} \le P \le 8.5 \mbox{ bar}$.

The night and day side temperatures are given by 
\begin{equation}
\frac{T_{\rm night}}{\mbox{1 K}} = 
\begin{cases}
1388.2145 + 267.66586 \tilde{P} -215.53357 \tilde{P}^2 + 61.814807\tilde{P}^3+  135.68661\tilde{P}^4+ 2.0149044 \tilde{P}^5 & \\
-40.907246\tilde{P}^6 -19.015628\tilde{P}^7 -3.8771634\tilde{P}^8 -0.38413901\tilde{P}^9-0.015089084\tilde{P}^{10}, & P \le \mbox{10 bar},\\
5529.7168 -6869.6504\tilde{P}+ 4142.7231\tilde{P}^2 -936.23053\tilde{P}^3 + 87.120975\tilde{P}^4, & \mbox{ otherwise},\\
\end{cases}
\label{eq:tnight}
\end{equation}
and
\begin{equation}
\frac{T_{\rm day}}{\mbox{1 K}} = 
\begin{cases}
2149.9581 + 4.1395571\tilde{P}-186.24851\tilde{P}^2 + 135.52524\tilde{P}^3 + 106.20433\tilde{P}^4 -35.851966\tilde{P}^5 & \\
-50.022826\tilde{P}^6 -18.462489\tilde{P}^7 -3.3319965\tilde{P}^8 -0.30295925\tilde{P}^9 -0.011122316\tilde{P}^{10}, & P \le \mbox{ 10 bar},\\
5529.7168 -6869.6504\tilde{P}+ 4142.7231\tilde{P}^2 -936.23053\tilde{P}^3 + 87.120975\tilde{P}^4, & \mbox{ otherwise},\\
\end{cases}
\label{eq:tday}
\end{equation}
respectively.  Our fits presented in equations (\ref{eq:tnight}) and (\ref{eq:tday}) are valid for $1 ~\mu\mbox{bar} \le P \le 3488 \mbox{ bar}$.

For completeness, we also provide a polynomial fit to the solid curve presented in Figure 1 of \cite{iro05}:
\begin{equation}
\begin{split}
\frac{T_{\rm Iro}}{\mbox{1 K}} &= 1696.6986 + 132.23180 \tilde{P} -174.30459 \tilde{P}^2 + 12.579612 \tilde{P}^3 + 59.513639 \tilde{P}^4 \\
& + 9.6706522 \tilde{P}^5 - 4.1136048 \tilde{P}^6 - 1.0632301 \tilde{P}^7 + 0.064400203 \tilde{P}^8 \\
& + 0.035974396 \tilde{P}^9 + 0.0025740066 \tilde{P}^{10}.\\
\end{split}
\end{equation}
This fit is valid for $1 ~\mu\mbox{bar} \le P \le 3488 \mbox{ bar}$.

\section{E-folding Time for Decay of Initial Grid-Scale Noise (B-grid core)}
\label{append:efold}

\begin{figure}
\begin{center}
\includegraphics[width=0.55\columnwidth]{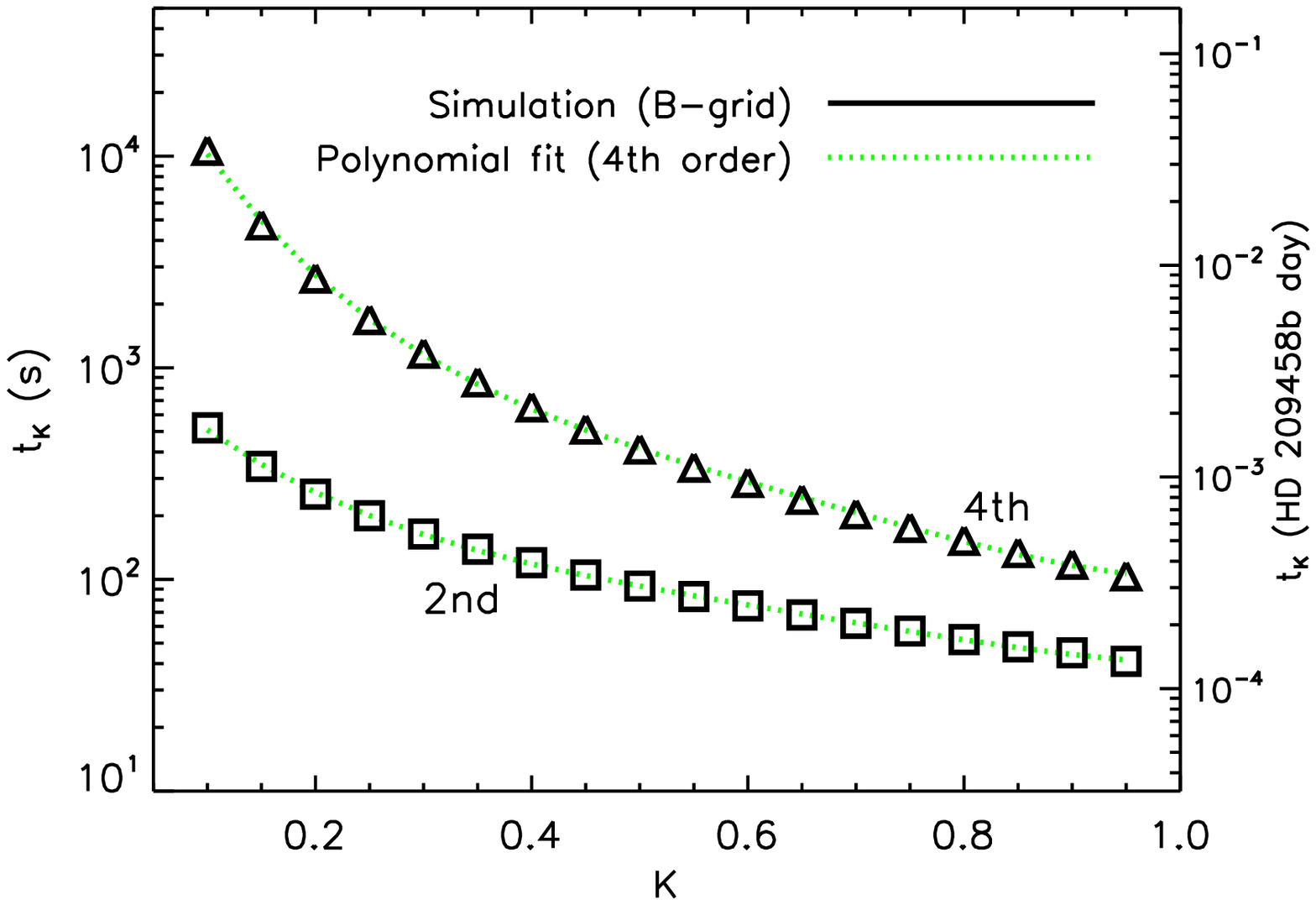}
\end{center}
\vspace{-0.2in}
\caption{E-folding time for initial grid-scale noise to decay away in the finite difference (B-grid) core of the \texttt{FMS}, $t_{\cal K}$, as a function of the horizontal mixing coefficient ${\cal K}$.  The measurements are identical for the G24, G36 and G72 simulations.  We show measurements (and fitting functions) from horizontal mixing schemes of both the second (squares) and fourth (triangles) order in wind, while retaining a fourth order temperature scheme, but present only results associated with the fourth order wind scheme in the paper.}
\label{fig:efold}
\end{figure}

A plausible way of determining the analogue of the numerical dissipation time $t_\nu$ (for the spectral core) in the case of the finite difference core --- we shall denote this by $t_{\cal K}$ --- is to turn off all of the dynamical terms in the simulation except for damping and record the e-folding time for initial grid-scale noise (in velocity) to decay away.  For completeness, we perform this task for a horizontal mixing scheme of both second and fourth order in wind, while retaining a fourth order temperature scheme at all times.  Note that the simulation results presented in the paper are all associated with a fourth order scheme in \emph{both} wind and temperature.  

Figure \ref{fig:efold} shows our measurements of $t_{\cal K}$ as a function of the horizontal mixing coefficient ${\cal K}$.  As expected, the horizontal mixing scheme of second order in wind is more dissipative (smaller $t_{\cal K}$ values) than the fourth order one.  For convenience, we provide fitting functions to these results:
\begin{equation}
\frac{t_{\cal K}}{1\mbox{ s}} = 
\begin{cases}
3.14235 -5.22936{\cal K} + 9.61774{\cal K}^2 -9.48662{\cal K}^3 + 3.55903{\cal K}^4 & \mbox{(2nd order wind)}, \\
4.86355 -10.2954 {\cal K} + 19.3846 {\cal K}^2 -19.1606 {\cal K}^3 + 7.21573 {\cal K}^4 & \mbox{(4th order wind)}.\\
\end{cases}
\end{equation}
The fits are valid for $0.1 \le {\cal K} \le 0.95$ and only for the deep model of HD 209458b (cf. \S\ref{subsect:hd209458b} and Table \ref{tab:params}).  For example, with the wind scheme being fourth order, we get $t_{\cal K} \approx 3 \times 10^{-3}$ HD 209458b day for the fiducial value of ${\cal K}=0.35$.

It is important to note that the measurements of $t_{\cal K}$ are identical for the G24, G36 and G72 simulations, implying that ${\cal K}$ is strictly speaking the analogue of the dissipation time $t_\nu$ and not the hyperviscosity $\nu$, further implying that comparing spectral and finite difference simulations with different values of ${\cal K}$ and $t_\nu$ (Figure \ref{fig:winds}) is a meaningful exercise.  Within the \texttt{FMS}, specifying ${\cal K}$ constitutes a ``resolution dependent" approach in that the smallest resolvable waves in the simulation are damped with a strength that is equal for every resolution.  Unfortunately, there is no way within the \texttt{FMS} B-grid core (\texttt{Memphis} release) to switch to a ``resolution independent" approach of specifying the magnitude of horizontal mixing, unlike in the case of the spectral core.  It is also not clear how to convert ${\cal K}$ into a quantity that is the true analogue of the hyperviscosity.  The implication is that comparing simulations at different resolutions, with the same value of ${\cal K}$, may not be a fair exercise since the analogue of the hyperviscosity is different in each of these runs (Figure \ref{fig:rm10_hd209458b_res2} and the right panel of Figure \ref{fig:winds2}).

Even if we take Figure \ref{fig:efold} at face value, the correspondence between $t_\nu$ and $t_{\cal K}$ is not straightforward.  For example, the pair of simulations in Figure \ref{fig:rm10_hd209458b_hs_adjusted} use $t_\nu = 10^{-6}$ HD 209458b day and ${\cal K}=0.9$, but Figure \ref{fig:efold} informs us that ${\cal K}=0.9$ corresponds to $t_{\cal K} \approx 4 \times 10^{-4}$ HD 209458b day (using the horiziontal mixing scheme of fourth order in wind).  If we insist on matching the ${\cal K}=0.9$ simulation with the $t_\nu=10^{-5}$--$10^{-4}$ HD 209458 day simulations, then Figure \ref{fig:winds} informs us that we are now faced with the original problem of the predictions for the maximum zonal wind speeds being discrepant.

\label{lastpage}


\begin{thebibliography}{99}

\bibitem[Adcroft et al.(2004)]{ad04} Adcroft, A., Campin, J.-M., Hill, C., \& Marshall, J. \ 2004, Monthly Weather Review, 132, 2845

\bibitem[Anderson et al.(2004)]{anderson04} Anderson, J.L., et al. \ 2004, Journal of Climate, 17, 4641

\bibitem[Asselin(1972)]{ass72} Asselin, R. \ 1972, Monthly Weather Review, 100, 487

\bibitem[Burrows et al.(2010)]{burrows10} Burrows, A., Rauscher, E., Spiegel, D.S., \& Menou, K. \ 2010, ApJ, 719, 341

\bibitem[Cho et al.(2003)]{cho03} Cho, J.Y.-K., Menou, K., Hansen, B.M.S., \& Seager, S. \ 2003, ApJ, 587, L117

\bibitem[Cho et al.(2008)]{cho08} Cho, J.Y.-K., Menou, K., Hansen, B.M.S., \& Seager, S. \ 2008, ApJ, 675, 817

\bibitem[Cooper \& Showman(2005)]{cs05} Cooper, C.S., \& Showman, A.P. \ 2005, ApJ, 629, L45

\bibitem[Cooper \& Showman(2006)]{cs06} Cooper, C.S., \& Showman, A.P. \ 2006, ApJ, 649, 1048

\bibitem[Dobbs-Dixon et al.(2010)]{dd10} Dobbs-Dixon, I., Cumming, A., \& Lin, D.N.C. \ 2010, ApJ, 710, 1395

\bibitem[Gordon \& Stern(1982)]{gs82} Gordon, C.T., \& Stern, W.F. \ 1982, Monthly Weather Review, 110, 625

\bibitem[Held \& Suarez(1994)]{hs94} Held, I.M., \& Suarez, M.J. \ 1994, Bulletin of the American Meteorological Society, 75, 1825

\bibitem[Held(2005)]{held05} Held, I.M. \ 2005, Bulletin of the American Meteorological Society, 86, 1609

\bibitem[Heng \& Spitkovsky(2009)]{hs09} Heng, K., \& Spitkovsky, A. \ 2009, ApJ, 703, 1819

\bibitem[Heng \& Vogt(2010)]{hv10} Heng, K., \& Vogt, S.S., 2010, preprint (arXiv:1010.4719v2)

\bibitem[Holton(2004)]{holton} Holton, J.R. \ 2004, An Introduction to Dynamic Meteorology, 4th edition (Massachusetts: Elsevier)

\bibitem[Iro et al.(2005)]{iro05} Iro, N., B\'{e}zard, B., \& Guillot, T. \ 2005, A\&A, 436, 719

\bibitem[Goodman(2009)]{goodman09} Goodman, J. \ 2009, ApJ, 693, 1645

\bibitem[Kundu \& Cohen(2004)]{kundu04} Kundu, P.K., \& Cohen, I.M. \ 2004, Fluid Dynamics, third edition (San Diego: Elsevier)

\bibitem[Langton \& Laughlin(2008)]{ll08} Langton, J., \& Laughlin, G. \ 2008, ApJ, 674, 1106

\bibitem[Li \& Goodman(2010)]{lg10} Li, J., \& Goodman, J. \ 2010, ApJ, 725, 1146

\bibitem[Longuet-Higgins(1968)]{lh68} Longuet-Higgins, M.S. \ 1968, Phil. Trans. Roy. Soc., 262, 511

\bibitem[Menou et al.(2003)]{menou03} Menou, K., Cho, J.Y.-K., Seager, S., \& Hansen, B.M.S. \ 2003, ApJ, 587, L113

\bibitem[Menou \& Rauscher(2009)]{mr09} Menou, K., \& Rauscher, E. \ 2009, ApJ, 700, 887

\bibitem[Merlis \& Schneider(2010)]{ms10} Merlis, T.M., \& Schneider, T. \ 2010, Journal of Advances in Modeling Earth Systems -- Discussion (JAMES-D), in press (arXiv:1001.5117v1)

\bibitem[Phillips(1957)]{p57} Phillips, N.A. \ 1957, Journal of Atmospheric Sciences, 14, 184

\bibitem[Rauscher \& Menou(2010)]{rm10} Rauscher, E., \& Menou, K. \ 2010, ApJ, 714, 1334

\bibitem[Roeckner \& von Storch(1980)]{rvs80} Roeckner, E., \& von Storch, H. \ 1980, Atmosphere-Ocean, 18, 239

\bibitem[Seager \& Deming(2010)]{sd10} Seager, S., \& Deming, D. \ 2010, In EXOPLANETS, Space Science Series of the University of Arizona Press (Tucson, AZ) (arXiv:1005.4037)

\bibitem[Shapiro(1970)]{shapiro70} Shapiro, R. \ 1970, Reviews of Geophysics and Space Physics, 8, 359

\bibitem[Shapiro(1971)]{shapiro71} Shapiro, R. \ 1971, Journal of Atmospheric Sciences, 28, 523

\bibitem[Showman \& Guillot(2002)]{sg02} Showman, A.P., \& Guillot, T. \ 2002, A\&A, 385, 166

\bibitem[Showman et al.(2008)]{showman08} Showman, A.P., Menou, K., \& Cho, J.Y.-K. \ 2008, Extreme Solar Systems, ASP Conference Series, Vol. 398, proceedings of the conference held 25-29 June, 2007, at Santorini Island, Greece. Edited by D. Fischer, F. A. Rasio, S. E. Thorsett, and A. Wolszczan, p.419 (arXiv:0710.2930)

\bibitem[Showman et al.(2009)]{showman09} Showman, A.P., Fortney, J.J., Lian, Y., Marley, M.S., Freedman, R.S., Knutson, H.A., \& Charbonneau, D. \ 2009, ApJ, 699, 564

\bibitem[Showman et al.(2010)]{showman10} Showman, A.P., Cho, J.Y.-K., \& Menou, K. \ 2010, Space Science Series of the University of Arizona Press (Tucson, AZ) (arXiv:0911.3170)

\bibitem[Simmons \& Burridge(1981)]{sb81} Simmons, A.J., \& Burridge, D.M. \ 1981, Monthly Weather Review, 109, 758

\bibitem[Smagorinsky(1963)]{sma63} Smagorinsky, J. \ 1963, Monthly Weather Review, 91, 99

\bibitem[Smagorinsky(1964)]{sma64} Smagorinsky, J. \ 1964, Quarterly Journal of the Royal Meteorological Society, 90, 1

\bibitem[Snellen et al.(2010)]{snellen10} Snellen, I.A.G., de Kok, R.J., de Mooij, E.J.W., \& Albrecht, S. \ 2010, Nature, in press (arXiv:1006.4364v1)

\bibitem[Stephenson(1994)]{s94} Stephenson, D.B. \ 1994, Q.J.R. Meteorol. Soc., 120, 699

\bibitem[Takacs \& Balgovind(1983)]{tb83} Takacs, L.L., \& Balgovind, R.C. \ 1983, Monthly Weather Review, 111, 2005

\bibitem[Thrastarson \& Cho(2010)]{tc10} Thrastarson, H.Th., \& Cho, J.Y.-K. \ 2010, ApJ, 716, 144

\bibitem[Udry \& Santos(2007)]{us07} Udry, S., \& Santos, N.C. \ 2007, ARA\&A, 45, 397

\bibitem[Vallis(2006)]{vallis06} Vallis, G.K. \ 2006, Atmospheric and Oceanic Fluid Dynamics: Fundamentals and Large-Scale Circulation (New York: Cambridge University Press)

\bibitem[Washington \& Parkinson(2005)]{wp05} Washington, W.M., \& Parkinson, C.L. \ 2005, An Introduction to Three-Dimensional Climate Modeling, second edition (Sausalito: University Science Books)

\bibitem[Wyman(1996)]{wyman96} Wyman, B.L. \ 1996, Monthly Weather Review, 124, 102

\end{thebibliography}
\end{document}